\documentclass[a4paper,11pt]{article}
\pdfoutput=1
\usepackage{amssymb}
\usepackage{textcomp}
\usepackage[utf8]{inputenc}
\usepackage{latexsym}
\usepackage{epstopdf}
\usepackage{epsfig}
\usepackage{graphicx}
\usepackage{amsmath}
\usepackage{multirow}
\usepackage{subfigure}
\usepackage{a4wide}
\usepackage{cite}
\usepackage{rotating}
\usepackage{hyperref}
\newcommand{\be}{\begin{equation}}
\newcommand{\ee}{\end{equation}}
\newcommand{\ba}{\begin{eqnarray}}
\newcommand{\ea}{\end{eqnarray}}
\newcommand{\bac}{\begin{array}{c}}
\newcommand{\eaa}{\end{array}}
\newcommand{\baz}{\begin{array}{cc}}
\newcommand{\mathsym}[1]{{}}

\newcommand{\bad}{\begin{array}{ccc}}

\newcommand{\bi}{\begin{itemize}}
\newcommand{\ei}{\end{itemize}}
\newcommand{\bmt}{\begin{pmatrix}}
\newcommand{\emt}{\end{pmatrix}}
\newcommand{\bt}{\begin{tabular}}
\newcommand{\et}{\end{tabular}}

\newcommand{\benu}{\begin{enumerate}}
\newcommand{\eenu}{\end{enumerate}}

\newcommand{\bav}{\begin{array}{cccc}}
\begin{document}
\title {\large\bf{ Triplet Leptogenesis, Type-II Seesaw Dominance,
    Intrinsic Dark Matter, Vacuum Stability and Proton Decay in
    Minimal SO(10)  Breakings}}

\author{\bf Mainak Chakraborty$^{**}$,  M.K. Parida $^{*}$, Biswonath Sahoo$^{\dagger}$    \\
 Centre of Excellence in Theoretical and Mathematical Sciences, \\
Siksha \textquoteleft O\textquoteright Anusandhan (Deemed to be University),\\ Khandagiri Square, Bhubaneswar 751030, 
Odisha,  India}                                                                                                              
\maketitle

\begin{abstract}
  We implement  type-II seesaw dominance for neutrino mass and baryogenesis through heavy scalar triplet leptogenesis in a class of minimal non-supersymmetric SO(10) models where matter parity as stabilising discrete symmetry as well as WIMP dark matter (DM) candidates are intrinsic predictions of the  GUT symmetry. We also find  modifications of relevant CP-asymmetry formulas in such models. Baryon asymmetry of the universe as solutions of Boltzmann equations is further shown to be realized  for both normal and inverted mass orderings  
 in concordance with cosmological bound  and best fit values of the neutrino oscillation data  including $\theta_{23}$ in the second octant and large values of leptonic Dirac CP-phases. Type-II seesaw dominance is at first successfully implemented in two cases of spontaneous SO(10)  breakings through SU(5) route  where
the presence of only one non-standard Higgs scalar of intermediate mass $\sim 10^9-10^{10}$ GeV  achieves unification. Lower values of the  SU(5) unification scales $\sim 10^{15}$ GeV are predicted to bring proton lifetimes to the accessible ranges of  Super-Kamiokande and Hyper-Kamiokande experiments. Our prediction of WIMP DM relic density in each model is due to a  $\sim {\cal O} (1)$ TeV mass matter-parity odd  real scalar
singlet ($\subset {16}_H \subset$ SO(10)) verifiable by LUX and
XENON1T experiments. This DM is also noted to resolve the vacuum
stability issue of the standard scalar potential. When applied to the
unification framework of M. Frigerio and T. Hambye, in addition to the
minimal fermionic triplet DM solution of $2.7$ TeV,
our mechanism  of type-II seesaw dominance and triplet leptogenesis is
also found to make an alternative prediction of the triplet fermion
plus the real scalar
singlet DM near the TeV scale.
\end{abstract}
\noindent{${}^{**}$email:mainak.chakraborty2@gmail.com}\\ 
\noindent{${}^{*}$email:minaparida@soa.ac.in}\\
\noindent{${}^\dagger$email:sahoobiswonath@gmail.com}

\section{Introduction}\label{sec:intr}
Despite the success of standard model (SM) paradigm based upon  $
SU(2)_L \times U(1)_Y\times SU(3)_C(\equiv G_{213})$  including the evidence of Higgs boson
\cite{Higgsexpt:2012}, the model fails to 
explain several issues like  neutrino oscillation \cite{nudata,Forero:2014bxa,Esteban:2018azc}, baryon asymmetry of
the universe (BAU) \cite{BAUexpt,Planck15}, nature of dark matter (DM)
and its stability \cite{DMexpt,GAMBIT}, the vacuum stability of scalar
potential, and the origin of disparate values of gauge couplings. Out of Grand unified theories (GUTs) \cite{Pati-Salam:1974,su5,so10,so10H,E6}
  suggested to alleviate  issues
 confronting the SM supersymmetric (SUSY) GUTs
\cite{Dimo-Raby-Wil:1981,Marciano-gs:1982,Amaldi:1991} have been considered
to be quite fundamental in understanding the stability of the standard
model vacuum and the origin of three forces of nature.  SUSY SO(10)\cite{so10} 
predicts  R-Parity \cite{RNM:1986,Krauss:1989,Martin:1992} for
stability of dark matter candidates, explains the  origin of
experimentally observed
parity violation in weak interaction
\cite{Pati-Salam:1974,rnmjcp:1975,Lee:1956,cmp-PRL:1984,Rizzo-gs:1981}
, and more interestigly it predicts the combined type-I \cite{type-I,Valle:1980}
plus type-II \cite{type-IIa,type-IIb,type-IIc,type-IId,type-IIe}
 hybrid seesaw formula \cite{Babu-RNM:1993,Akhmedov2}  
 for neutrino mass matrix $m_{\nu}$. Among
a large number of investigations in  neutrino physics\cite{nurev0},
 SUSY GUTs with type-II seesaw
dominance have the special advantage \cite{Altarelli:2011}  that the observed pattern of
quark and lepton mixings at low energies can be explained without using
any flavor symmetry  \cite{Bajc-gs-Vissani:2003,Goh-RNM-Ng:2003} from the fact
that the b-quark and the $\tau$-lepton masses are nearly
equal at the GUT scale \cite{dp:2001}. 

However, in the absence of any experimental evidence of SUSY,
very recently non-SUSY SO(10) has been  in the spot light
because of its inherent and intrinsic capability to predict DM
candidates and  their stability.  In particular, an intrinsic gauged discrete symmetry \cite{Hambye:Rev}, surviving as
 matter parity in SO(10) breakings \cite{Kadastic:dm1,Kadastic:dm2,Frig-Ham:2010,psb:2010,mkp:2011,Ferrari:2018,Mambrini:2013,pnsa:2016,spc:2017} 
\be
Z_{\rm MP}= {(-1)}^{3(B-L)}, \label{eq:MP} 
\ee
has been found to stabilise the non-standard scalars and fermions 
which are also predicted to originate intrinsically from this non-SUSY GUT representations as  possible
DM candidates leading to a number of attractive predictions \cite{Kadastic:dm1,Kadastic:dm2,Frig-Ham:2010,psb:2010,mkp:2011,Ferrari:2018,Mambrini:2013,pnsa:2016,spc:2017}. It is quite
interesting to note that the matter parity conservation down to low energies naturally
emerges whenever the popular Higgs representation ${126}_H$
 is used to break the underlying $U(1)_{B-L}$ gauge symmetry and ${10}_H$ is used to break the
  electroweak symmetry materialising in type-I$\oplus$type-II seesaw
  formula for neutrino masses discussed in Sec.\ref{sec:t2mnu}. So far the SO(10) based matter parity conserving DM models have been largely exploiting the  
type-I seesaw dominance in non-SUSY SO(10)
\cite{Kadastic:dm1,Kadastic:dm2,Frig-Ham:2010,Ferrari:2018,Mambrini:2013,spc:2017}. On
the other hand it
is interesting to note that whenever type-II  dominance occurs in
 a minimal SUSY or non-SUSY SO(10) GUT, it  predicts the following constraint
relation among  mass eigen values $m_i$($M_{N_i}$) of left-handed (right-handed)  neutrinos 
\ba
 m_1:m_2:m_3&::&
  M_{N_1}: M_{N_2}: M_{N_3}. \label{eq:t2con}
\ea
All neutrino mass generation mechanisms are
also expected to respect the cosmological bound \cite{Planck15} $
\sum_i {\hat m}_{i}\le 0.23\,\, {\rm eV}.$

Alternative to RHN decay leptogenesis 
in SM extensions\cite{Fuku-Yana:1986}, interesting models of triplet
leptogenesis have been also proposed
\cite{Ma-Sarkar:1998,Hambye-gs:2003,Hambye:2012,Sierra:2011ab,Sierra:2014tqa} where
the heavy scalar triplet $\Delta_L(3,-1,1)$ decay generates the desired CP-asymmetry that
leads to baryogenesis through sphaleron interaction \cite{sphaleron} near the
electroweak scale. With such  nontrivial constraints  and the triplet 
leptogenesis requirement \cite{Hambye-gs:2003,Hambye:2012,Sierra:2014tqa} that one of
the heavy neutrinos mediating the loop for CP-asymmetry generation should be around the $\Delta_L$ mass,
in the present work we find that the triplet leptogenesis theory  \cite{Hambye-gs:2003,Hambye:2012} is ideally suited
for a general class of three different non-SUSY SO(10)  models. This class of models with SM paradigm, besides
successfully generating BAU, also explains neutrino oscillation data
via type-II seesaw dominance and predicts matter parity stabilised
weakly interacting massive particle (WIMP) \cite{WIMP} dark matter from within the non-SUSY SO(10) consistent with precision unification of gauge couplings leading to verifiable proton decay. These models are further noted to resolve the issue of vacuum stability of the SM Higgs potential.
In the presence of type-II seesaw dominance in  matter parity
conserving SO(10) breakings, we further find  
 modifications of relevant  CP-asymmetry formulas of
\cite{Hambye-gs:2003,Hambye:2012,Sierra:2014tqa} based upon SM extensions.
These objectives have been realized through type-II seesaw dominance induction in a general class of three non-SUSY  SO(10) Models: (i) Model-I with $\kappa (3,0,8)$- unification framework,
(ii) Model-II with $\eta(3,-1/3,6)$- unification framework, and 
(iii) Model-III with Frigerio-Hambye type unification framework
\cite{Frig-Ham:2010}.\\
Our successful realization of heavy triplet leptogenesis consistent
with $\theta_{23}$ in the second octant, large Dirac phases and
other neutrino mixings  in concordance with most recent neutrino
oscillation data \cite{Esteban:2018azc} uncovers a new set of important results
in the area of thermal leptogenesis. This is especially so in view of an
  interesting recent conclusion
that strongly thermal SO(10)-inspired leptogenesis (through type-I seesaw) can be hardly
compatible with atmospheric neutrino mixing angle in the second octant
\cite{Bari:2019,Chianese-Bari:2018}. \\
Highlights of this work are 
\begin{itemize}
\item{Implementation of Type-II seesaw dominance and precision
  coupling unification in a class of three different  non-SUSY SO(10) breakings
   to SM via SU(5) route with only the scalar singlet/fermionic triplet dark
  matter  at the TeV scale verifiable by LUX and XENON1T
  experiments.}
\item{ Precision coupling unification ensured by the
presence of a single nonstandard scalar/fermionic multiplet of
intermediate mass $\sim 10^{9.2}-10^{10.4}$ GeV
  \cite{spc:2017,Frig-Ham:2010} and type-II dominance
  ensured by higher scale insertion of a complete ${15}_H \subset $ 
  SU(5) at $M_{\Delta}$ with much larger SO(10) unification scale,
  $M_{SO(10)}\gg M_{SU(5)} > M_{\Delta}$.}
\item{ Lower values of SU(5) GUT scale, $M_{SU(5)}\sim 10^{15}$ GeV,
  triggering verifiable proton
  decay by Super-Kamiokande and Hyper-Kamiokande search experiments
  \cite{SuperK,HyperK} in all the
  three models.}
\item{ Prediction of a class of new  CP-asymmetry  formulas for
  leptogenesis compared to existing models
  \cite{Hambye-gs:2003,Hambye:2012,Sierra:2014tqa} in minimal matter parity
  conserving SO(10) breakings in the presence of type-II seesaw dominance.}

\item{First successful minimal non-SUSY SO(10) realization of baryogenesis via
heavy scalar triplet decay leptogenesis 
with the $\Delta_L-$ dilepton coupling matrix determined from best fit values
of the neutrino oscillation data in concordance with both normally
ordered (NO) and
invertedly ordered (IO) masses  with $\theta_{23}$ in the second octant including
large Dirac CP-phases.}

\item{These realizations of type-II seesaw and heavy triplet decay thermal leptogenesis 
  with $\theta_{23}$ in the second octant to be  contrasted with  SO(10)-inspired
   strongly thermal RHN decay leptogenesis which are
   hardly compatible with atmospheric mixing angle in the second
  octant \cite{Bari:2019,Chianese-Bari:2018}.} 

\item{Intrinsic matter parity stabilised TeV mass real scalar singlet
  ansatz for both WIMP  DM relic
  density and
   SM scalar potential vacuum stability  verifiable by  LUX and XENON1T.}

\item{Results indicating non-SUSY SO(10) as  self
  sufficient theory for neutrino masses, baryon asymmetry, dark
  matter, vacuum stability of SM scalar potential, origin of three gauge forces, and  observed
  proton stability.} 
\item{All other predictions remaining similar to Model-I and Model-II, two different paths
  for TeV scale DM through Model-III utilizing Frigerio-Hambye type unification framework \cite{Frig-Ham:2010}: 
\begin{enumerate}
\item{Triplet fermionic DM mass same as $m_{\Sigma} \sim 2.7$
TeV in the minimal scenario of \cite{Frig-Ham:2010}, a new heavy Higgs scalar singlet of intermediate mass ($=10^{9}$ GeV)  resolves the vacuum
stability issue by threshold effects on Higgs quartic coupling
\cite{Espinosa:2012}.}\\
\item{A combined DM scenario comprising of both the fermionic triplet
  and real scalar singlet near TeV scale in
  concordance with LUX and XENON1T bounds. In this case the
  non-perturbative Sommerfeld enhancement effect can be dispensed with.}
\end{enumerate}}
\end{itemize} 
 This work is organized in the following manner. 
In Sec.\ref{sec:unifew} we discuss gauge coupling unification and GUT
scale predictions in new models of type-II seesaw
dominance emerging from matter parity conserving minimal non-SUSY
SO(10) breakings. Prediction of the light real scalar singlet DM  is also noted in this section with necessary derivation. 
Proton lifetime predictions including GUT threshold effects have been
discussed in Sec.\ref{sec:taupth}. 
In Sec.\ref{sec:t2mnu} we discuss
 neutrino mass fitting by type-II seesaw dominated formula and derivation of relevant
Yukawa couplings as inputs to leptogenesis. Our predictions on WIMP
DM and resolution of vacuum stability are discussed in
Sec.\ref{sec:relstab}. Solutions to vacuum stability problem in the
triplet fermionic DM model and its modification are discussed in
Sec. \ref{sec:vsfdm}. Baryogenesis via heavy scalar triplet decay is
discussed in Sec.\ref{sec:baulept}. We summarize and conclude in Sec.\ref{sec:sumconc}.

\section{Type-II Seesaw Dominance Models from Non-SUSY SO(10)}\label{sec:unifew}

\subsection{ New TeV Scale Non-SUSY Dark Matter  Models }\label{sec:t2susy}
For type-II seesaw dominance in SUSY SO(10) two essential ingredients
have been suggested \cite{Goh-RNM-Nasri:2004}: 
 (i).  SUSY SO(10) must break into minimal
supersymmetric standard model (MSSM) in two steps: SUSY
SO(10) $\to$ SUSY SU(5) $\to$ MSSM with  the mass scale in the first step being
 heavier enough than the SUSY SU(5) breaking scale:
$M_{SUSY SO(10)} >> M_{SUSY SU(5)} $, (ii). To maintain pre-existing SUSY 
unification, the fine-tuned mass of the full Higgs multiplet
${15}_H \subset$  SU(5)
 containing the LH triplet
$\Delta_L(3,-1,1)$ must be lighter than the SUSY unification scale.
Clearly the implementation needs  MSSM like TeV scale particle spectrum \cite{Marciano-gs:1982} and the
SUSY GUT scale of eq.(\ref{eq:SUSYMU}) derived from RG extrapolation
of the CERN-LEP data \cite{Amaldi:1991}.
\be 
M_{SUSY\, SU(5)} \sim 10^{16} {\rm GeV}, \label{eq:SUSYMU}
\ee
 
In purely non-SUSY SM paradigm, the well known failure of  coupling
unification is restored \cite{mkp:2011,RNM-mkp:2011} through a
 TeV scale particle spectrum  which is the MSSM spectrum
minus the superpartners
\ba
&&\phi^{S}_1(2,1/2, 1),\phi^{S}_2(2,-1/2, 1),
    \phi^{F}_1(2,1/2,1),\Phi_2^{F}(2,-1/2,1),\,\, \nonumber\\
&&F_{\sigma}(3, 0,1),F_{8}(1,0,8) \label{eq:spec}  
\ea
where $\phi^{S}_1(2,1/2, 1),\phi^{S}_2(2,-1/2, 1)$ represent two scalars
of a two-Higgs doublet model. In the  SO(10) context \cite{mkp:2011}
 $\phi^S_{1}\subset {10}_{H}$, $\phi^S_{2}\subset {16}_{H}$, $\phi^{F}_{1,2} \subset
{10}_{F}$ and $(F_{\sigma},F_{8}) \subset {45}_F$
of SO(10). In attributing non-SUSY SU(5) origin of scalar doublet DM and radiative seesaw
\cite{Ma:2006}, a somewhat different TeV scale spectrum has been also suggested \cite{Ma-Suematsu:2008}. In the absence of any  evidence in favour of SUSY \cite{LHCsusy} or MSSM like
TeV spectrum of eq.{\ref{eq:spec}}, in this work we complete
unification for
type-II seesaw dominance through only one non-standard
scalar $\kappa (3,0,8)$ as in Model-I, or $\eta (3,-1/3,6)$ as in
Model-II.
In Model-III we implement type-II seesaw dominance using unification
framework of \cite{Frig-Ham:2010} where a fermionic triplet DM
{$\Sigma_F(3,0,1)$} of mass 2.7 TeV has been suggested. 
 A real scalar
singlet dark matter $\chi_S^R(1,0,1)$ is predicted to complete vacuum
stability in all the three models. Thus, in contrast to the extended particle
spectrum of eq.(\ref{eq:spec}), our TeV scale particles accessible to
LUX and XENON1T are
\be
\chi_S^R(1,0,1)\,\,{\rm OR}\,\, {\Sigma_F(3,0,1)}.\,\,\label{eq:specnew}
\ee    
All the three models also predict  one order lighter unification scale
\be
M_{SU(5)} \simeq 10^{15}\,\, {\rm GeV} \label{eq:newMU}
\ee 
 with corresponding reduction in  proton lifetime 
 accessible to Super-Kamiokande and Hyper-Kamiokande \cite{SuperK,HyperK}. 
  
\subsubsection{General Considerations with Matter Parity}\label{sec:mpgen}
In SM extended theories \cite{GAMBIT} the candidates of DM are
externally added with an ad-hoc assumption on their stabilising
  discrete symmetry. Such theories are naturally expected to be incomplete
in the absence of any fundamental origin of such DM candidates and their stabilising symmetry. 

But in  non-SUSY $SO(10)$ this stabilising
 symmetry
\cite{Hambye:Rev,Kadastic:dm1,Kadastic:dm2} called matter parity
($Z_{MP}$) defined through eq.(\ref{eq:MP}) of Sec.\ref{sec:intr}), automatically
emerges as
intrinsic gauged discrete symmetry whenever the neutral component of
the right-handed (RH)
Higgs triplet $\Delta_R(1, 3, -2, 1) \subset {126}_H$ of SO(10)
is assigned appropriate  VEV to break $SU(2)_R\times U(1)_{B-L}$ gauge
symmetry  leading to the
SM Lagrangian either directly or through intermediate left-right gauge
symmetries\cite{psb:2010,mkp:2011,Ferrari:2018,Mambrini:2013,pnsa:2016}, or through SU(5) route. 
Here the quantum numbers of RH Higgs triplet $\Delta_R(1,3,-2, 1)$
are defined under left-right gauge symmetry $SU(2)_L \times SU(2)_R
\times U(1)_{B-L}\times SU(3)_C$ ($\equiv G_{2213}$)\cite{rnmjcp:1975}.  
 Such breakings also enable to identify the SO(10) representations
with odd value of matter parity $Z_{MP}=-1$ for $16$, $144$,$560$,....but
with even value of $Z_{MP}=+1$ for
$10$, $45$, $54$, $120$, $126$, $210$, ${210}'$,$660$ ..... Then it
turns out that 
the would-be DM scalars are predicted to be in the non-standard scalar
representations ${16}_H,
{144}_H...$ while the DM fermions are predicted  to be in the non-standard fermionic
representations ${10}_F,{45}_F,{54}_F,{120}_F,{126}_F,
{210}_F$..... of even matter parity. All the SM fermions contained in
the spinorial representation ${16}_F$ of SO(10) thus possess odd
matter parity $Z_{MP}=-1$ while the standard model Higgs contained in ${10}_H$
carries even value of $Z_{MP}=+1$.
All the models discussed below rely upon the spontaneous symmetry
breaking pattern
\be
SO(10) \longrightarrow SU(5)\times U(1)_{X} \longrightarrow SU(5) 
\longrightarrow SM. \label{eq:chain1}
\ee
where
\be
X=4 T_{3R}+ 3( B-L). \label{eq:defX}
\ee
In eq.(\ref{eq:defX}) $T_{3R}=$ third component of RH isospin and $B-L=$ baryon minus
lepton number. We will assume the two large mass scales for the first two-steps to
be identical to $M_{SO(10)}$ so that no generality is lost by considering the
effective symmetry breaking chain

\ba
SO(10) \longrightarrow SU(5) \longrightarrow SM. \label{eq:chain2}
\ea
For the sake of utilisation in
 Model-I, Model-II and  Model-III
discussed below we need the following SO(10) branching rules
\cite{Slansky:1979}\\
\par\noindent{\bf SO(10) $\supset$ SU(5)$\times U(1)_X$:}\\
\ba
&&{210}_H=[\Omega= 1_H(0)] + {75}_H (0)+ {24}_H (0)+ ...., \nonumber\\
&&{45}_H= 1_H (0)+ {24}_H(0)++ {10}_H(4)+ {\overline {10}}_H(-4)+...,\nonumber\\ 
&&{126}_H=[\Delta^0_R= 1_H(-10)]+ {\overline {50}}_H(2)+ {15}_H(6)+ ...,\nonumber\\
&&{16}_H= [\chi_S=1_H(-5)] + {10}_H (-1)+ {\overline {5}}_H (3)....,\nonumber\\
&&{10}_H= 5_H(2) + {\overline {5}_H}(-2) + ....\label{eq:10to5}
\ea
The SU(5) singlet $1_H\subset
{126}_H$ noted above is the neutral component of the right-handed (RH) triplet
$\Delta_R(1,3,-2,1)$ under $G_{2213}$  \cite{rnmjcp:1975}. The
left-handed (LH)
scalar triplet $\Delta_L(3,1,-2,1)$ responsible for type-II seesaw and
the RH triplet $\Delta_R(1,3,-2,1)$ that generates RHN masses through
its $U(1)_X$ breaking VEV are noted to originate from the same representation ${126}_H$. 
We further need the following branching rules for SU(5)$\supset$ SM 
\par\noindent{\bf  SU(5) $\supset SU(2)_L\times U(1)_Y\times
  SU(3)_C (=G_{213})$:}
\ba
&&{5}_H=\phi(2,1/2,1)+(1,2/3,3),\nonumber\\
&&{\overline 5}_H=\phi^{\prime}(2, -1/2,1)+(1,-2/3,{\bar 3}),\nonumber\\
&&{24}_H=S_{24}{(1,0,1)}+(3,0,1)+(1,0,8)+(2,-5/6,3) +(2,5/6, {\bar 3}),\nonumber\\    
&&{15}_H=\Delta_L(3,-1, 1)+(2,1/6,3)+(1,2/3,6),\nonumber\\
&&{50}_H=\eta (3, -1/3,6) +.....................,\nonumber\\
&&{75}_H=S_{75}(1,0,1) + \kappa (3, 0, 8)+.........\label{eq:5tosm}  
\ea
where the quantum numbers noted in the parentheses are with respect to
 the SM gauge
group  $G_{213}$. 
We find that either ${210}_H\oplus {126}_H$  or ${45}_H \oplus {126}_H$ can be used
  to break SO(10)$\to$ SU(5)$\to$ SM by preserving matter parity. In
  the last step ${5}_H$ containing the standard Higgs doublet can break 
SM $\to U(1)_{em}\times SU(3)_C$  also without breaking matter parity.
As the minimal SU(5) itself can not unify gauge couplings or resolve
issues like neutrino masses, baryon asymmetry, WIMP DM and vacuum
stability of SM, we discuss below how these problems are addressed in
each of the
three different SO(10) models.     
In all the three models, the first step of symmetry breaking
in eq.(\ref{eq:chain2})
is implemented by the use of the Higgs representations ${210}_H$ and
${126}_H$ to break SO(10) $\to$ SU(5) by assigning VEVs
$\langle {\bf \Omega} \rangle (=V_{\Omega})$ and $ \langle {\bf
  \Delta_R} \rangle (V_{\Delta_R})$ with $V_{\Omega}=V_{\Delta_R}=
V_{B-L} \simeq M_{SO(10)}  \ge 10^{17}$ GeV.

\subsection{ Model-I: $\kappa(3,0,8)$ Assisted Unification and
  Scalar Singlet
 DM}\label{sec:t2kappa}
   In this case the  breaking SU(5)$\to$
SM is realized by the SU(5) Higgs representation ${75}_H \subset
{210}_H$ of SO(10) when a VEV $\langle {\bf S}_{75} \rangle (=V_{S_{75}})
=M_{SU(5)}$ is
assigned in the SM singlet direction.  Finally the breaking of SM to the low energy symmetry
$U(1)_Q\times SU(3)_C$ is implemented by the VEV of the standard Higgs
doublet $\phi(2,1/2,1)\subset {5}_H$ of SU(5) which originates from the scalar representation ${10}_H$ of SO(10).
 Since all the Higgs representations associated with symmetry breaking
 carry even matter parity, this intrinsic matter parity as gauged discrete
symmetry $Z_{MP}$ of eq.(\ref{eq:MP}), remains unbroken  down to low energies.
Unlike the MSSM inspired TeV scale spectrum of eq.(\ref{eq:spec}), essentially needed in the
non-SUSY or split-SUSY type-II seesaw dominance model \cite{RNM-mkp:2011}, the minimal models
belonging to this new class have the same spectrum as the minimal SM
plus a TeV scale DM candidate which may be a fermion or scalar.  

\subsubsection{Intermediate Mass of $\kappa (3,0,8)$}\label{sec:kmass}
 
 We present relevant derivations briefly that allow the scalar
 component $\kappa (3,0,8)$ and the DM scalar $\xi(1,0,1)\subset {16}_H$ to acquire
 desired masses at intermediate and TeV scales, respectively.\\

 At first we note that a mild fine-tuning is needed up to one part in
$10^2$ to keep the  SU(5) representation ${75}_H \subset
{210}_H$  at the SU(5) breaking scale.
For this purpose we consider the SO(10) invariant potential due to
${210}_H$
\be
V_{210}^{(10)}=M_{210}^2{210}_H^2+m_{210}{210}_H^3+\lambda_{210}{210}_H^4, 
\ee
Using the decompositions of this SO(10) Higgs representation under
SU(5) given in eq.(\ref{eq:10to5} and assigning the largest VEV $\langle \Omega
\rangle =V_{\Omega}$ gives  the mass-squared term for ${75}_H$
\be
M_{75}^2=M_{210}^2+m_{210}V_{\Omega}+\lambda_{210}V_{\Omega}^2.\label{eq:M75}
\ee
Thus it is possible to get $M_{75}/M_{210}=10^{-2}-10^{-4}$ by finetuning any
one of the four parameters in eq.(\ref{eq:M75}).
Now using the decomposition of ${75}$ under the standard model from eq.(\ref{eq:5tosm}) and assigning SU(5)
breaking VEV $\langle S_{75} \rangle=V_{S_{75}}$, we get the $(mass)^2$ for
$\kappa(3,0,8)$
\be
M_{\kappa}^2=M_{75}^2+m_{75}V_{S_{75}}+\lambda_{75}V_{S_{75}}^2. \label{eq:MK}
\ee
By fine tuning any one of the four parameters in the above relation it
is possible to get $\kappa(3,0,8)$ at the desired mass scale \cite{kynshi-mkp:1993} . 

\subsubsection{Prediction of a Low-Mass Real Scalar Singlet
Dark Matter}\label{sec:redm}

We consider the SO(10) invariant Higgs potential
\ba
V_{(16,126)}&=&M_{16}^2{16}_H^{\dagger}{16}_H+m_{(16,126)}\left[{126}^{\dagger}_H{16}_H{16}_H+{126}_H{16}_H^{\dagger}{16}_H^{\dagger}\right]
\nonumber\\
&+&\lambda_{(16,126)}{126}_H^{\dagger}{126}_H{16}_H^{\dagger}{16}_H+m_{(16,210)}{210}_H{16}_H^{\dagger}{16}_H+\lambda_{(16,210)}{210}_H^2{16}_H^{\dagger}{16}_H
...\label{eq:v16-126}
\ea
Using the component notation from eq.(\ref{eq:10to5}) and assigning
SO(10) breaking VEVs $\langle \Omega \rangle=V_{\Omega}$, $\langle \Delta_R \rangle=V_{B-L}$, 
\ba
V_{(16,126)}&=&M_{16}^2\chi_S^{\dagger}\chi_S++m_{(16,210)} \Omega\chi_S^{\dagger}\chi_S+
m_{(16,126)}\left[\Delta_R^{\dagger}\chi_S\chi_S+\Delta_R\chi_S^{\dagger}\chi_S^{\dagger}
  \right] \nonumber\\
&+&\lambda_{(16,126)}{\Delta_R}^{\dagger}\Delta_R{\chi_S}^{\dagger}\chi_S
+\lambda_{(16,210)}V_{\Omega}^2{\chi_R}^{\dagger}\chi_S, \label{eq:V16126}
\ea
which leads to the singlet scalar potential for $\chi_S$
\ba 
V_{\chi}&=&\left[M_{16}^2+m_{(16,210)}V_{\Omega}+\lambda_{(16,210)}V_{\Omega}^2  +\lambda_{(16,126)}V_{B-L}^2\right]\chi_S^{\dagger}\chi_S \nonumber\\
&+& m_{(16,126)}V_{B-L}\left[\chi_S^{\dagger}\chi_S^{\dagger}+\chi_S\chi_S\right]. \label{eq:vchi}
\ea
where we have used $\langle \Delta_R\rangle =V_{B-L}\simeq M_{GUT}^{10}$. 
In the expression for  $V_{\chi}$ we decompose the complex scalar singlet $\chi_S$ into its real
and imaginary parts 
\be
\chi_S=\chi_S^R+i\chi_S^I, \label{eq:reim}
\ee
leading to
\ba
 V^{\prime}_{\chi}&=&\left [M_{16}^2 +m_{(16,210)}V_{\Omega}+\lambda_{(16,210)}V_{\Omega}^2+
\lambda_{(16,126)}V_{B-L}^2+ 2 m_{(16,126)}V_{B-L}\right](\chi_S^R)^2  \nonumber\\
&+&\left[ M_{16}^2+m_{(16,210)}V_{\Omega}+\lambda_{(16,210)}V_{\Omega}^2+
\lambda_{(16,126)}V_{B-L}^2-2 m_{(16,126)}V_{B-L}\right](\chi_S^I)^2.  \label{eq:ripot}
\ea
This expression clearly predicts different masses $M_{\rm Re}$ and $M_{\rm Im}$
for the real and the imaginary parts of the scalar singlet $\chi_S$, respectively
\ba
M_{\rm Re}^2&=&M_{S}^2+2 m_{(16,126)}V_{B-L}, \nonumber\\
M_{\rm Im}^2&=&M_{S}^2 -2 m_{(16,126)
}V_{B-L}, \label{eq:rimass}
\ea
where
\be
M_S^2=M_{16}^2+m_{(16,210)}V_{\Omega}+\lambda_{(16,210)}V_{\Omega}^2+
\lambda_{(16,126)}V_{B-L}^2.
\label{eq:ms2}
\ee
We find from eq.(\ref{eq:rimass}) that it predicts the real and imaginary
parts of the same complex scalar singlet to have widely different
masses. Taking the  natural values of the parameters 
\be
M_{16}\sim m_{(16,210)}\sim m_{(16,210)}\sim V_{B-L} \sim V_{\Omega} \sim
M_{GUT}^{(10)},\label{eq:M16etc}
\ee 
 we find from eq.(\ref{eq:ms2}) and eq.(\ref{eq:M16etc}) that $M_S \sim M^{(10)}_{GUT}$. Then it is possible to fine tune the
parameters to
make only one of the two values $M_{\rm Re}$ or $M_{\rm Im}$ to
remain light while the other mass is at the
GUT scale. 

In order to check the applicability of derivation  in the presence of
SU(5), we note that all the scalar interactions after SO(10) breaking
involve SU(5) singlets:
 $\chi_S(1,0,1)\subset {\bf 1}^{(5)}_H\subset
{16}_H^{(10)}$ with $B-L=-1$,$\chi_S^{\dagger}(1,0,1)\subset [{\bf 1}^{(5)}_H]^{\dagger}\subset
[{16}_H^{(10)}]^\dagger$ with $B-L=+1$, $\Delta_R \subset  {\bf 1}^{(5)}_H\subset
{126}_H^{(10)}$ with $B-L=-2$,$\Delta_R^{\dagger}\subset [{\bf 1}^{(5)}_H]^{\dagger}\subset
[{126}_H^{(10)}]^\dagger$ with $B-L=+2$. Then
eq.(\ref{eq:v16-126})-eq.(\ref{eq:M16etc}) hold for these SU(5)
singlets leading to any desired mass of the real part (or the
imaginary part) of the scalar singlet.
  In  Sec.\ref{sec:relstab}
 we have utilized this prediction to accommodate a real scalar
singlet WIMP DM $\xi \equiv \chi_S^I$  by matching the observed relic density. We have further noted that the presence of this real scalar singlet resolves the issue of vacuum stability of the SM scalar potential. 

\subsubsection{Gauge Coupling Unification}\label{sec:kgu}
\begin{figure}[h!]
\begin{center}
\includegraphics[scale=0.8]{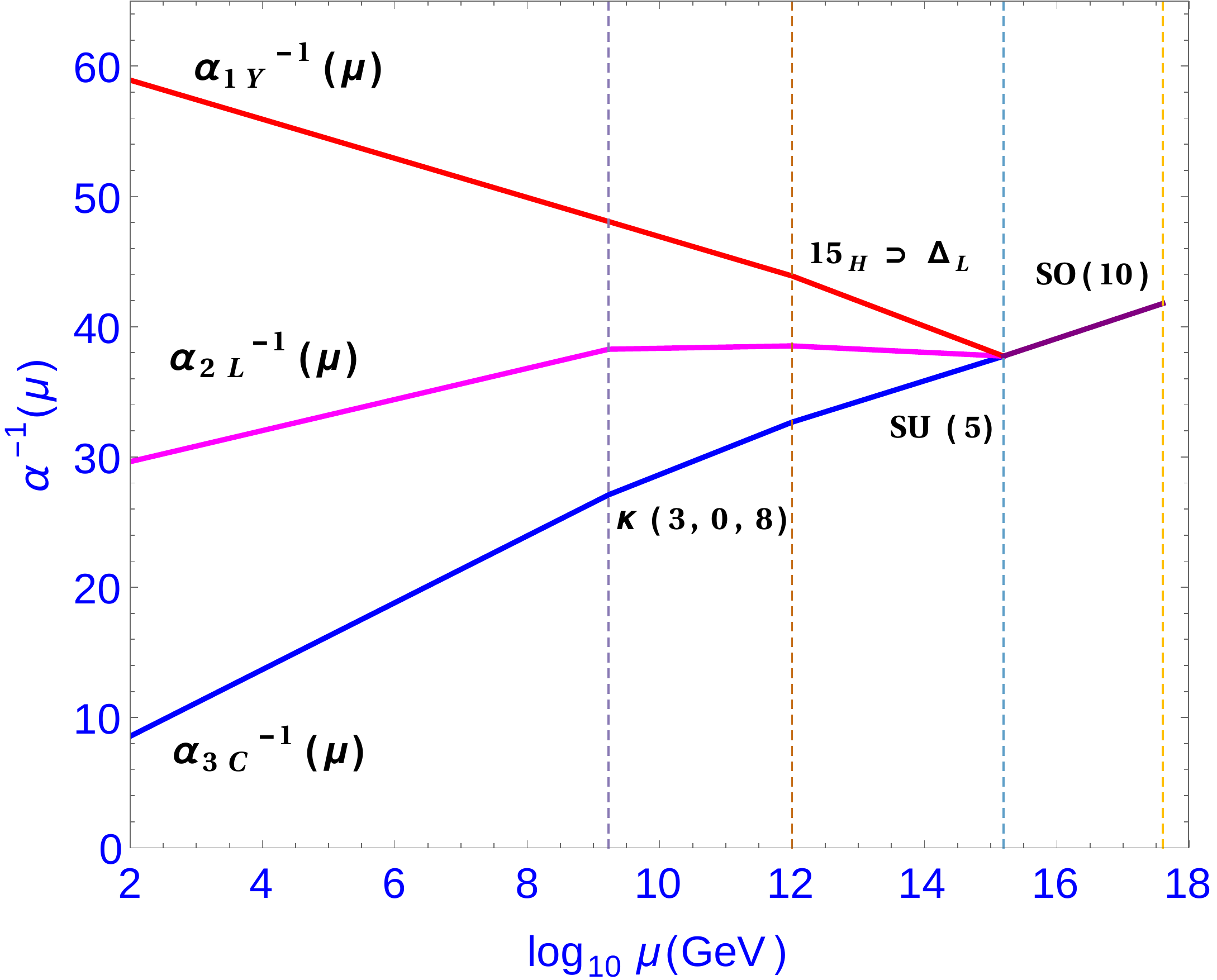}
\end{center}
\caption{ Evolution of gauge couplings with $\kappa (3,0,8)$ at
  $M_{\kappa}=10^{9.2}$ GeV and all the members of ${15}_H$ at  $M_{\Delta_L}=10^{12}$ GeV shown by the first
  and the second vertical dashed lines, respectively. The third
  (fourth) vertical dashed line 
  indicates the SU(5) (SO(10)) unification scale $M_{SU(5)}=10^{15.2}$ GeV ($M_{SO(10)}=10^{17.6}$ GeV). } 
\label{fig:cu}
\end{figure}

The renormalisation group (RG) evolutions \cite{GQW:1974,Jones:1982,Lang:1994} for the three gauge
couplings $ g_{2L}, g_Y$,  and $g_{3C}$ of  $G_{213}$
 have been defined through
eq.(\ref{eq:tl}) and eq.(\ref{eq:RGSM}) in  Sec.\ref{sec:AppTh} of  Appendix.   
Denoting the three  fine structure constants at any mass
scale $\mu\ge M_{Z}$ by
$\alpha_i(\mu)= \frac{ g_i^2(\mu)}{4\pi}(i=1Y,2L,3C)$ evolution of
gauge couplings  is shown in Fig. \ref{fig:cu}. Although
Fig. \ref{fig:cu} has been displayed for one particular choice of
$M_{\Delta_L}=M_{{15}_H}=10^{12}$ GeV, such a type-II seesaw
  dominance model is realizable for any value of
  $M_{\Delta_L}=M_{{15}_H}= 10^{9.3}-10^{15.2}$ GeV.   
 The predicted  mass scales and the GUT fine structure constant in Model-I
 are
\ba
M_{\kappa}&=& 10^{9.2}~~{\rm GeV}, \nonumber\\
M_{\Delta_L}&=&M_{{15}_H}= 10^{12}~~{\rm GeV},\nonumber\\ 
M^0_{U}&=& 10^{15.22} {\rm GeV}, \nonumber\\
\alpha_G^{-1}&=&41.79,\label{minsolk}
\ea

\subsection{ Model-II: $\eta (3,-1/3,6)$ Assisted Unification with
  Scalar Singlet DM}\label{sec:t2eta}  
In this case of Model-II, the popular and  useful representation
${126}_H$ that breaks  SO(10) and $U(1)_X$ with VEV $\langle
\Delta_R^0 \rangle=V_{B-L}=M_{SO(10)}\ge 10^{17}$ GeV also leads to
 matter parity conservation down to low energies. Precision gauge
 coupling unification of the SM gauge couplings are achieved  provided the SM multiplet
$\eta (3,-1/3,6)$ originating from the RH scalar triplet
$\Delta_R(3,1,{\bar {10}})\subset {126}_H$ of SO(10) has an
 intermediate mass $M_{\eta}=10^{10.3}$ GeV. We discuss below how this
 lighter mass of $\eta$ is derived from SO(10) invariant scalar potential. 
\subsubsection{\bf Intermediate Mass of $\eta (3,-1/3,6)$}\label{sec:emass}  
We consider the relevant part of the Higgs potential 
\ba
V_{\kappa}&=&M_{126}^2{126}_H^{\dagger}{126}_H
+m_{126,210}{210}_H{126}_H^{\dagger}{126}_H+\lambda_{(126,210)}{210}_H^2 {126}_H^{\dagger}{126}_H\nonumber\\
&+&\lambda_{(126)}{126}_H^{\dagger}{126}_H{126}_H^{\dagger}{126}_H
\nonumber\\       
&\supset&
V_{\eta}=\left[M_{126}^2+m_{(126,210)}\langle {\bf \Phi}_{(210)}\rangle +\lambda_{(126,210)}\langle
  {\bf \Phi}_{210}\rangle^2 +\lambda_{(126)}V_R^2\right]\eta^{\dagger}{\eta},
\label{eq:etamass}
\ea
where the relevant VEVs have been defined in Sec.\ref{sec:t2kappa}.
Thus, by fine tuning only few of the six quantities $M_{126},
m_{(126,210)},\langle {\bf \Phi}_{(210)}\rangle,V_R, \lambda_{(126)}$ and
$\lambda_{(126,210)}$ occurring in 
eq.(\ref{eq:etamass}), it is possible to achieve an intermediate mass for
$\eta (3,-1/3,6)$.\\ 

\begin{figure}[h!]
\begin{center}
\includegraphics[scale=0.7]{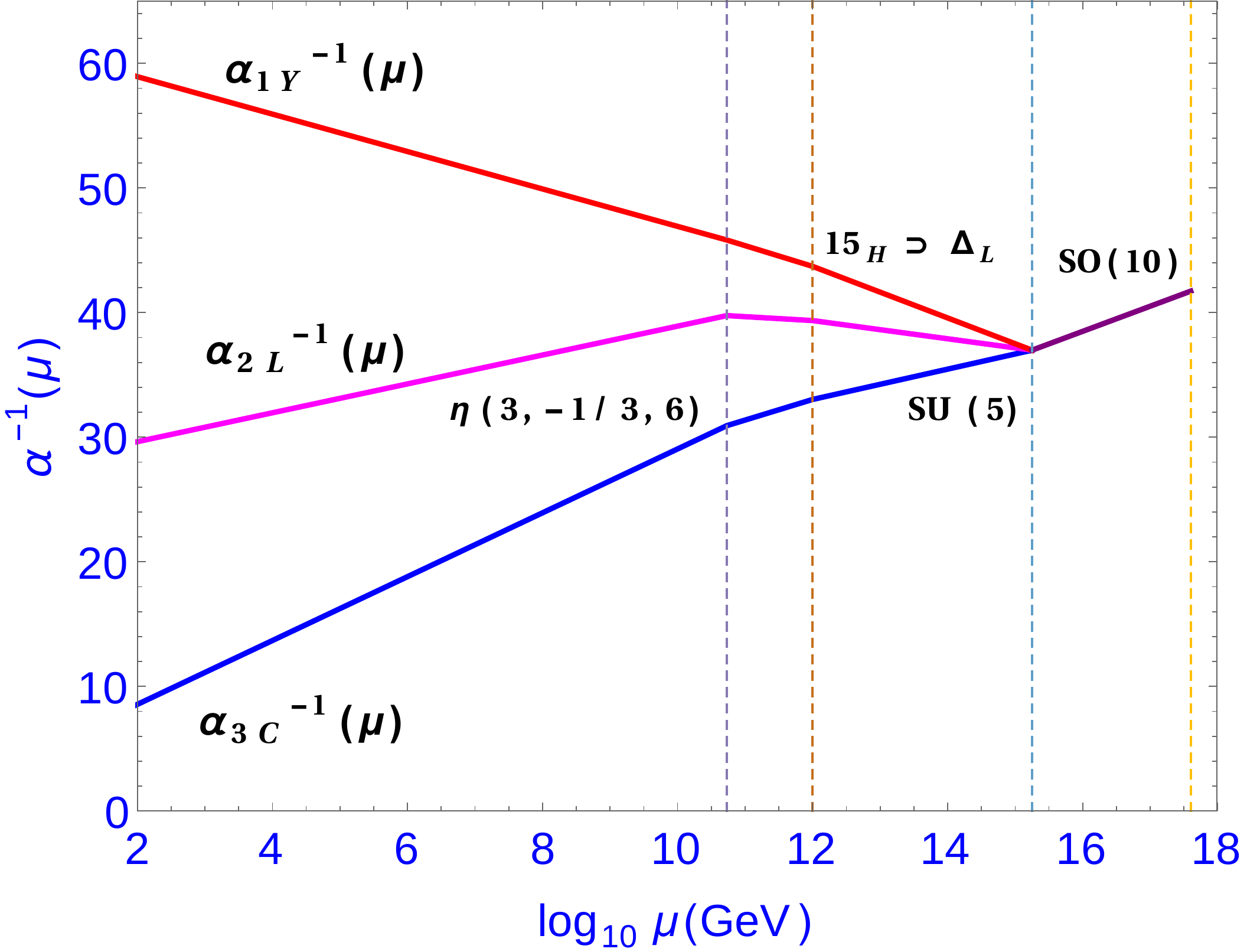}
\end{center}
\caption{ Evolution of gauge couplings with $\eta (3,-1/3,6)$ at
  $M_{\eta}=10^{10.3}$ GeV and all the members of ${15}_H$ at  $M_{\Delta_L}=10^{12}$ GeV shown by the first
  and the second vertical dashed lines, respectively.
 The third
  (fourth) vertical dashed line 
  indicates the SU(5) (SO(10)) unification scale $M_{SU(5)}=10^{15.2}$ GeV ($M_{SO(10)}=10^{17.6}$ GeV). } 
\label{fig:cup}
\end{figure}

We further note that for the first step breaking, SO(10)
$\to$ SU(5),  ${45}_H$ can be also used instead of ${210}_H$ in this
Model-II. Further the well known SU(5) scalar representation ${24}_H$
driving the second step of breaking SU(5) $\to$ SM is  also contained
in ${45}_H$ of SO(10). 
Then we have the following SM symmetric potential after SU(5) breaking through the
VEV  $\langle \Phi_{24}\rangle \subset {24}_H$ of SU(5) 
\ba
V^{(5)}_{\eta}&=&\left[M_{50}^2+m_{(50,24)}\langle {\bf \Phi}_{(24)}\rangle +\lambda_{(50,24)}\langle
  {\bf \Phi}_{(24)}\rangle^2 \right]\eta^{\dagger}{\eta} \nonumber\\
&=& M_{\eta}^2 \eta^{\dagger}{\eta}.
\label{eq:etamass5}
\ea
Thus fine tuning the relevant parameters in eq.(\ref{eq:etamass5}) can
also lead to the desired intermediate mass of $\eta (3,-1/3,6)$.
The RGEs for gauge couplings
\cite{GQW:1974,Jones:1982,Lang:1994}, their solutions including
 the values of beta function coefficients in different ranges of
mass scales are given in  Appendix Sec.\ref{sec:AppTh}.
When this lone submultiplet $\eta (3,-1/3,6)$ in the grand desert has the intermediate mass, unification of
gauge couplings is achieved which has been shown in Fig. \ref{fig:cup}.

The mass scales and the GUT gauge coupling in this case are
\ba
M_{\eta}&=& 10^{10.3}~~{\rm GeV}, \nonumber\\
M_{\Delta_L}&=&M_{{15}_H}= 10^{12}~~{\rm GeV},\nonumber\\ 
M^0_{U}&=& 10^{15.3} {\rm GeV}, \nonumber\\
\alpha_G^{-1}&=&38.39,\label{minsole}
\ea
As in the case of Sec.\ref{sec:t2kappa}, all the derivations leading to
eq.(\ref{eq:v16-126}),  eq.(\ref{eq:vchi}), eq.(\ref{eq:reim}),
eq.(\ref{eq:rimass}), and eq.(\ref{eq:ms2}) are also applicable in this
case. Thus a real scalar DM possessing $Z_{MP}=-1$ and originating from ${16}_H$ of SO(10) is also predicted in this case.   
 
The requirement of type-II seesaw dominance with precision coupling
unification in this model places the common mass of ${15}_H$ heavier
 than  $M_{\eta}$ with $M_{\Delta_L}=M_{{15}_H} \ge 2\times 10^{10} {\rm GeV}$.

\subsection{Model-III via Frigerio-Hambye Type Unification \cite{Frig-Ham:2010}}\label{sec:t2frigham} 

\begin{figure}[h!]
\begin{center}
\includegraphics[scale=0.8]{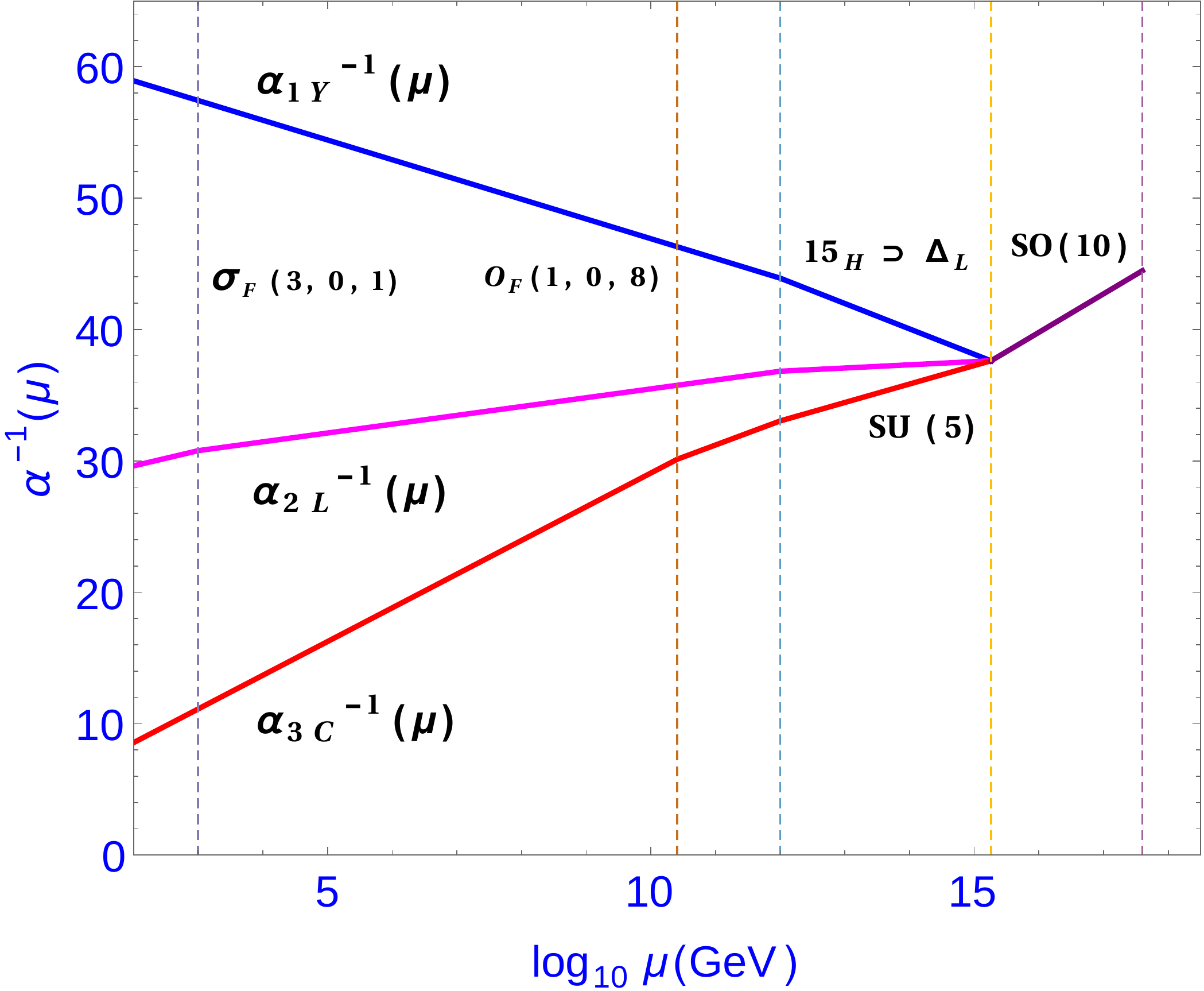}
\end{center}
\caption{ Evolution of gauge couplings 
  through the unification framework \cite{Frig-Ham:2010} with  the
  fermionic DM  $\Sigma_F (3,0,1)$ at
  $m_{\Sigma}=10^{3}$ GeV and fermionic color octet at
  $m_{O_F}=10^{10.4}$ GeV  indicated by the first and the second vertical dashed
  lines, respectively. The mass scales for ${15}_H$ at $10^{12}$ GeV, SU(5)
  unification  at $M_{SU(5)}=10^{15.26}$ GeV, and SO(10) unification 
at $M_{SO(10)}=10^{17.6}$ GeV are indicated by the third, fourth, and
the fifth vertical dashed lines, respectively.} 
\label{fig:ham}
\end{figure}

Recently, Frigerio and Hambye \cite{Frig-Ham:2010} have predicted precision coupling
unification with fermionic color
octet at $\ge 10^{10.3}$ GeV and a 2.7 TeV mass
fermionic triplet DM $\Sigma_F(3,0,1)$  which may have  
direct and indirect
experimental detection possibilities
\cite{Frig-Ham:2010,Hambye:2008,Hryczuk:2014}.
The requirement of type-II seesaw dominance as outlined above with precision
unification in this model also places the lower bound on the triplet Higgs
mass  $M_{\Delta_L}=M_{{15}_H}\ge 10^{10.3}~~{\rm GeV}$.
 The resulting evolution of gauge couplings for
$M_{\Delta}=M_{{15}_H}=10^{12}$ GeV is shown in Fig. \ref{fig:ham}. 

The mass scales and the GUT coupling in this case are
\ba
m_{\Sigma}&=& 10^3~~{\rm GeV}, \nonumber\\ 
m_{O_F}&=& 10^{10.3}~~{\rm GeV}, \nonumber\\
M_{\Delta_L}&=&M_{{15}_H}= 10^{12}~~{\rm GeV},\nonumber\\ 
M^0_{U}&=& M_{U}^5= 10^{15.26} {\rm GeV}, \nonumber\\
\alpha_G^{-1}&=&33.78. \label{minsolf}
\ea

\subsection{Coupling Evolution Above SU(5) Unification Scale}
In each of the three different models discussed above we have
estimated one-loop evolution for the inverse
fine-structure constant using integral form of RG equation,
\be
\frac{1}{\alpha_G^{(i)}(\mu)}=\frac{1}{\alpha_G(M_U)}-\frac{a^{(i)}_5}{2\pi}\ln (\frac{\mu}{M_U})\,\,(i= I, II, III)\,\label{eq:su5evo}
\ee
Including the contributions of three standard fermion generations,
super heavy gauge bosons (V), scalar(H) and non-standard fermion (F) representations, we
have estimated the one-loop value of the SU(5) beta function coefficient
$a^{(i)}_5(i=I,II,III)$ in each of the three models, Model-I,
Model-II, and Model-III   .\\
\par\noindent{Model-I:}\\
\ba
&&{24}_V, 3\times ({10}_F+{\bar 5}_F), {75}_H, {15}_H, {5}_H:\nonumber\\
&&a_5^{(I)}=-14/3. \,\,                                 \label{eq:cI}
\ea
\par\noindent{Model-II:}\\ 
\ba
&&{24}_V, 3\times ({10}_F+{\bar 5}_F), {50}_H, {24}_H, {15}_H, {5}_H:\nonumber\\
&&a_5^{(II)}=-11/2. \,\,                                 \label{eq:cII}
\ea
\par\noindent{Model-III:}\\
\ba
&&{24}_V, 3\times ({10}_F+{\bar 5}_F), {24}_F, {24}_H, {15}_H, {5}_H:\\
&&a_5^{(III)}=-8. \,\,                                 \label{eq:cIII}
\ea
These evolutions have been shown by single curves for respective
unified gauge couplings for $\mu \ge M_U$ in Fig. \ref{fig:cu},
Fig. \ref{fig:cup}, and Fig.\ref{fig:ham} leading to\\
\ba
&&(\frac{1}{\alpha_G})_{SO(10)}=40.2,\,40.19\,\,{\rm and}~~40.45,\label{eq:so10alpha}
\ea
for Model-I,
  Model-II, and Model-III, respectively,  at $M_{SO(10)}\simeq 10^{17}$ GeV.

\section{ Threshold  Effects and Proton Lifetime Predictions} \label{sec:taupth}
\subsection{ Threshold Corrections on the GUT Scale}
Threshold effects due to small log corrections near the GUT scale \cite{Weinberg:1980,Hall:1981,Ovrut:1981,mkp:1987,RNM-mkp:1993,lmpr:1995,Lang:1994} have
been discussed in the Appendix in Sec. \ref{sec:AppTh}.
 In order to evaluate these corrections, at
 first we evaluate the three  matching functions
 $\lambda_i(M_U)(i=1Y,2L,3C)$ representing threshold effects of 
   each model as shown   in Sec.\ref{sec:kgu}, Sec.\ref{sec:emass}, and
 Sec.\ref{sec:t2frigham}
\be
\alpha^{-1}_i (M_U)=\alpha^{-1}_G-\frac {\lambda_i(M_U)}{12\pi},\,\,
(i=1Y,2L,3C). \label{eq:matchi} 
\ee
 Analytic expression for threshold effect on the
unification scale noted as $\Delta [\ln(M_U/M_Z)]$  in
terms of matching functions have been also explicitly derived in
Sec.\ref{sec:AppTh}. Estimation of each matching function in terms of
small logarithmic correction of different superheavy components needs
their beta function coefficients which have been also given in Table
\ref{tab:decomp} and Table \ref{tab:bth}. Then threshold effects on unification 
scale have been explicitly derived  as shown in detail in Sec.\ref{sec:AppTh}
for Model-I, Model-II and Model-III.

Using partial degeneracy approximation under which all superheavy
component masses belonging to the same GUT representation are assumed
to be
identical \cite{RNM-mkp:1993,lmpr:1995}, 
 threshold effects due to each SU(5) representation and their modification to the GUT scale are presented in Table \ref{tab:MUth} for Model I, Model-II and Model III.  
Contributions of different superheavy
scalars (S),fermions (F) and gauge bosons (V), are denoted by subscripts S, F, and V
respectively. These are 
from SU(5) representations ${5}_H,{24}_H,{75}_H,{24}_F$, and
${24}_V$ which have their origins in SO(10) representations
${10}_H,{45}_H,{210}_H,{45}_F$ and ${45}_V$, respectively. In our
notation the small log contribution is proportional to $\eta_i
=\ln(M_i/M^0_U),(i=5,24,50,75,24_F,24_V)$ and
$\eta^{\prime}_X=\log_{10}(M_X/M_U) (X=S,F,V)$. All the superheavy scalar thresholds proportional to $\eta^{\prime}_S=\log_{10} (M_S/M_U^0)$ have been derived by  maximising their contributions.
\begin{table}[!h]
\caption{ Predictions of threshold effects on the SU(5) unification
   scales in  Model-I, Model-II and Model-III defined in the text.} 
\begin{center}
\begin{tabular}{|c|c|c|c|}
\hline
{ } & {Model-I} &{Model-II}& {Model-III}\\
&&&\\
\hline
&&&\\
$ M_U^0$(GeV) & $10^{15.244}$ & $10^{15.280}$ & $10^{15.310}$\\
&&&\\
$\Delta\ln\left({M_U \over M_Z}\right) $ & $= -0.0196078\eta_5$&$=
-0.02024\eta_5$&$= -0.011\eta_5$\\
& $+0.044563\eta_{75}$ &$-0.0445\eta_{24}$&$-0.0549\eta_{24}$ \\
&$+0.9358\eta_V $& $-0.31255\eta_{50} $ & $ -0.1318\eta_{24_F} $\\
&&$+0.9382\eta_V$&$+1.1538\eta_V$\\
&&&\\
\hline
&&&\\
$M_U$(GeV) & $10^{15.244\pm 0.11} \times $ & $10^{15.28\pm 0.133} \times $ & $10^{15.31\pm 0.03} \times $ \\
&&&\\
& $ 10^{\pm 0.0642\eta^{\prime}_S\pm 0.9358\eta^{\prime}_{V}}$ & $10^{\pm 0.3773\eta^{\prime}_S\pm 0.9352\eta^{\prime}_{V}}$
& $ 10^{\pm 0.0659\eta^{\prime}_S\pm 0.13186\eta^{\prime}_F\pm 1.1538\eta^{\prime}_{V}}$\\
&&&\\
\hline
$ \alpha_G^{-1}$ & $41.77 $ & $ 38.39 $ & $33.78$ \\
&&&\\
\hline
\end{tabular}
\end{center}
\label{tab:MUth}
\end{table}

\subsection{Proton Lifetime Predictions}\label{sec:taup}
Currently, experimentally measured  lower limits on the proton  life
time for the decay modes $p\to e^+\pi^0$  and $p\to \mu^+\pi^0)$ are\cite{SuperK}
\begin{eqnarray}
&&\tau_p^{expt.}(p\to e^+\pi^0)~\ge ~1.67\times 10^{34}~~{\rm yrs},\nonumber\\
&&\tau_p^{expt.}(p\to \mu^+\pi^0)~\ge ~7.7\times 10^{33}~~{\rm yrs}.  \label{taupexpt}       
\end{eqnarray}

 Including strong and electroweak renormalization
effects on the ${\rm d}=6$ operator and taking into account quark mixing, chiral symmetry breaking
effects, and lattice gauge theory estimations, the decay rates
are \cite{Wise:1980,Buras:1978,Babu-Pati:2010},
\be  
\Gamma(p\rightarrow e^+\pi^0)
=(\frac{m_p}{64\pi f_{\pi}^2}
\frac{{g_G}^4}{{M_{X^0}}^4})|A_L|^2|\bar{\alpha_H}|^2(1+D'+F)^2\times R,
\label{eq:width}
\ee
where $ R=[A_{SR}^2+A_{SL}^2 (1+ |{V_{ud}}|^2)^2]$ for $SU(5)$, $V_{ud}=0.974=$ 
 the  $(1,1)$ element of $V_{CKM}$ for quark mixings, and
$A_{SL}(A_{SR})$ is the short-distance renormalization factor in the
left (right) sectors.  In eq.(\ref{eq:width}) $A_L=1.25=$
long distance renormalization factor. But the short distance
renormalisation factor  
$A_{SL}\simeq A_{SR}$ has been estimated for each model in the Appendix.
 These are  estimated by
evolving the ${\rm dim.} 6$ operator for proton decay by using the
anomalous dimensions \cite{Wise:1980,Buras:1978,Babu-Pati:2010,Bajc,Nath:2007,Langacker:1981} and the beta function
coefficients for gauge couplings of each  model. In eq.(\ref{eq:width})
 $\bar\alpha_H =$
hadronic matrix elements, $m_p =$proton mass
$=938.3$ MeV, $f_{\pi}=$ pion decay 
constant $=139$ MeV, and the chiral Lagrangian parameters are $D=0.81$ and
$F=0.47$. Here $\alpha_H= {\bar{\alpha_H}}(1+D'+F)$ which
has been estimated from 
lattice
gauge theory computations  \cite{Aoki:2007,Aoki:2014}. We have used the
hadronic matrix element  \cite{Aoki:2014}
\be
\frac{{\bar{\alpha}}_H(1+D^{\prime}+F)}{{\sqrt 2} f_{\pi}}=\langle \pi^0
|(ud)_R u_L|p\rangle=-0.103(23)(34)\,\, {\rm GeV^2},  \label{hadm}
\ee
where the first (second) uncertainty is statistical (systematic). 
We have estimated values of $A_{SR}$ for different models in 
 Sec.\ref{sec:dim6} of Appendix. The corresponding numerical values
 are given in Table \ref{tab:SRLR}.
 In both the eq.(\ref{eq:width}) and
eq.(\ref{eq:taup}) $M_{X^0}$ stands for the degenerate mass of superheavy $X^{\pm
  4/3}$ and $Y^{\pm 1/3}$ gauge bosons mediating proton decay. Then the expression for the
 inverse
decay rate or proton lifetime is 
\begin{equation}
\Gamma^{-1}(p\rightarrow e^+\pi^0) 
 =
  \frac{4}{\pi}\frac{f_{\pi}^2}{m_p}\frac{M_{X^0}^4}{\alpha_G^2}\frac{1}{\alpha_H^2
    A_R^2}\frac{1}{F_q},\label{eq:taup}
\end{equation}
where the GUT-fine structure constant $\alpha_G=0.0263$ and the
factor  $F_q=(1+(1+|V_{ud}|^2)^2)\simeq 4.8$. 
This formula has 
the same form as given in \cite{Babu-Pati:2010} which has been
modified here for the SU(5) case.\\
Using the analytic formula of eq.(\ref{eq:taup}) and the threshold
corrected GUT scales presented in Table \ref{tab:MUth} we now present
proton lifetime predictions analytically for the three models using
$M_{X^0}=M_U$ for which threshold effects have been given Table \ref{tab:MUth}.

\par\noindent{\bf Model I}\\ 
\begin{eqnarray}
\tau_P\simeq 10^{33.24\pm 0.44 \pm 0.256 \eta^{\prime}_S\pm
    3.743\eta^{\prime}_V}~~{\rm yrs},\label{eq:tau1}
\end{eqnarray}
\par\noindent{\bf Model II}\\
\begin{eqnarray}
&&\tau_p\simeq  10^{33.11\pm 0.5335 \pm
    1.51\eta^{\prime}_S\pm 3.74\eta^{\prime}_V}~~{\rm yrs},\label{eq:tau2}
\end{eqnarray}
\par\noindent{\bf Model III}\\
\begin{eqnarray}
&&\tau_p\simeq  10^{33.42\pm 0.120 \pm
    0.2636\eta^{\prime}_S\pm 0.5272 \eta^{\prime}_F \pm 4.610\eta^{\prime}_V}~~{\rm
    yrs},\label{eq:tau3}
\end{eqnarray}
where $\eta^{\prime}_X =\log_{10}(M_X/M_U) (X=S,F,V)$.
Despite the superheavy gauge boson contributions being from the same
SM multiplets $(2,5/3,3)_V\oplus (2,-5/3, {\bar 3})_V \subset
{24}_V\subset SU(5)$ in all the three models, their threshold
corrections to proton lifetime are significantly different because of
differing renormalization group effects in  Model I, Model II and Model
III.  

Numerical estimations on proton lifetime  for Model-I are shown in Table
\ref{tab:taukdeg1} for different splitting factors of superheavy
masses. In all the three models, when threshold and coupling constant
uncertainties are absent, the predicted $\tau_P$  corresponds to the
minimum value of the hadronic matrix element in eq.(\ref{hadm}) \cite{Aoki:2014}.
\begin{table}[h!]
\caption{ Numerical upper limits on predicted proton lifetime in Model-I as a function of superheavy scalar (S) and gauge boson
  (V) mass splittings as defined in the text and Appendix. The factor
  $10^{\pm 0.44}$ is due to
  uncertainty in input parameters of eq.(\ref{eq:inputpara}).}
\centering
\begin{tabular} { |p{1.0 cm} |p{1.0 cm }|p{2.5 cm}|p{1.0 cm }|p{1.0
      cm}|p{2.5 cm}|}
 \hline
${M_S \over M_U} $ & ${M_V \over M_U}$& $\tau_P (yrs)$ & ${M_S \over M_U} $
 & ${M_V \over M_U} $ & $\tau_P (yrs)$\\  \hline
$10$ & $2$ & $4.19\times 10^{34\pm 0.44}$ & $3$ & $2$ & $3.08\times 10^{34\pm 0.44}$ \\
\hline
 $1$ & $2$ & $2.32\times10^{34\pm 0.44}$ & $1$ & $3$ & $1.06\times 10^{35\pm 0.44}$ \\ \hline
\end{tabular} 
\label{tab:taukdeg1}
\end{table}

Numerical values of  proton lifetime  predictions for Model-II for different
superheavy masses are presented in Table \ref{tab:taupdeg2}.

\begin{table}[h!]
\caption{Numerical upper limits on predicted proton lifetime in Model-II as a function of
  superheavy scalar (S) and gauge boson (V) mass splittings  defined
  in the text and Appendix. The factor $10^{\pm 0.53}$  is due to
  uncertainty in input parameters of eq.(\ref{eq:inputpara}).}
\centering
\begin{tabular} {|p{1.5 cm} |p{1.5 cm }|p{2.7 cm}|p{1.5 cm }|p{1.5 cm}|p{2.7 cm}|}  
 \hline
 ${M_S \over M_U} $ & ${M_V \over M_U}$& $\tau_P (yrs)$ & ${M_S \over M_U} $
 & ${M_V \over M_U} $ & $\tau_P (yrs)$   \\ \hline
 $10$ & $2$ & $4.1\times 10^{34\pm 0.53}$ & $1$ & $4$ & $2.29\times 10^{35\pm 0.53} $ 
\\ \hline
 $1$ & $3$ & $7.84\times 10^{34\pm 0.53} $ & $4$ & $3$ & $1.32\times10^{35\pm 0.53}$
 \\ \hline
\end{tabular} 
\label{tab:taupdeg2}
\end{table}

For the type-II seesaw induced Model-III  numerical values  of proton
lifetime predictions are presented in Table \ref{tab:taupdegf1} excluding
superheavy fermion effects, and in Table \ref{tab:taupdegf2} including
these fermion effects.
\begin{table}[h!]
\caption{ Numerical upper limits on predicted proton lifetime in Model-III as a function of superheavy scalar (S) and gauge boson(V) mass splittings as defined in the text and assuming superheavy fermions at unification scale. The factor $10^{\pm 0.12}$ is due to
  uncertainty in input parameters of eq.(\ref{eq:inputpara}).}
\centering
\begin{tabular} {|p{1.5 cm} |p{1.5 cm }|p{2.7 cm}|p{1.5 cm }|p{1.5 cm}|p{2.7 cm}|}  
 \hline
 ${M_S \over M_U} $ & ${M_V \over M_U}$& $\tau_P (yrs)$ & ${M_S \over M_U} $
 & ${M_V \over M_U} $ & $\tau_P (yrs)$   \\ \hline
 $10$ & $2$ & $1.17\times 10^{35\pm 0.12}$ & $1$ & $3$ & $4.16\times 10^{35\pm 0.12}$ 
\\ \hline
 $3$ & $2$ & $8.57\times 10^{34\pm 0.12}$ & $3$ & $3$ & $5.55\times 10^{35\pm 0.12}$
 \\ \hline
 $1$ & $2$ & $6.42\times 10^{34\pm 0.12}$ & $2$ & $2$ & $7.7\times 10^{34\pm 0.12}$
 \\ \hline
 \end{tabular} 
\label{tab:taupdegf1}
\end{table}

\begin{table}[h!]
\caption{Numerical upper limits on predicted proton lifetime in
  Model-III as a function of superheavy fermion (F) and gauge boson(V)
  mass splittings as defined in the text assuming superheavy scalars
  at unification scale. The factor $10^{\pm 0.12}$ is due to
  uncertainty in input parameters of eq.(\ref{eq:inputpara}).}
\centering
\begin{tabular} {|p{1.5 cm} |p{1.5 cm }|p{2.7 cm}|p{1.5 cm }|p{1.5 cm}|p{2.7 cm}|}  
 \hline
 ${M_F \over M_U} $ & ${M_V \over M_U}$& $\tau_P (yrs)$ & ${M_F \over M_U} $
 & ${M_V \over M_U} $ & $\tau_P (yrs)$   \\ \hline
 $10$ & $2$ & $2.16\times 10^{35\pm 0.12}$ & $2$ & $2$ & $9.25\times 10^{35\pm 0.12}$ 
\\ \hline
 $6$ & $2$ & $1.65\times 10^{35\pm 0.12}$ & $1$ & $2$ & $6.42\times 10^{34\pm 0.12}$
 \\ \hline
 $3$ & $2$ & $1.14\times 10^{35\pm 0.12}$ & $25$ & $1$ & $1.43\times 10^{34\pm 0.12}$
 \\ \hline
 \end{tabular} 
\label{tab:taupdegf2}
\end{table}
On proton lifetime predictions the following distinguishing features
are noted: Although the uncertainty due to input parameters  is lowest 
,  the threshold uncertainty  due to superheavy gauge bosons
($X^{\pm 4/3},Y^{\pm 1/3}$) is nearly one order larger in
Model-III that uses Frigerio-Hambye unification framework \cite{Frig-Ham:2010}.
It is quite interesting to note that for natural values of these
superheavy gauge boson
masses only few times
heavier or lighter than the respective GUT scales, the predicted
proton lifetimes for $p\to e^+\pi^0$ are accessible to ongoing
searches at Super-Kamiokande and Hyper-Kamiokande \cite{SuperK,HyperK}.  
\section{Type-II Seesaw Fit to  Neutrino Oscillation
  Data}\label{sec:t2mnu}
A hall mark of type-II seesaw dominance is its capability to fit 
neutrino oscillation data with high precision for all values of
mixings and phases.  

\subsection{ Type-II Seesaw Dominance in SO(10) from General Seesaw}
Suppressing generation indices,  the SO(10) invariant Yukawa Lagrangian 
\be
{\cal  L}_{yuk}=Y{16}_F{16}_F{126}_H^{\dagger}+\lambda {16}_F{16}_F{10}_H ,\label{yuk10} 
\ee
predicts the same Majorana Yukawa  coupling Y of LH scalar triplet
$\Delta_L$ to LH
dileptons and the RH triplet scalar $\Delta_R$ to RHNs. This is due to
the fact that both the LH and the RH scalar triplets are contained in 
${126}_H$. After SO(10)
breaking accompanied by $U(1)_X$ breaking by $\langle \Delta_R \rangle
=V_{B-L}\ge 10^{17}$ GeV, the heavy RHN mass matrix is predicted
\be
M_N=YV_{B-L}.                         \label{MN}
\ee     
Below the SU(5) breaking scale, we have  SM gauge  theory with 
 three additional heavy right-handed neutrinos
(RHNs)  $N_{i}, (i=1,2,3)$ and a heavy Higgs scalar triplet
$\Delta_L (3, -1, 1)\subset {15}_H$. 
The Yukawa Lagrangian encoding type-I+type-II seesaw  can be presented as
\begin{equation}
\mathcal{L}=\mathcal{L}^{(I)}+\mathcal{L}^{(II)}
\end{equation}
where $\mathcal{L}^{(I)}$ and $\mathcal{L}^{(II)}$ are the additional
terms arising due to right handed neutrinos and scalar triplet,
respectively. Specifically they are  given by
\begin{eqnarray}
&&-\mathcal{L}^{(I)}=\overline{N_i}\lambda_{ij}\tilde{\phi}^\dagger l_{L_j} +\frac{1}{2}\overline{N_i} Y_{ij} C \overline{N_j}^T V_{B-L}+h.c ~~,\label{l1}\\
&&-\mathcal{L}^{(II)}=l_{L_i}^T C i \tau_2 Y_{ij}  (\frac{\vec{\tau}.\vec{\Delta_L}}{\sqrt{2}}) l_{L_j}+M_{\Delta}^2 Tr [\Delta_L^\dagger \Delta_L] +
\mu_\Delta
\tilde{\phi}^\dagger(\frac{\vec{\tau}.\vec{\Delta_L}}{\sqrt{2}})^\dagger
\phi +h.c, \label{l2}
\end{eqnarray}
where $l_{L_i}^T=(\nu_{L_i},~e_{L_i})$~($i$ is the lepton generation
index) and the standard Higgs doublet components
$\phi^T=(\phi^+,\phi^0)\subset {10}_H$ of SO(10).
Here $\tilde{\phi}=i\tau_2\phi^\ast$, $\vec{\tau}=(\tau_1,\tau_2,\tau_3)$ ($\tau_i$ are the $2\times2$ Pauli spin matrices). Similarly the scalar triplet
$\Delta_L$ in the adjoint representation of $SU(2)_L$ is expressed as $\vec{\Delta_L}=(\Delta_L^1,\Delta_L^2,\Delta_L^3)$. The Yukawa couplings $Y$ and $\lambda$
are both $3\times3$ matrices in flavour space and $C$ is the charge
conjugation matrix. Thus the term
$(\frac{\vec{\tau}.\vec{\Delta_L}}{\sqrt{2}})$ can be written explicitly as
\be
(\frac{\vec{\tau}.\vec{\Delta_L}}{\sqrt{2}})
                                            =\left( \begin{array}{cc}
     \frac{\Delta^+}{\sqrt{2}} & \Delta^{++} \\ \Delta^0 & -\frac{\Delta^+}{\sqrt{2}}  \\ 
    \end{array}\right)_L
\ee                              
where $\Delta_L^0=\frac{1}{\sqrt{2}}(\Delta_L^1+i\Delta_L^2),~~\Delta_L^+=\Delta_L^3,~~\Delta_L^{++}= \frac{1}{\sqrt{2}}(\Delta_L^1-i\Delta_L^2)$.
Compared to SM extended theory with added $\Delta_L$+RHNs
\cite{Hambye-gs:2003,Sierra:2014tqa}, these Lagrangians have the
following  distinctions. The RHN mass of SO(10) breaking origin
$M_N=YV_{B-L}$
  appearing
in the second term of eq.(\ref{l1}) has the same Majorana coupling
matrix $Y$ that also defines the type-II seesaw term
$m_{\nu}^{II}$ of eq.(\ref{t2app}) originating from first term of
eq.(\ref{l2}). As a result, the freedom of choosing a RHN diagonal basis
\cite{Hambye-gs:2003,Sierra:2014tqa} is lost these SO(10) models.
On the other hand the same unitary  
PMNS matrix $U$ that diagonalises neutrino mass matrix under type-II
dominance approximation  also diagonalises $M_N$. This has been shown below 
to lead to an inevitable  transformation on the Dirac neutrino Yukawa coupling
$\lambda \to \lambda U^\ast$ and, therefore, to  new class of CP-asymmetry formulas for leptogenesis. Another important difference
from SM extension is that the Dirac neutrino Yukawa coupling matrix $\lambda$ 
is predicted to be known from SO(10) symmetry.
 Further in the SM extension the
trilinear coupling $\mu_{\Delta}$ in the third term of eq.(\ref{l2}) is a lepton number 
conserving bare mass term. On the other hand in these SO(10) breaking
models, the $U(1)_X$ breaking VEV predicts 
$\mu_{\Delta}$ as the product of a quartic coupling and $V_{B-L}$ 
\ba
V_{(126,10)}&=&\lambda_{(126,10)}{126}_H^{\dagger}{126}_H{10}_H{10}_H 
\supset  \mu_{\Delta}\Delta_L\phi\phi+ h.c., \nonumber\\
\mu_{\Delta}&=&\lambda_{(126,10)}V_{B-L}. \label{v12610}
\ea
As shown below in Fig.\ref{feyn1}, this interaction is responsible for the second vertex generating type-II seesaw contribution.


For the type-II seesaw mass term the Feynman diagram is shown in
Fig.\ref{feyn1}. 
\begin{figure}[h!]
\begin{center}
\includegraphics[width=6cm,height=5cm,angle=0]{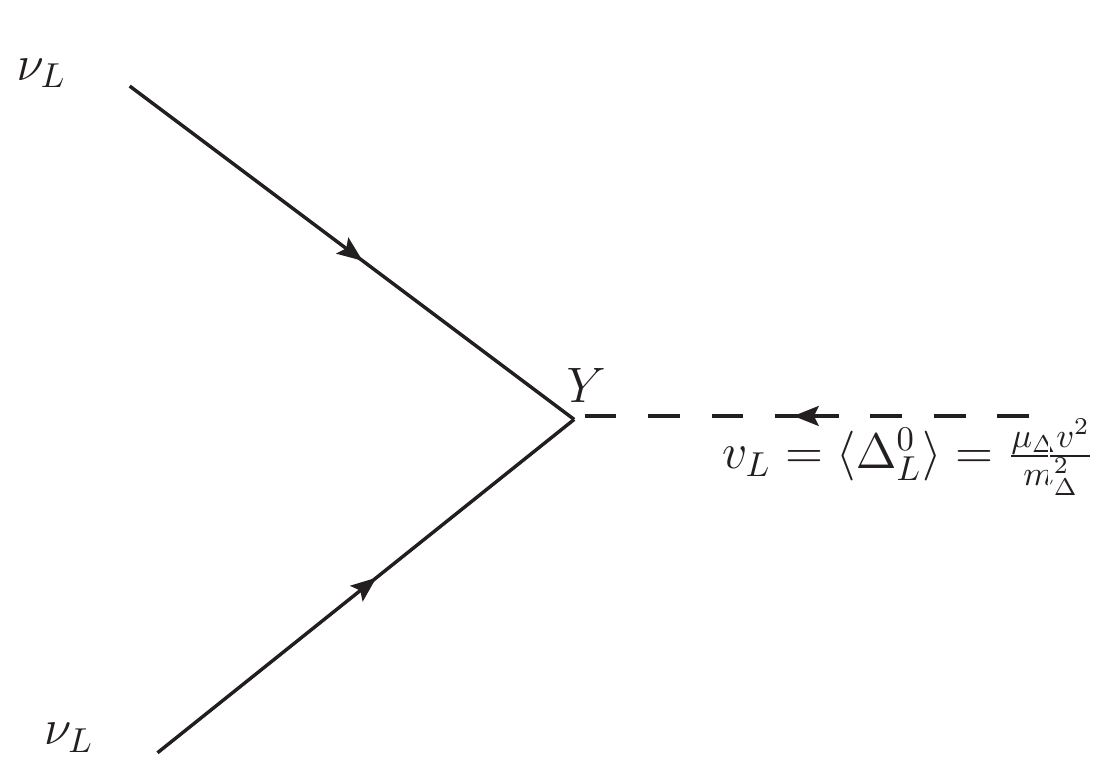}
\hspace{1.0cm}
\includegraphics[width=6cm,height=5cm,angle=0]{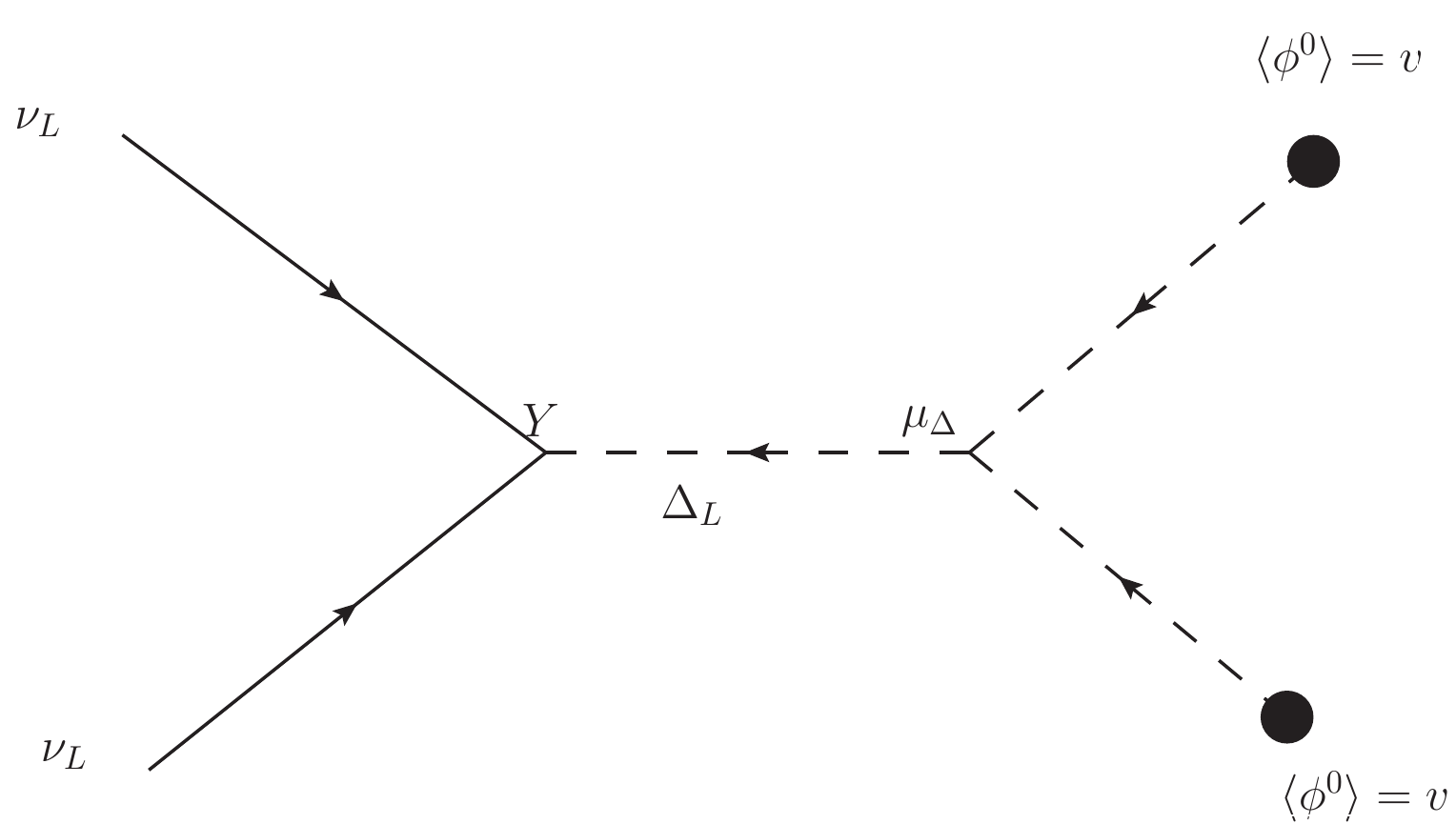}
\caption{Schematic representation of generation of Type-II seesaw term
  comprised of first and third terms of eq.(\ref{l2})(left panel). The
  same quantity after $\Delta_L$ gets  induced VEV in the SM extension (right panel). }
\label{feyn1}
\end{center}
\end{figure}

The SO(10) predicted interaction potential of eq.(\ref{v12610})
in conjunction with $Y\Delta_L l^c l^c$ Yukawa term leads to the 
 induced vacuum expectation value of $\Delta_L$  whenever the
SM doublet $\phi\subset {10}_H$ acquires electroweak VEV $v=246$ GeV,
\be
v_L=\frac{\mu_{\Delta} v^2}{2 M_\Delta^2}, \label{vl1}
\ee
leading to the type-II seesaw formula
\begin{equation}
m_\nu^{II}=Yv_L \label{t2} .
\end{equation}
 
In the SO(10) case $\mu_{\Delta}=\lambda_{(126,10)} V_{B-L}$ which originates from the scalar interaction term $\lambda_{(126,10)}{126}_{\rm H}^{\dagger}{126}_H{10}_H{10}_H$ and the generalised
expression for the induced VEV is 
\be
v_L=\frac {\lambda_{(126,10)} V_{B-L}v^2}{2M_{\Delta}^2}. \label{vl2}
\ee
However when the self energy correction to the triplet mass is taken as
$\lambda_{(126,10)} V_{B-L} \simeq M_{\Delta}$ leading to a simpler form of the
induced VEV
\be
v_L=\frac {v^2}{2M_{\Delta}^2}.           \label{vl3}
\ee
Thus the general form of seesaw formula predicted by SO(10) is
 \begin{eqnarray}
m_\nu & = & m_\nu^{II} +m_\nu^{I} =  Yv_L -M_D^T M_N^{-1} M_D \nonumber\\ 
  & = & Y\mu_{\Delta}v^2/{(2M_{\Delta}^2)} -\frac{v^2}{2V_{B-L}}\lambda \frac{1}{Y}\lambda^T.  \label{mnu_full} 
\end{eqnarray}
It is crucial to note that the same Majorana coupling matrix $Y$ occurs in
the type-II as well as type-I seesaw term which is a general prediction
of any SO(10) with Higgs representations ${126}_H$ and ${10}_H$. 

In comparison the general seesaw formula resulting from SM extension
with $\Delta_L$+RHNs is \cite{Sierra:2014tqa} 
\ba
m_\nu & = & {\cal M}_\Delta^{\nu} +{\cal M}_{N}^{\nu} =  Yv_L -M_D^T
M_N^{-1} M_D \nonumber\\ 
 & = & Y\mu_{\Delta}v^2/{(2M_{\Delta}^2)}-\frac{v^2}{2}\lambda
\frac{1}{M_N}\lambda^T \label{gensm}
\ea
where the Majorana coupling $Y$ occurs only in the type-II seesaw term.
Thus, in general, the  active neutrino mass matrix consists of both Type-I and Type-II seesaw terms.
Depending upon the magnitude of the two terms, both the contributions
are included as in the case of  hybrid seesaw
\cite{Akhmedov2,pnsa:2016}. In  
the specifically designed SO(10) breakings, the models  allow Type-II seesaw
dominance \cite{Goh-RNM-Nasri:2004,RNM-mkp:2011}.
It is clear from eq.(\ref{mnu_full}) that type-II seesaw dominance
 occurs when 
\be
M_{\Delta} \ll M_N,\,\, {\rm or}\,\, M_{\Delta}\ll V_{B-L} \label{t2cond1} 
\ee
In  the three models discussed in Sec.\ref{sec:unifew}, 
theoretically allowed values of $M_{\Delta}$ consistent with precision
gauge coupling
unification are $M_{\Delta}=10^{9.3}-10^{15}$ GeV in Model-I and
$M_{\Delta}=10^{10.3}-10^{15}$ GeV in both Model-II and Model-III whereas
in all the three models $V_{B-L}\ge 10^{17}$ GeV.  Thus the condition
of type-II seesaw dominance is satisfied to an excellent
approximation. We, therefore, use the type-II dominance approximation
\be
m_{\nu}= m_{\nu}^{II}=Yv_L,      \label{t2app}
\ee
 where $v_L$ is given in eq.(\ref{vl1}), to parametrise neutrino data
 as discussed in the next Sec.\ref{sec:osc}.

\subsection{ Neutrino Mass Matrix from Oscillation
  Data}\label{sec:osc}
Using standard parametrisation of PMNS matrix $U_{PMNS}(\equiv U)$
 \cite{Beringer:1900zz}, the $3\times 3$ neutrino mass matrix is
represented as
\begin{equation}
m_\nu = U_{PMNS}~diag(m_1, m_2, m_3) U_{PMNS}^T \label{mnu}
\end{equation}
where $m_i (i=1,2,3) =$ mass eigen values  and using the  PDG convention \cite{Beringer:1900zz} 
\begin{equation}
 U_{\rm{PMNS}}= \left( \begin{array}{ccc} c_{12} c_{13}&
                      s_{12} c_{13}&
                      s_{13} e^{-i\delta}\cr
-s_{12} c_{23}-c_{12} s_{23} s_{13} e^{i\delta}& c_{12} c_{23}-
s_{12} s_{23} s_{13} e^{i\delta}&
s_{23} c_{13}\cr
s_{12} s_{23} -c_{12} c_{23} s_{13} e^{i\delta}&
-c_{12} s_{23} -s_{12} c_{23} s_{13} e^{i\delta}&
c_{23} c_{13}\cr
\end{array}\right) 
diag(e^{\frac{i \alpha_M}{2}},e^{\frac{i \beta_M}{2}},1)
\end{equation}
where $s_{ij}=\sin \theta_{ij}, c_{ij}=\cos \theta_{ij}$ with
$(i,j=1,2,3)$, $\delta$ is the Dirac CP phase and $(\alpha_M,\beta_M)$
are Majorana phases. During our actual calculation 
the mass eigenvalues and mixing angles are taken to be the best fit values of the present oscillation data\cite{Forero:2014bxa,Esteban:2018azc}
presented  in Table \ref{osc} below.
\begin{table}[!h]
\caption{Input data from neutrino oscillation experiments \label{osc} \cite{Forero:2014bxa,Esteban:2018azc}}
\label{input}
\begin{center}
\begin{tabular}{|c|c|c|}
\hline
{ Quantity} & {best fit values} &{ $3\sigma$ ranges}\\
\hline
$\Delta m_{21}^2~[10^{-5}eV^2]$ & $7.39$ & $6.79-8.01$\\
$|\Delta m_{31}^2|~[10^{-3}eV^2](NO)$ & $2.52$ & $2.427-2.625$\\
$|\Delta m_{32}^2|~[10^{-3}eV^2](IO)$ & $2.51$ & $2.412-2.611$\\
$\theta_{12}/^\circ$ & $33.82$ & $31.61-36.27$\\
$\theta_{23}/^\circ (NO)$ & $49.6$ & $40.3-52.4$\\
$\theta_{23}/^\circ (IO)$ & $49.8$ & $40.6-52.5$\\
$\theta_{13}/^\circ (NO)$ & $8.61$ & $8.22-8.99$\\
$\theta_{13}/^\circ (IO)$ & $8.65$ & $8.27-9.03$\\
$\delta/^\circ (NO)$ & $215$ & $125-392$\\
$\delta/^\circ (IO)$ & $284$ & $196-360$ \\
\hline
\end{tabular}
\end{center}
\end{table}
We now determine the $m_\nu$ matrix for the normally
ordered (NO) masses. For the sake of convenience we take the mass of the lightest neutrino as $m_1=0.001$ eV.
Then using the solar and atmospheric mass squared differences for NO from Table \ref{osc}, the other two neutrino mass eigenvalues are 
computed with $m_2=0.00865$ eV and $m_3=0.0502$ eV. Using these mass eigenvalues and the best fit values of the mixing angles and Dirac CP 
phases,  eq.(\ref{mnu}) gives   
\begin{equation}
m_\nu^{NO}(\rm eV)=  \left( \begin{array}{ccc}
              0.00367-0.00105i  & -0.00205+0.00346i & -0.00634+0.00294i  \cr
               -0.00205+0.00346i & 0.03154+0.00034i & 0.02106-0.0001i \cr
                -0.00634+0.00294i & 0.02106-0.0001i   &  0.02383-0.00027i\cr
                    \end{array}\right) \label{bfp_nh} ~~.
\end{equation}
Keeping all the other parameters  at their best fit values, we derive
the numerical structure of $m_\nu$ for another value of the Dirac CP
phase $\delta=170^\circ$ (which is within the given $3\sigma$ range) 

\begin{equation}
m_\nu^{NO}(\rm eV)=  \left( \begin{array}{ccc}
              0.00435-0.00038i  & -0.00293-0.00104i & -0.00708-0.00089i  \cr
               -0.00293-0.00104i & 0.03165-0.00010i & 0.02107+1.95\times10^{-6}i \cr
               -0.00708-0.00089i &  0.02107+1.95\times10^{-6}i &  0.02377+0.00007i\cr
                    \end{array}\right) \label{bfp_nh1}~~.
\end{equation}
\paragraph{}
For inverted ordering (IO),  setting the lightest mass eigenvalue 
$m_3=0.001$ eV and using the 
mass squared differences  from Table \ref{osc}, we get  $m_1=0.04938$ eV, $m_2=0.0501$ eV.
Then  the value of  
 $m_\nu$  corresponding to best fit values of parameters turns out to be
\begin{equation}
m_\nu^{IO}(\rm eV)=  \left( \begin{array}{ccc}
                  0.0484-0.00001i & -0.001122+0.0055i & -0.00137+0.00471i \cr
                   -0.001122+0.0055i  &  0.02075-0.00025i& -0.02459-0.00026i \cr
                    -0.00137+0.00471i & -0.02459-0.00026i &  0.02910-0.00026i\cr
                    \end{array}\right). \label{bfp_ih}
\end{equation}
Similarly the structure of the mass matrix for another value of
$\delta=200^{\circ}$, which is in a different quadrant, is estimated
\begin{equation}
m_\nu^{IO}(\delta=200^\circ)(\rm eV)=  \left( \begin{array}{ccc}
                0.04849-0.00001i & 0.005399+0.00196i & 0.00413+0.00166i \cr
                0.005399+0.00196i &  0.02189+0.00043i& -0.02369+0.000352i \cr
                0.00413+0.00166i & -0.02369+0.000352i  &  0.029816+0.000285i\cr
                    \end{array}\right). \label{bfp_ih1} ~~
\end{equation}
Using the up-quark diagonal basis  the Dirac neutrino
mass matrix  is taken to be nearly equal to the 
up quark mass matrix\cite{pnsa:2016} 
\begin{equation}
M_D(\rm GeV) \simeq {\rm diag}( .00054, .26302, 81.99 ).\label{MD}         
\end{equation}
from which the coupling matrix is
$\lambda \simeq\frac{M_D}{v/\sqrt 2}$.
\paragraph{}
 The masses of the heavy
neutrinos are generated at very high  
 $SO(10)$ breaking scale which also induces the trilinear coupling $\mu_{\Delta}$ as noted above
\ba
M_N&=&YV_{B-L}, \nonumber\\
\mu_{\Delta}&=&\lambda_{(126,10)}V_{B-L}, \label{eq:MNLAMBdef}
\ea
where $V_{B-L}=\langle \Delta_R^0 \rangle \ge 10^{17}$ GeV and
$\lambda_{126,10}{126}_H{126}_H^{\dagger}{10}_H{10}_H$ is the SO(10)
invariant scalar interaction term. 
Using eq.(\ref{MN}) and eq.(\ref{t2app}) we have 
\begin{eqnarray}
M_N &=& m_\nu \frac{V_{B-L}}{v_L} \nonumber\\
{\hat M}_N  &=& diag(m_1,m_2,m_3)\frac{V_{B-L}}{v_L}, \label{t2con2}
\end{eqnarray}
which predicts the relation in eq.(\ref{eq:t2con}).
Among all the variables in the RHS of eq.(\ref{t2con2}), only the
numerical value of $v_L$ is unknown. Again $v_L$ itself contains two unknowns
  $\mu_{\Delta}$  and $M_\Delta$ which are varied in suitable ranges
to generate a parameter space where we  implement leptogenesis in Sec.\ref{sec:baulept}.


\section{Scalar Dark Matter Prediction with Vacuum Stability}\label{sec:relstab}
In this section we show how WIMP (weakly interacting massive
particles) DM  and vacuum stability
\cite{Espinosa:2012} issues are reconciled in Model-I and Model-II
using our derivation
discussed in Sec. \ref{sec:redm}. In Sec.\ref{sec:vsfdm} we also
discuss  two different ways for confronting the vacuum stability issue
in Model-III. 
\subsection{Intrinsic Dark Matter from SO(10) }\label{sec:intdm}
The scalar dark matter candidate that we utilise here can be identified with the imaginary part of the SM singlet component of ${16}_H\subset$ SO(10) which carries 
odd matter parity \cite{Kadastic:dm1,Kadastic:dm2,Frig-Ham:2010}
\be
\xi \equiv \chi_S^I =Im (\chi_S)\subset {16}_H \subset {\rm SO(10)}. \label{eq:dmid}
\ee
In Sec.\ref{sec:redm} we have shown how this imaginary part of the real scalar singlet  carrying odd matter parity can be as light as desired while keeping the real part at the GUT scale, or vice versa.

It is well known that  the standard model Higgs potential
\begin{equation}
V_{SM}=-{\mu}^2 \phi^{\dagger}\phi+ \lambda_{\phi}(\phi^{\dagger}\phi)^2, \label{eq:smpot}
\end{equation}
develops instability as the Higgs quartic coupling $\lambda_{\phi}$ runs 
negative at an energy scale $10^9-10^{10}$ GeV. In the last step of our symmetry breaking chain of
eq.(\ref{eq:chain2}), the SM gauge symmetry is broken spontaneously to
$U(1)_{em}\times SU(3)_C$ by the electroweak VEV of the standard Higgs
doublet $\phi \subset {5}_H \subset$ SU(5). As a result, only the SM
Higgs $\phi$ remains light at the electroweak scale while the color
triplet in ${5}_H$  acquires mass at the SU(5) scale
\cite{Aguilla:1981,RNM-gs:1983}.  But, as shown in Sec.\ref{sec:redm}, the matter parity odd real scalar singlet $\xi$ could be as light as $\sim (100-1000)$ GeV in each of the three models  leading to the modified scalar potential  
 
\begin{equation}
V=-\mu^2 \phi^\dagger \phi +\mu_\xi^2 \xi^\dagger \xi  + \lambda_\phi (\phi^\dagger \phi)^2 +\lambda_\xi (\xi^\dagger \xi)^2 +
2 \lambda_{\phi \xi}  (\phi^\dagger \phi)(\xi^\dagger \xi). \label{pot_dm}
\end{equation}

In eq.(\ref{pot_dm}) $\lambda_{\xi}\simeq \lambda_{16}\equiv$  dark matter self-coupling associated with the interaction $\lambda_{16}[{16}_H^{\dagger}{16}_H]^2$ and $\lambda_{\phi \xi}\simeq \lambda_{10,16}$ which is associated with $\lambda_{10,16}{16}_H^{\dagger}{16}_H{10}_H^2$. This latter type SO(10) invariant interaction has induced the $\phi-\xi$ Higgs portal interaction of the 
SM  effective gauge theory.
 Also $\mu_{\xi}^2\equiv M_{\rm Im}^2$ as
defined through eq.(\ref{eq:rimass}) of Sec. \ref{sec:redm}. 
 The VEV of the standard Higgs doublet redefines the DM mass
 parameter        
\begin{eqnarray}
&& M_{DM}^2 = 2(\mu_\xi^2 +\lambda_{\phi \xi}^2 v^2), \nonumber\\
&& m_\phi^2 =2 \mu^2= 2 \lambda_\phi v^2. 
\end{eqnarray}
In SM extensions 
\cite{Garg:2017,gond:2010,chen:2012,Khan:2014} the origins of the
scalar DM, its mass, and the DM stabilising discrete symmetry are
unknown apriori. On the other hand, in this work, all these quantities including the SM Higgs are intrinsic to self sufficient SO(10) theory.    
In order to  constrain the  Higgs portal coupling $\lambda_{\phi \xi}$ we use recent results of DM direct detection experiments like LUX-2016\cite{Akerib:2016vxi},
XENON1T\cite{Aprile:2017iyp,Aprile:2018dbl}, PANDA-X-II\cite{Cui:2017nnn}. Bounds on DM relic density $(\Omega_{\rm DM}h^2=0.1172-0.1224)$ as reported by WMAP\cite{wmap} and Planck\cite{Planck15}
are also taken into account. We first proceed to calculate the relic
density for different combinations of dark matter mass $(m_\xi)$ and
the  Higgs portal
coupling $(\lambda_{\phi \xi})$. It is then easy to restrict the values of $m_\xi$ and $\lambda_{\phi \xi}$ using the bound on relic density as 
quoted above. In direct detection experiments it is assumed that WIMPs passing through earth scatter elastically off the target material of the detector.
The energy transfer to the detector nuclei  can be measured through various types of signals. All those direct detection experiments provide
DM mass vs DM-nucleon scattering cross section plot which clearly separates two regions: allowed (regions below the curve) and forbidden (regions above the curve). 
Our aim is to constrain the model parameters $(\lambda_{\phi \xi},m_\xi)$ using bounds on relic density as well as exclusion plots from several
DM direct detection experiments.
\subsubsection{Estimation  of Relic Density:}\label{sec:xirelic}
Defining $\Gamma $ (H) as the particle decay rate (Hubble parameter),  
at a certain stage of evolution of the Universe a particle species is
said to be  coupled  if  $\Gamma > H$  or decoupled if $\Gamma < H$. The WIMP
DM particle has been decoupled from the thermal bath at some early
epoch and has remained as a thermal relic. 
\paragraph{}
In order to calculate the relic density of the scalar singlet DM, we  solve the Boltzmann equation\cite{Kolb:1990vq,Bertone:2004pz} for the corresponding 
particle species which is given by
\begin{equation}
\frac{d n}{d t}+3Hn =-\langle \sigma v \rangle (n^2 -n_{eq}^2) \label{boltz}
\end{equation}
where $n=$  actual number density  at a certain
instant of time and $n_{eq}=$  equilibrium number density of scalar DM.
Here $v=$ velocity and $\langle \sigma v \rangle=$  thermally averaged
annihilation crosssection. Approximate solution of Boltzmann equation gives the expression of relic density\cite{Bertone:2004pz,Gondolo:1990dk,Banik:2013zla} 
\begin{equation}
\Omega_{\rm DM} h^2= \frac{1.07\times10^9 x_F}{\sqrt{g_\ast} M_{ pl}\langle \sigma v \rangle} \label{relic}
\end{equation}
where $x_F=m_\xi/T_F$,  $T_F=$ freezeout temperature, $g_\ast=$  effective number of massless degrees of freedom and  $M_{pl}=1.22\times10^{19}$ GeV. $x_F$ can be computed by iteratively solving the equation
\begin{equation}
x_F=\ln \left(\frac{m_\xi}{2\pi^3}\sqrt{\frac{45 M_{pl}^2}{8 g_\ast x_F}} \langle\sigma v \rangle \right) .\label{xf}
\end{equation}
In  eq.(\ref{relic}) and eq.(\ref{xf}), the only particle physics
input is  the thermally averaged annihilation
cross section.
The total annihilation cross section is obtained by summing over all the annihilation channels of the singlet DM which are $\xi\xi \rightarrow f\bar{f},W^+W^-,ZZ,hh$
where $f$ is symbolically used for all the fermions. Using the expression of total annihilation 
cross section\cite{McDonald:1993ex,Guo:2010hq,Biswas:2013nn,Biswas:2011td} in eq.(\ref{xf}) at first we compute
$x_F$ which is then plugged into eq.(\ref{relic}) to yield the relic density. Two free parameters involved in this computation are mass of the 
 DM particle $m_\xi$ and the Higgs portal coupling $\lambda_{\phi \xi}$. The relic density has been estimated for a wide range of values of the 
DM matter mass ranging from few GeVs to few TeVs while the coupling $\lambda_{\phi \xi}$ is also varied simultaneously in the range $(10^{-4}-1)$.
The parameters $(m_\xi,\lambda_{\phi\xi})$ are constrained by using the bound on the relic density  reported by WMAP and Planck. In Fig.\ref{cons_para}
we show only those combinations of $\lambda_{\phi \xi}$  and $m_\xi$ which are capable of producing relic density in the experimentally observed range.

\subsubsection{Bounds from Direct Detection Experiments}\label{sec:bound}
\begin{figure}[h!]
\begin{center}
\includegraphics[scale=0.5,angle=270]{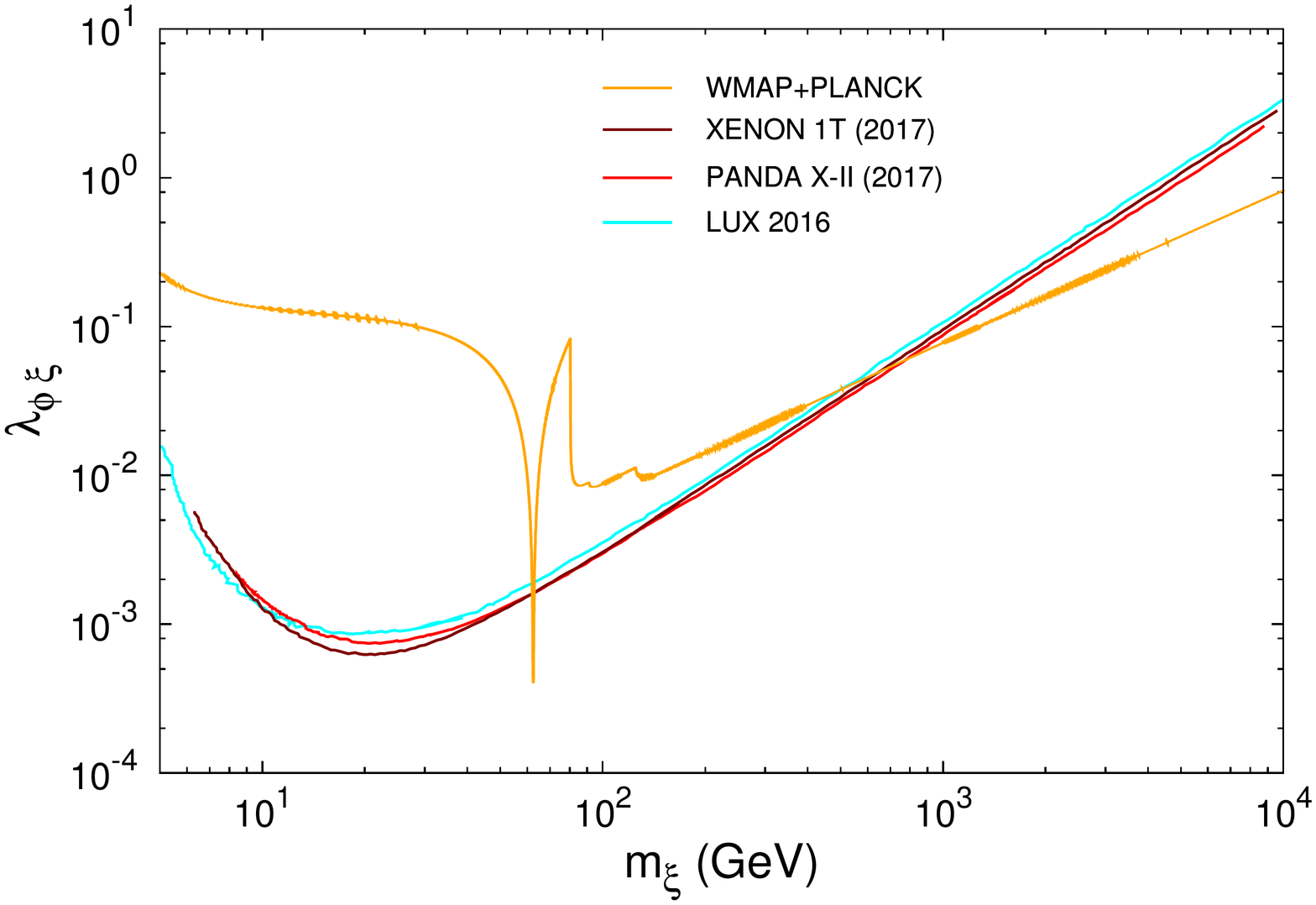}
\caption{The yellow curve denotes the values of the parameters $(\lambda_{\phi\xi},m_\xi)$ allowed by the relic density bound $(\Omega_{\rm DM}h^2=0.1172-0.1224)$ as 
observed by Planck  and WMAP. The cyan, brown and red curves are exclusion plots obtained from dark matter direct detection experiments LUX-2016, XENON1T(2017)
and Panda-XII(2017), respectively. Any point below those plots are allowed by direct detection experiments. It is to be noted that the Panda-XII 
experiment has provided the most stringent bound till date.}
\label{cons_para}
\end{center}
\end{figure}

 We get exclusion plots of DM-nucleon scattering cross section and DM mass from different direct
detection experiments. The spin independent scattering cross section of singlet DM on nucleon is given by\cite{Cline:2013gha}
\begin{equation}
\sigma^{\rm SI}=\frac{4 f^2\lambda_{\phi \xi}^2 \mu^2 m_N^2}{\pi m_\xi^2 m_h^4} ~~({\rm  cm}^2) \label{sigsi}
\end{equation}
where $m_h=$ mass of the SM Higgs ($\sim 125$ GeV), $m_N=$ nucleon mass $\sim 939$ MeV, $\mu=(m_\xi m_N)/(m_\xi+m_N)=$  reduced 
DM-nucleon mass and the factor $f \sim 0.3$. Using eq.(\ref{sigsi}) the exclusion plots of $\sigma-m_\xi$ plane can be easily brought to $\lambda_{\phi\xi}-m_\xi$
plane. We superimpose the $\lambda_{\phi\xi}-m_\xi$ plots for different experiments on the plot of allowed parameter space constrained by relic density bound. 
So the parameter space $(\lambda_{\phi\xi}~vs~m_\xi)$ constrained by both the relic density bound and the direct detection experiments can be obtained from Fig.\ref{cons_para}.
\paragraph{}
In Fig.\ref{cons_para} the points on the yellow curve which are also below the exclusion lines of the direct detection experiments are allowed by both the relic density
as well as the upper bounds on the DM-nucleon annihilation cross section as reported by the direct detection experiments. From the Fig.\ref{cons_para} it is clear 
that scalar singlet dark matter with mass below $\sim 750$ GeV is ruled out by direct detection experiments. Although few allowed points can be obtained around
$m_\xi \sim 62$ GeV, for those points the Higgs portal coupling is
very low and the mass of the DM particle is also highly fine tuned.
\paragraph{}
\subsubsection{Resolution of Vacuum Stability}\label{sec:vstab}
\begin{figure}[h!]
\begin{center}
\includegraphics[scale=0.5,angle=270]{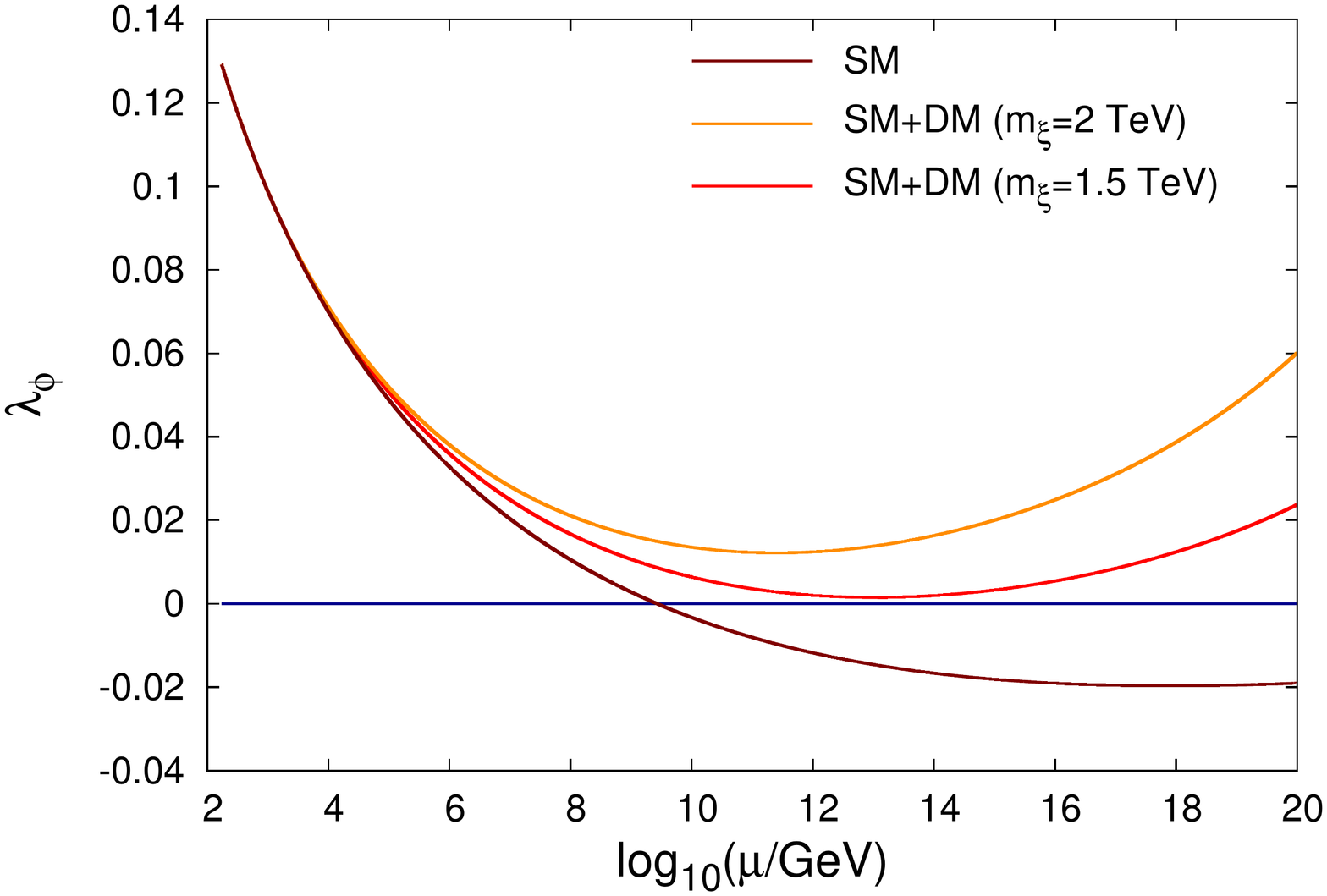}
\caption{Running of SM Higgs quartic coupling in presence of scalar singlet WIMP dark matter. The yellow and red curves are for dark matter masses $2$ TeV and $1.5$ TeV respectively
whereas the brown curve is for standard model alone.}
\label{dm_stab}
\end{center}
\end{figure}

To resolve the vacuum instability problem we choose few points on the yellow curve (allowed by relic density bound) 
of Fig.\ref{cons_para} at high mass region. We examine whether the vacuum has now become stable upto the Planck scale after the
addition of the scalar singlet WIMP DM to SM. To trace the evaluation of the SM  Higgs quartic coupling upto higher energy
scales, we solve the corresponding set of renormalisation group equations
eq.(\ref{rge}) which are given in Appendix.\ref{rgeq}. The corresponding values of the Higgs portal 
coupling $(\lambda_{\phi\xi})$ and self coupling of the scalar singlet DM $(\lambda_\xi)$ at the electroweak scale
which are taken to be the initial value of our analysis are stated in Table.\ref{tab:cc}. 
Thorough analysis of vacuum stability using those chosen points $(\lambda_{\phi\xi},m_\xi)$ 
reveals that it is not  possible to cure the vacuum instability problem of SM by adding a scalar singlet WIMP DM of 
mass below $1.4$ TeV. The evolution of SM Higgs quartic coupling $(\lambda_\phi)$ with the energy scale $(\mu)$
for two different values of DM masses, $1.5$ and $2$ TeV, is depicted in Fig.\ref{dm_stab} which clearly indicates that this
vacuum instability is indeed resolved for DM mass $1.5$ TeV. 

\begin{table}[h!]
\caption{Initial values of coupling constants at top quark mass for two different values of the dark matter mass.
The values of $\lambda_{\phi \xi}$ and $m_\xi$ are obtained from the
plot of constrained parameter space of Fig.\ref{cons_para}.}
\begin{tabular}{|c|c|c|c|c|c|c|c|}  
 \hline
   $m_\xi$ ( TeV ) & $\lambda_{\phi \xi}(m_t) $ &  $\lambda_{\xi}(m_t) $ & $\lambda_{\phi}(m_t)$ & $g_{1Y}(m_t)$ & $g_{2L}(m_t)$ &  $g_{3C}(m_t)$ &  $y_{t}(m_t)$   \\ 
   & & & & & & & \\ \hline
  $1.5$ &  $0.118$ & $0.22$ &  &  & &  &    \\
  &  &  & $0.129$ & $0.35$ & $0.64$ & $1.16$ & $0.94$\\
  $2$& $0.158$ & $0.1$ &  &  &  &  &
\\ \hline
\end{tabular} 
\label{tab:cc}
\end{table}
\section{Vacuum Stability with Triplet Fermionic Dark Matter}\label{sec:vsfdm}
We point out below two different ways to resolve the vacuum stability
problem of the SM scalar potential existing in the original model \cite{Frig-Ham:2010} or its type-II seesaw dominance induction carried out through the Model-III.  
\subsection{Intermediate Mass Scalar Singlet Threshold
  Effect}\label{sec:vstabthrs}
Even after the implementation of type-II seesaw dominance through
Frigerio-Hambye framework,  an interesting solution to the vacuum
stability problem applies as noted in \cite{Espinosa:2012}. In this
mechanism a Higgs scalar singlet originating from any one of SO(10) scalar
representations like ${45}_H,{54}_H,{210}_H,...$ is made light to have
mass around $10^8-10^9$ GeV. Then vacuum expectation value of this
scalar singlet generates threshold effects that prevents the SM Higgs quartic coupling from being negative at 
higher scales. In this case the triplet fermion DM mass remains at
$\sim 2.7$ TeV \cite{Frig-Ham:2010} and needs non-perturbative
Sommerfeld enhancement to account for relic density.  
\subsection{ Through the Added Presence of Light Scalar Singlet Dark Matter}\label{sec:compdm}
    
\subsubsection{Combined Effect on Relic Density}\label{sec:combrel}
It has been shown  \cite{Frig-Ham:2010} that the neutral component of
the triplet fermion $\Sigma (3,0,1)$ can play the role of WIMP dark matter and the 
correct amount of relic abundance is generated when the mass of the triplet is $\sim 2.4$ TeV. This mass value arrived through perturbative calculation
 is further shifted to $2.7$ TeV when the non perturbative Sommerfeld
effect is also taken into account. The  triplet fermionic DM  model seems to be somewhat constrained from various direct detection experiments. It is found that the 
DM-nucleon scattering cross section\cite{DM-nuclX} at $m_\Sigma \sim 1$ TeV nearly touches the upper bound given in the most recent XENON1T result.
 Thus, there may be possibility of facing more severe constraint in near future
unless the presence of the $\sim 2.7$ TeV triplet fermionic DM is confirmed through precision measurements.
In the indirect detection experiment, signals are produced in the annihilation process: DM DM $\rightarrow$ SM particles. Here we have tree level annihilation only to $W^+W^-$ channel. The annihilation cross section to
$W^+W^-$ shows a peak  near $m_\Sigma \sim 2.7$ TeV when 
nonperturbative effects are taken into account. But this high value of annihilation cross section exceeds the upper limit given by the 
combined analysis \cite{Comb-ana} of Fermi LAT and MAGIC. So it can be
inferred that when the dark matter is composed of only the fermion triplet, it fails to 
satisfy all the experimental constraints (relic density, direct and
indirect detection bounds) simultaneously. In other words the triplet
fermionic dark matter appears to be strongly constrained by indirect detection experiments. Even after resolution of the instability through such threshold
effect\cite{Espinosa:2012} , the $2.7$ TeV mass triplet fermionic DM
model may have to face the constraints due to relic density and direct
and indirect detection experiments.

Now we modify the Model-III in such a way that all the above mentioned difficulties are evaded without disturbing the gauge coupling unification or type-II seesaw driven leptogenesis.
We utilise the real gauge singlet scalar $(\xi \subset 16_H^\dagger,
m_\xi=500-2000 {\rm GeV})$ which acts a WIMP dark matter in the
modified scenario. Thus the total relic
abundance is now shared by the fermion triplet \cite{Frig-Ham:2010} and the real  scalar
singlet. Since the dark matter has now two components, the constraint relations from direct
and indirect detection experiments will also have a small change. 

We keep the mass of the triplet fermionic dark matter at $m_\Sigma=1$ TeV. The interesting point about the choice of this triplet mass is that at
this lower mass $\sim 1$ TeV,  we need not include the non-perturbative 
Sommerfeld enhancement  to match
the relic density and the annihilation crosssection to $W^+W^-$ as a perturbative calculation 
gives us a fair estimation of the cross section. But the relic abundance produced only due to the triplet
DM ($m_\Sigma=1$ TeV) is much less than the experimental value quoted
by Planck\cite{Planck15} and WMAP\cite{wmap}. This disagreement is
compensated  through the intervention  of the real singlet DM
candidate $\xi$. Since the scalar singlet can not have any
renormalisable interaction with the fermionic triplet DM $\Sigma_F$, we can 
estimate the relic density for fermionic triplet ($(\Omega h^2)_1$)
and scalar
 singlet $((\Omega h^2)_2)$ separately and  add them up to get the total 
relic density 
\begin{equation}
 (\Omega h^2)_t=(\Omega h^2)_1 + (\Omega h^2)_2 ~.
\end{equation}
To estimate the relic abundance of the neutral component $\Sigma^0$ that  acts as dark matter,
we have to take into account the annihilations and co-annihilations of $\Sigma^0$ itself and $\Sigma^+\Sigma^-$. Calculating all such contributions
$(\sigma(\Sigma^0,\Sigma^0),\sigma(\Sigma^0,\Sigma^{\pm})$
$,\sigma(\Sigma^+,\Sigma^-),\sigma(\Sigma^{\pm},\Sigma^{\pm}))$ and adding them up after multiplying by
suitable weightage factor\cite{Ma-Suematsu:2008} we get the effective cross section
$\langle \sigma_{eff} |v| \rangle$ which is thereafter plugged into the equation
\begin{equation}
 (\Omega h^2)_1 = \frac{ 1.07 \times 10^9 x_F }{ \sqrt{g_\ast} M_{pl} \langle \sigma_{eff} |v| \rangle }
\end{equation}
to obtain the relic density due to the triplet fermion DM. It is to be
noted that $x_F=m_\Sigma/T_F$  in the above equation is related to the
freezeout temperature $T_F$ and  is determined by the iterative solution of
the equation \cite{Ma-Suematsu:2008} 
\begin{equation}
x_F = \ln \left[ \frac{1}{ 2 \pi^3} \sqrt{\frac{45}{2}} \frac{M_{pl} m_\Sigma \langle \sigma_{eff} |v| \rangle}{ \sqrt{g_\ast x_F}} \right].
\end{equation} 
The computational procedure for relic density $(\Omega h^2)_2$ due to scalar singlet has already been discussed in Sec.\ref{sec:xirelic}. In the estimation of the total relic density 
there are three free parameters: mass of the fermion triplet $(m_\Sigma)$, mass of the scalar singlet $(m_\xi)$ and the Higgs portal coupling $(\lambda_{\phi\xi})$.
The quantity $(\Omega h^2)_1(m_\Sigma=1)$ TeV is found to be much less than the experimental value. The other part of the relic density required to meet the experimental constraint
$( 0.1172<(\Omega h^2)_t <0.1224)$ is generated by the real scalar
singlet. It is worthwhile to mention that we have varied
$\lambda_{\phi\xi}$ and $m_\xi$ over a wide
range of values and $(\lambda_{\phi\xi},m_\xi)$ get their first round of constraint from the relic density bound given above. With the set of values of $(\lambda_{\phi\xi},m_\xi)$
already restricted by relic density bound, we proceed further to constrain them by direct detection bound. For this two component dark matter, the constraint 
relation appears as \cite{Ma-Suematsu:2008,Cao:2007fy}
\begin{equation}
\frac{\epsilon_\Sigma}{m_\Sigma} \sigma_{\Sigma N} + \frac{\epsilon_\xi}{m_\xi} \sigma_{\xi N} < \frac{\sigma_0}{m_0} 
\end{equation}
where the symbols have the following meanings. $\sigma_0,m_0$ are DM-nucleon scattering cross section and mass of the dark matter, respectively, for single component
dark matter scenario. Now for the two component dark matter scenario under consideration $\sigma_{\Sigma N}(\sigma_{\xi N})$ is the scattering cross section
of $\Sigma(N)$ with detector nucleon and $m_{\Sigma}(m_\xi)$ is its mass. The factor $\epsilon_i$ designates the fraction of  density of the $i$th dark matter particle in a
certain model: $\epsilon_i=\rho_i/\rho_0$ which can also be expressed in terms of thermally averaged annihilation cross sections as 
\begin{eqnarray}
&&\epsilon_\Sigma=\frac{\langle \sigma v \rangle_\xi}{ \langle \sigma v \rangle_\xi +\langle \sigma v \rangle_\Sigma } \nonumber\\
&&\epsilon_\xi=\frac{\langle \sigma v \rangle_\Sigma}{ \langle \sigma v \rangle_\xi +\langle \sigma v \rangle_\Sigma } ~.
\end{eqnarray}
To find the ratio $\sigma_0/m_0$ we have used the latest result of XENON1T experiment\cite{Aprile:2017iyp,Aprile:2018dbl}. The lowest value of the scalar singlet mass for which 
both the constraints are satisfied simultaneously is $m_\xi=1.13$ TeV with $\lambda_{\phi\xi}=0.09$. 
\subsubsection{Effects on Vacuum Stability}
\begin{figure}[h!]
\begin{center}
\includegraphics[scale=0.5,angle=0]{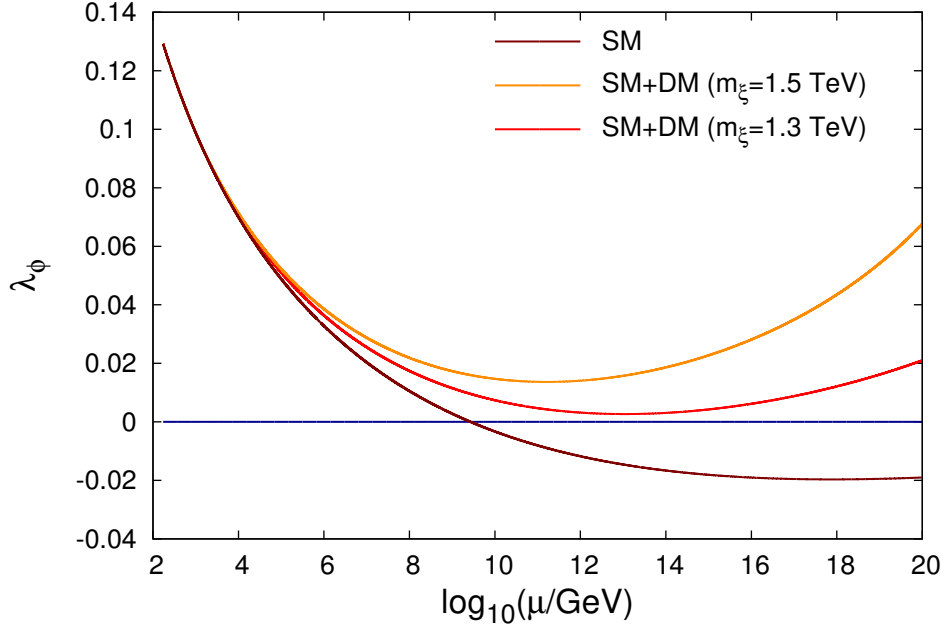}
\caption{Running  SM Higgs quartic coupling in the presence of scalar singlet and fermionic triplet WIMP dark matter as denoted by curves above the horizontal line. 
The yellow and red curves are for scalar singlet dark matter masses
$1.5$ TeV and $1.3$ TeV, respectively. The brown curve is for the SM alone.}
\label{vac_stab_trip}
\end{center}
\end{figure}

We now study whether the addition of this scalar singlet can make the
vacuum stable. We solve the corresponding
RG equations, given in Appendix \ref{rgeq}, using Table \ref{tab:cc} and
with the allowed sets of values of $\lambda_{\phi\xi}$ and $m_\xi$.
This exercise is repeated for many sets of values of
$(\lambda_{\phi\xi},m_\xi)$. It is found that the lowest mass of $\xi$
needed to overcome the vacuum instability 
 is $m_\xi=1.3$ TeV with the corresponding value $\lambda_{\phi
   \xi}=0.13$. This has been  shown graphically in Fig.\ref{vac_stab_trip}. Another example for 
$m_\xi=1.5$ TeV, $\lambda_{\phi\xi}=0.16$ has  been also presented in
 the same plot. For both the cases self coupling of dark matter $\xi$ is kept fixed at $\lambda_\xi=0.1$.

\section{Baryogenesis Through Leptogenesis}\label{sec:baulept}
Currently the standard approach towards understanding 
baryon asymmetry of the Universe (BAU) requires fulfillment of 
  Sakharov\cite{Sakharov} conditions: (i) baryon number violation,
  (ii) C and CP violations, and (iii) departure from thermal equilibrium.
 BAU is defined
as 
\begin{equation}
 Y_B=\frac{n_B-n_{\overline{B}}}{s}.
\end{equation}
where $n_B,n_{\overline{B}}$ are number densities of baryons and anti-baryons, respectively, and $s$ is the entropy density.
Another equivalent definition of BAU is
\begin{equation}
\eta_B=\frac{n_B-n_{\overline{B}}}{n_\gamma}.
\end{equation}
 where $n_\gamma=$  photon density.
 Planck satellite experimental values  are\cite{Planck15}
\ba
&&Y_B=8.66\pm 0.11 \times 10^{-11},\label{YB}\\
&&\eta_B=6.10\pm 0.08 \times 10^{-10}.\label{etaB}
\ea
 Out of various possible mechanisms of baryogenesis  such as GUT
 baryogenesis,
electroweak baryogenesis, Affleck-Dyne 
mechanism, and baryogenesis via leptogenesis \cite{one,one2,khlopov,Pilaftsis:1997,Buchmuller:2004nz,Buchmuller:2005,Abada:2006ea,Davidson:2008bu,King:2015}
we follow leptogenesis path where analogues of Sakharov's conditions
are satisfied.
\subsection{CP Asymmetry and Leptogenesis}
From the interaction Lagrangian in eq.(\ref{l1}) and
eq.(\ref{l2}), it is clear that lepton number violation is possible
due to the coexistence of the Dirac neutrino Yukawa matrix
$\lambda$  and the Higgs-triplet-bilepton Yukawa matrix $Y$ along with
$\mu_\Delta$.
In general in this model there are two sources of 
CP asymmetry: (i) decay of RHN to lepton and Higgs pair and 
(ii) decay of scalar triplet ($\Delta_L$) to lepton pair. 
We point out that the type-II seesaw dominance in SO(10) predicts a
class of
modified formulas in both these cases where PMNS matrix ($U$)  occurs quadratically in the expressions for corresponding CP-asymmetries. 
\subsubsection{RHN Decay and SO(10) Modified Formula}
\begin{figure}[h!]
\begin{center}
\includegraphics[width=5cm,height=4cm,angle=0]{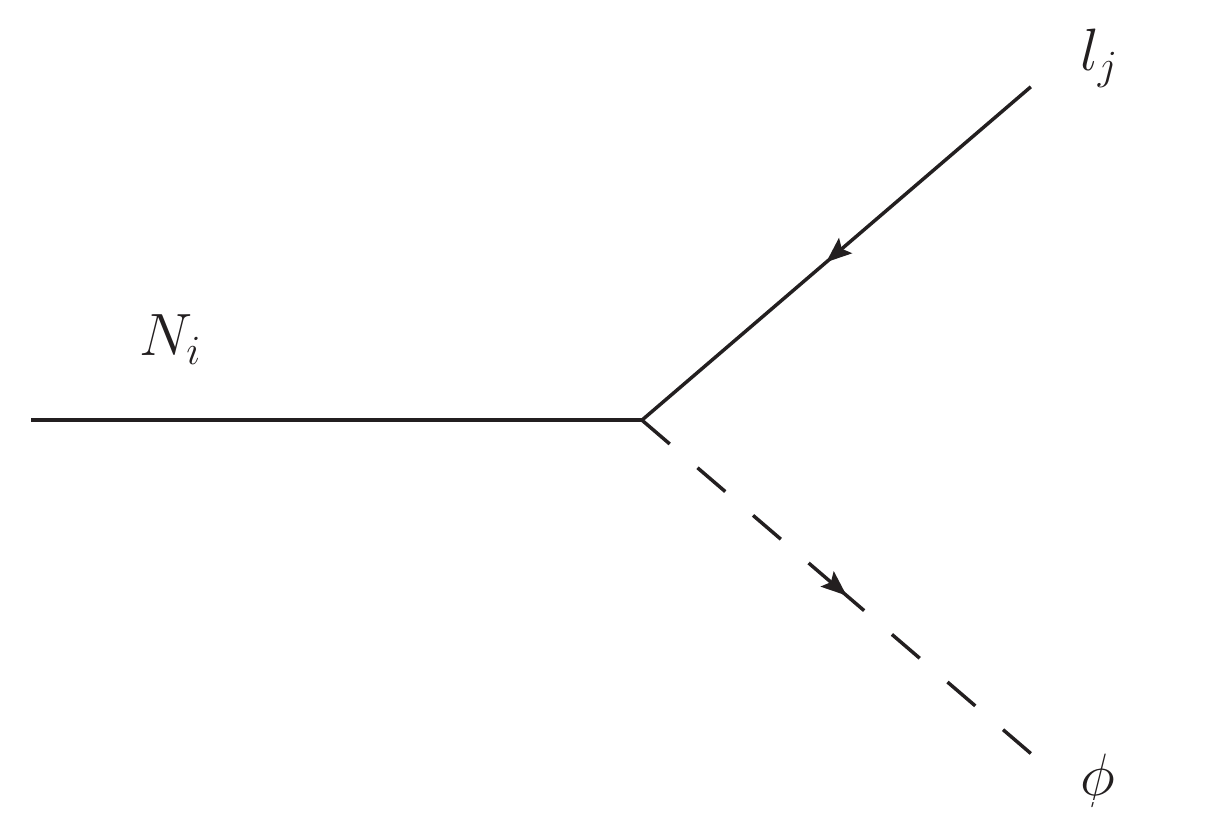}
\caption{Tree level decay of RHN to lepton and scalar Higgs}
\label{n_t}
\end{center}
\end{figure}

\begin{figure}[h!]
\begin{center}
\includegraphics[width=14cm,height=3cm,angle=0]{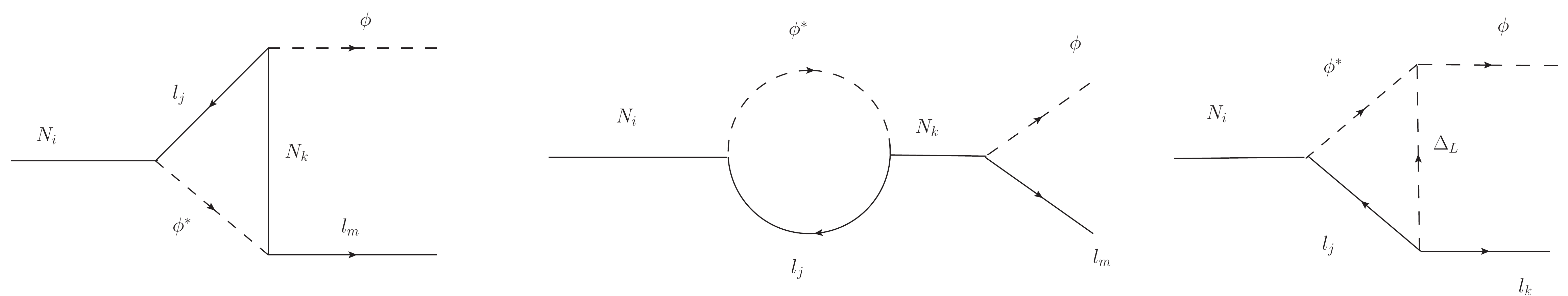}
\caption{One loop decay diagrams of RHN: (left) vertex correction, (middle) self energy correction, (right) vertex correction due to scalar triplet.}
\label{n_l}
\end{center}
\end{figure}

In the  case of  RHN decay, the CP asymmetry arises due the interference of the  tree level 
diagram with that of the one loop vertex and self energy diagrams shown in Fig. \ref{n_t} and Fig.\ref{n_l}.
The CP asymmetry \cite{Hambye-gs:2003} produced due to RHN decay is
given by 
\begin{equation}
\epsilon^{l_m}_{N_k}=\frac{1}{8\pi(\lambda \lambda^\dagger)_{kk}}\sum_{j\neq k} \Big\{ {\cal I}m[(\lambda\lambda^\dagger)_{kj} \lambda_{km}\lambda_{jm}^\ast] g(x_{ij}) +
{\cal I}m[(\lambda\lambda^\dagger)_{jk} \lambda_{km}\lambda_{jm}^\ast] \frac{1}{1-x_{ij}} \Big \}
\label{epsN1}
\end{equation}
where $x_{ij}=\frac{M_{N_j}^2}{M_{N_i}^2}$ and
$g(x)=\sqrt{x}\Big[\frac{1}{1-x}+1-(1+x)\ln\Big(\frac{1+x}{x}\Big)\Big]$. The
right most diagram of Fig.\ref{n_l} is a new 
contribution to the CP asymmetry due to loop mediation by $\Delta_L$. The corresponding CP asymmetry
is given by
\begin{equation}
\epsilon_{N_k}^{\Delta^{l_m}}=-\frac{1}{2 \pi (\lambda \lambda^\dagger)_{kk} M_{N_k}} \sum_i {\cal I}m[\lambda_{km} \lambda_{ki} Y_{mi}^\ast \mu_\Delta]
\Big \{ 1- \frac{M_\Delta^2}{M_{N_k}^2} \ln \Big( 1+ \frac{M_{N_k}^2}{M_\Delta^2} \Big ) \Big \}.
\label{epsN2}
\end{equation}
The total CP asymmetry produced due to  RHN decay is given by $\epsilon_{N_k}^{l_m~tot}= \epsilon^{l_m}_{N_k} +\epsilon_{N_k}^{\Delta^{l_m}}$. For hierarchical
RHNs it is sufficient to consider the decay of the  lightest RHN neutrino only for $\epsilon_{N_1}^{l_m~tot}$.
\paragraph{}
The formulas for CP asymmetries presented in  eq.(\ref{epsN1}) and eq.(\ref{epsN2}) are usually derived \cite{Hambye-gs:2003,Sierra:2014tqa} under the assumption that the RHNs  are in their  diagonal basis . 
But in our  type-II seesaw dominance  $SO(10)$ models, the light
neutrino mass matrix $m_\nu \simeq m_\nu^{\rm II}=  Y v_L$ and the RHN mass matrix  $M_N=Y V_{B-L}$. Then these models  dictate that the diagonalising matrices for
 the RHN neutrino mass  and the light neutrino mass matrices are
 one and the same. Thus it forbids a RHN mass diagonal basis of the type
 \cite{Hambye-gs:2003,Sierra:2014tqa}.  But while calculating the CP asymmetry parameters as we need the  RHN masses in their mass basis, this can be realised by the rotation of RHN 
fields through the mixing matrix $U\simeq U_{\rm PMNS}$). In other words in type-II seesaw dominance SO(10) models the Dirac neutrino Yukawa coupling matrix $\lambda$ will be modified as
$\lambda \rightarrow \lambda U^\ast$. After inserting this
transformation relation, the conventional CP asymmetry formulas of eq.(\ref{epsN1}) and eq.(\ref{epsN2})
are modified to assume their new forms
\begin{eqnarray}
\epsilon^{l_m}_{N_k}&=&\frac{1}{8\pi(\lambda
    \lambda^\dagger)_{kk}}\sum_{j\neq k} \Big\{ {\cal
    I}m[(\lambda\lambda^\dagger)_{kj} (\lambda
    U^\ast)_{km}(\lambda^\ast U)_{jm}] g(x_{ij}) \nonumber\\
    &+& {\cal I}m[(\lambda\lambda^\dagger)_{jk} (\lambda U^\ast)_{km}(\lambda^\ast U)_{jm}] \frac{1}{1-x_{ij}} \Big\},\nonumber\\
\epsilon_{N_k}^{\Delta^{l_m}}&=&-\frac{1}{2 \pi (\lambda
    \lambda^\dagger)_{kk} M_{N_k}} \sum_i {\cal I}m[(\lambda
    U^\ast)_{km} (\lambda U^\ast) _{ki} Y_{mi}^\ast \mu_\Delta] \nonumber\\
& \times& \Big \{ 1- \frac{M_\Delta^2}{M_{N_k}^2} \ln \Big( 1+ \frac{M_{N_k}^2}{M_\Delta^2} \Big ) \Big \}.
\label{epsn_u}
\end{eqnarray}
In the standard model extensions \cite{Hambye-gs:2003,Sierra:2014tqa}, the trilinear scalar coupling  $\mu_{\Delta}$ occurring in CP-asymmetry formulas of
eq.(\ref{epsN2}) and eq.(\ref{epsn_u}) is an apriori unknown parameter. But, in our SO(10) models, the origin of this trilinear coupling is traced
back to the Higgs scalar quadrilinear interaction between ${10}_H$ and ${126}_H$
as shown in eq.(\ref{v12610}).
\subsubsection{Scalar Triplet Decay and SO(10) Modified Formula}\label{sec:modcp}
\begin{figure}[h!]
\begin{center}
\includegraphics[width=9cm,height=3cm,angle=0]{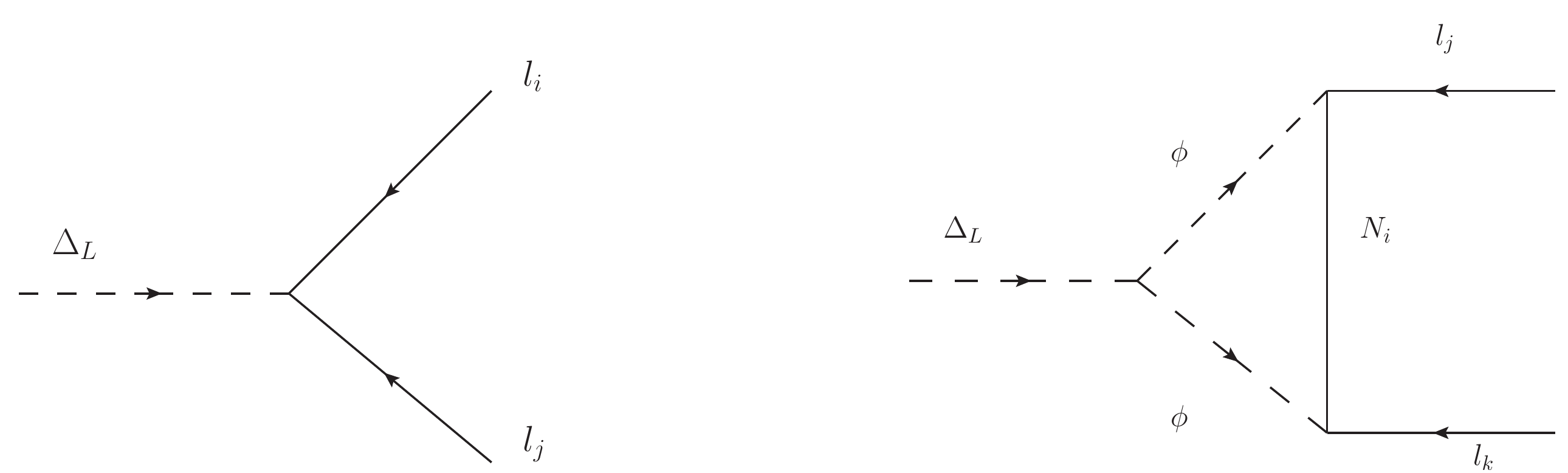}
\caption{(left panel)tree level and one loop vertex(right panel) diagram of $\Delta_L$ decay to lepton pair}
\label{Delta}
\end{center}
\end{figure}

The presence of the $Y$ coupling term in eq.(\ref{l2}) allows
$\Delta_L$ to decay to a lepton pair at the tree
level shown in Fig.\ref{Delta}.  This decay process is also possible at one
loop level through vertex  correction in the
presence of a RHN as shown in the right panel of Fig.\ref{Delta}.
As we confine to minimal SO(10) models with only one $\Delta_L$ we do not discuss  one loop self
energy correction mediated by two different Higgs scalar triplets \cite{Ma-Sarkar:1998}.
\paragraph{}
In this work our aim is to carry out  leptogenesis  at different temperature regimes where  lepton flavours may be fully or partly
distinguishable making the  CP asymmetry parameter to explicitly contain lepton
flavour indices. We also consider the possibility that the lepton
flavours may lose their distinguishability at some higher
temperatures. The flavour dependent CP asymmetry which arises due to the interference of the tree level and the vertex correction terms is given by \cite{Hambye-gs:2003,Sierra:2014tqa} 
\begin{equation}
\epsilon_\Delta^{l_i}=-\frac{1}{4\pi}\sum_{k,j}M_{N_k}\frac{{\cal I}m [\mu_\Delta Y^{ij}\lambda_{ki}^\ast \lambda_{kj}^\ast]}{M_\Delta^2 Tr[YY^\dagger]+\mu_\Delta^2}
\ln \Big (1+\frac{M_\Delta^2}{M_{N_k}^2}\Big). \label{eq:cpsm} 
\end{equation}
In the type-II seesaw limit $M_{\Delta}\ll M_{N_k}$ this formula has been shown to assume a simpler form
\be
 \epsilon_\Delta^{l_i}= \frac{M_\Delta}{2 \pi v^2} \frac{\sqrt{B_l B_\phi}}{\tilde{M_\Delta}}{\cal I}m \Big [ ({\cal M}_{\Delta}^{\nu} {{\cal M}_N^{\nu}}^{\dagger} )_{ii} \Big],
\label{eq:cpsmt2}
\ee
where $B_l,B_\phi$, defined through eq.\ref{Br}, are branching ratios of triplet decay to bileptons and two SM Higgs dublets, respectively, and ${\cal M}_{\Delta}^{\nu}$ and ${\cal M}_N^{\nu}$ are Type-II, Type-I seesaw mass matrices \cite{Sierra:2014tqa}. 
Since both the LH$\nu$s and the RHNs are diagonalised by the same unitary
matrix $U$ in  type-II seesaw dominance  SO(10) models, the formula in
eq.(\ref{eq:cpsm}) is also
 modified    
\begin{equation}
\epsilon_\Delta^{l_i}=-\frac{1}{4\pi}\sum_{k,j} \sum_{m,n} M_{N_k}\frac{{\cal I}m [\mu_\Delta Y^{ij}\lambda_{km}^\ast \lambda_{kn}^\ast U_{mi}U_{nj}]}{M_\Delta^2 Tr[YY^\dagger]+\mu_\Delta^2}
\ln \Big (1+\frac{M_\Delta^2}{M_{N_k}^2}\Big),
\label{epsd_u}
\end{equation}
where $\mu_{\Delta}$ is defined through eq.(\ref{v12610}).
A further necessary consequence is the proportionality ratio between the
LH$\nu$ and RHN mass eigen values given in eq.(\ref{eq:t2con}) of
Sec.\ref{sec:intr}: ${\hat m}_{\nu_1}:{\hat m}_{\nu_2}:{\hat m}_{\nu_3}::{\hat
  M}_{N_1}:{\hat M}_{N_2}:{\hat M}_{N_2}$.
We emphasize that these modifications  are  consequences of
 type-II seesaw dominance in  SO(10). 

In the limit when the mass of the scalar triplet is much lower than the RH neutrino mass eigenvalues $(M_\Delta \ll M_{N_i})$, the CP asymmetry formula can be simplified a step further to give
\begin{equation}
\epsilon_\Delta^{l_i}= \frac{M_\Delta}{2 \pi v^2} \frac{\sqrt{B_l B_\phi}}{\tilde{M_\Delta}} {\cal I}m \Big [ (m_\nu^{II} U^T {m_\nu^I}^\dagger U)_{ii} \Big] ,\label{epsdt2_u}
\end{equation}
where  $m_\nu^I$ and $m_\nu^{II}$ are Type-I, Type-II seesaw mass matrices
given in eq.(\ref{mnu_full}) and eq.(\ref{t2app}). Explicit form of the parameter $\tilde{M_\Delta}$ is given in eq.(\ref{mdt_ex}) .

We further  emphasize that  eq.(\ref{epsn_u}),
eq.(\ref{epsd_u}) and eq.(\ref{epsdt2_u}) are the correct new formulas of CP asymmetry which are valid in the presence
of type-II seesaw dominance in SO(10). In the three models being
discussed here with RHN masses heavier that $M_{\Delta}$, we have used
 eq.(\ref{epsd_u}) throughout this work for the prediction of BAU from heavy scalar triplet leptogenesis. 
\subsection{Boltzmann Equations}
In order to get BAU by solving  Boltzmann equation  we need to take into account 
only the reactions in the hot plasma whose decay rates at that temperature
are comparable to the Hubble rate  
$\Gamma (T) \sim H(T)$. 
\paragraph{}
 The interaction Lagrangian shown in eq.(\ref{l1}) and eq.(\ref{l2}) clearly contains
lepton number violating terms. Whenever $N$ decays to $(l,\phi)$ pair
or $\Delta_L$ decays to $(l_i,l_j)$, lepton number is violated by one or two
units, respectively. Even though  they conserve
baryon number, $(B-L)$ is  violated in these processes. Our aim is to find out the evolution of $(B-L)$ abundance which at later time gets converted into
baryon number through sphaleron interaction. It is worthwhile to mention that for unflavoured leptogenesis $(B-L)$ is conserved by sphalerons whereas for 
flavoured leptogenesis the conserved quantity in the sphaleron process is $(\frac{B}{3}-L_i)$. Therefore, for flavoured leptogenesis, we focuss on 
the evolution of $(\frac{B}{3}-L_i)$ rather than $(B-L)$.  The
evolution of $(B-L)$ (or $(\frac{B}{3}-L_i)$) can not be computed
independently but  includes evolution of other parameters. As
 the relevant Boltzmann equations are a set of coupled differential equations,
 they have to be solved 
simultaneously in order to get solution for any of the
variables. In general
  the asymmetry can be produced by the decay of both $N$
and $\Delta$ and the set of Boltzmann equations contain first order
differentials of RHN density, scalar triplet density and scalar
triplet asymmetry. This scalar triplet asymmetry is arising due
to the fact that, unlike RHNs, triplets are not self conjugate.
  The right hand side of relevant Boltzmann equations consists of interaction terms that tend to change the density of the corresponding variables. Taking into account all such 
interactions, the network of lepton flavour dependent  coupled Boltzmann equations are \cite{Sierra:2011ab,Sierra:2014tqa}
\begin{eqnarray}
&& \dot{Y}_{N_1}=-\Big(\frac{Y_{N_1}}{Y_{N_1}^{eq}}-1 \Big ) \gamma_{D_{N_1}} \label{boltz_big1},\\
&& \dot{Y}_\Sigma=-\Big(\frac{Y_\Sigma}{Y_\Sigma^{eq}}-1 \Big )\gamma_D -2\Big[\Big(\frac{Y_\Sigma}{Y_\Sigma^{eq}}\Big )^2-1 \Big ]\gamma_A \label{boltz_big2},\\
&& \dot{Y}_{\Delta_\Delta}=-\Big[\frac{Y_{\Delta_\Delta}}{Y_\Sigma^{eq}}-\sum_{k} \Big(\sum_i B_{l_i}C_{ik}^l -B_\phi C_k^\phi \Big)\frac{Y_{\Delta_k}}{Y_l^{eq}}\Big ]\gamma_D\label{boltz_big3},\\
&& \dot{Y}_{\Delta_{B/3-L_i}}= -\Big[\Big(\frac{Y_{N_1}}{Y_{N_1}^{eq}}-1 \Big ) \epsilon_{N_1}^{l_i~tot}+\Big( \sum_k C^l_{ik}\frac{ Y_{\Delta_k}}{Y_l^{eq}} +
\sum_k C^\phi_k \frac{Y_{\Delta_k}}{Y_l^{eq}} \Big) K^0_i \Big] \gamma_{D_{N_1}} \nonumber\\
&&\hspace{1.95cm} - \Big[ \Big( \frac{Y_\Sigma}{Y_\Sigma^{eq}} -1 \big) \epsilon^{l_i}_\Delta -2 \sum_j \Big( \frac{Y_{\Delta_\Delta}}{Y_\Sigma^{eq}} -
\frac{1}{2} \sum_k C^l_{ijk} \frac{Y_{\Delta_k}}{Y_l^{eq}}\Big)B_{l_{ij}} \Big]\gamma_D \nonumber\\
&&\hspace{1.95cm}-2 \sum_{j,k}\Big(C^\phi_k +\frac{1}{2}C^l_{ijk}\Big)\frac{Y_{\Delta_k}}{Y^{eq}_l}\Big( \gamma^{\prime \phi \phi}_{l_i l_j}
+ \gamma^{\phi l_j}_{\phi l_i} \Big)
 -\sum_{j,m,n,k} C^l_{ijmnk} \frac{Y_{\Delta_k}}{Y^{eq}_l}\Big( \gamma^{\prime l_n l_m}_{l_il_j} 
+\gamma^{l_m l_j}_{l_il_n} \Big)~~.\nonumber\\
\label{boltz_big4}
\end{eqnarray}
To specify our notational conventions, $Y_{\Delta_X}$ stands for the ratio of number density (or difference of number density) to the entropy density, i.e
$Y_{\Delta_X}=\frac{n_X-n_{\bar{X}}}{s}$, where $n_X~(n_{\bar{X}})$ is
the $X~({\bar X})$ number density. Expressions for number densities of
different particle species $X~(\bar{X})$ are given in 
the Appendix \ref{ap1}. 
All the variables of the differential equations $(Y_{N_1},Y_{\Delta_\Delta},Y_\Sigma,Y_{\Delta_{ B/3-L_i}})$ are functions of $z=M_\Delta/T$.
Therefore, generically, $\dot{Y}_X$ denotes $\dot{Y}_X\equiv\dot{Y}_X(z)=s(z)H(z)\frac{dY_X(z)}{dz}$. The scalar triplet density and asymmetry are denoted as 
$\Sigma=\Delta+\Delta^\dagger$ and $\Delta_\Delta=\Delta-\Delta^\dagger$, respectively. Superscript \textquoteleft$eq$\textquoteright~ denotes the equilibrium value
of the corresponding quantity. Functional forms of all such equilibrium densities are presented in the Appendix \ref{ap1}. Here $B_l$ and $B_\phi$ stand for branching ratio of
$\Delta$ decaying to leptons and $\phi\phi$, respectively and $\Gamma_\Delta^{tot}$ is the total decay width of $\Delta_L$
\cite{Sierra:2014tqa} 
\begin{eqnarray}
&& B_l=\sum_{i=e,\mu,\tau} B_{l_{i}} = \sum_{i,j=e,\mu,\tau} B_{l_{ij}} =\sum_{i,j=e,\mu,\tau} \frac{M_\Delta}{8\pi \Gamma_\Delta^{tot}} |Y^{ij}|^2 ,\\
&& B_\phi =\frac{|\mu_\Delta|^2}{8\pi M_\Delta \Gamma_\Delta^{tot}}, \nonumber\\
&&\Gamma_\Delta^{tot}=\frac{M_\Delta}{8\pi}\Big( \sum_{i,j} |Y_{ij}|^2 +\frac{|\mu_\Delta|^2}{M_\Delta^2} \Big).                     \label{Br}
\end{eqnarray}
It is obvious that the two branching ratios satisfy $B_l+B_\phi=1$.
 Similarly a quantity related to the decay of 
RHN has been defined, $K_i^0=\frac{\Gamma (N_1 \longrightarrow l_i,\phi)}{\sum_i\Gamma (N_1 \leftrightarrow l_i,\phi)}$. Here
$\gamma_D$ is the total reaction density of the triplet including its decay and inverse decay to lepton pair or scalars, and  
$\gamma_{D_{N_1}}$ is the reaction density related to
the lightest RHN ($N_1$).
In this notation 
$\gamma_A$ signifies the gauge induced $2\leftrightarrow2$ scattering of triplets to fermions, scalars and gauge bosons.
Lepton flavour and number $(\Delta L=2)$ violating Yukawa scalar induced $s-$ channel $(\phi \phi \leftrightarrow \bar{l_i} \bar{l_j})$ and $t-$ channel
$(\phi l_j \leftrightarrow \bar{\phi}\bar{l_i} )$ scattering related reaction densities are denoted as 
$\gamma^{\phi \phi}_{l_i l_j}$ and $\gamma^{\phi l_j}_{\phi l_i}$, respectively. Similarly reaction densities related to
Yukawa induced triplet mediated lepton flavour violating $2\leftrightarrow2$ $s-$ channel and $t-$ channel processes are 
given by $\gamma^{l_nl_m}_{l_i l_j}$ and $\gamma^{l_j l_m}_{l_il_n}$, respectively. The primed $s-$ channel reaction densities
are given by $\gamma^\prime=\gamma-\gamma^{\rm on~ shell}$.
Explicit expressions of 
these reaction densities are presented in the Appendix \ref{ap2}. Now the matrices $C^l_{ijk}$ and $C^l_{ijmnk}$ are defined as\cite{Sierra:2014tqa}
\begin{eqnarray}
&&C^l_{ijk}=C^l_{ik} + C^l_{jk},\nonumber\\
&&C^l_{ijmnk}=C^l_{ik} + C^l_{jk}-C^l_{mk}-C^l_{nk},
\end{eqnarray}
where $C^l$ matrix relates the asymmetry of lepton doublets with that of $B/3-L_i$. Similarly $C^\phi$ establishes the relation between the asymmetry of scalar triplet and
$B/3-L_i$, 
\begin{eqnarray}
&& Y_{\Delta_{l_i}}=-\sum_k C^l_{ik} Y_{\Delta_k}\nonumber\\
&& Y_{\Delta_\phi}=-\sum_k C^\phi _k Y_{\Delta_k}
\end{eqnarray}
where $Y_{\Delta_k}$ is the $k-th$ component of the asymmetry vector $\vec{Y}_{\Delta}$ which can be represented as
\begin{equation}
\vec{Y}_{\Delta} \equiv  (Y_{\Delta_\Delta}, Y_\Delta{_{B/3-L_k}} )^T .
\end{equation}
In the above equations   $k=1,2,3$ is the
generation index for fully (three) flavoured leptogenesis whereas  $k=1,~2$ for 
two flavoured leptogenesis and, therefore, the corresponding $\vec{Y}_{\Delta}$ will be a column matrix with four or three entries. The structure of $C^l$
and $C^\phi$ matrices are determined from different chemical equilibrium conditions. Their detailed structure and dimensionality in different energy regimes are discussed in 
Appendix \ref{ap3}. 
\paragraph{}
 As the RHN masses in our models are sufficiently heavier than $M_{\Delta}$, their deacy asymmetries are washed out.  
 Thus, neglecting  the RHN related quantities, the set of Boltzmann equations are  reduced to a simplified form\cite{Sierra:2014tqa}
\begin{eqnarray}
&& \dot{Y}_\Sigma=-\Big(\frac{Y_\Sigma}{Y_\Sigma^{eq}}-1 \Big )\gamma_D -2\Big[\Big(\frac{Y_\Sigma}{Y_\Sigma^{eq}}\Big )^2-1 \Big]\gamma_A ,\\ \label{s_b1}
&& \dot{Y}_{\Delta_\Delta}=-\Big[\frac{Y_{\Delta_\Delta}}{Y_\Sigma^{eq}}-\sum_{k} \Big(\sum_i B_{l_i}C_{ik}^l -B_\phi C_k^\phi \Big)\frac{Y_{\Delta_k}}{Y_l^{eq}}\big]\gamma_D,\\ \label{s_b2}
&& \dot{Y}_{\Delta_{B/3-L_i}}= - \Big[ \Big( \frac{Y_\Sigma}{Y_\Sigma^{eq}} -1 \big) \epsilon^{l_i}_\Delta -2 \sum_j \Big( \frac{Y_{\Delta_\Delta}}{Y_\Sigma^{eq}} -
\frac{1}{2} \sum_k C^l_{ijk} \frac{Y_{\Delta_k}}{Y_l^{eq}}\Big)B_{l_{ij}} \Big]\gamma_D \nonumber\\
&&\hspace{1.95cm}-2 \sum_{j,k}\Big(C^\phi_k +\frac{1}{2}C^l_{ijk}\Big)\frac{Y_{\Delta_k}}{Y^{eq}_l}\Big( \gamma^{\prime \phi \phi}_{l_i l_j}
+ \gamma^{\phi l_j}_{\phi l_i} \Big)
 -\sum_{j,m,n,k} C^l_{ijmnk} \frac{Y_{\Delta_k}}{Y^{eq}_l}\Big( \gamma^{\prime l_n l_m}_{l_il_j} 
+\gamma^{l_m l_j}_{l_il_n} \Big)~~.\nonumber\\
\label{s_b3}
\end{eqnarray}
Simultaneous solution of above three equations enable us to know the value of the asymmetry parameters $(Y_{\Delta_{B/3-L_i}})$ at high value of $z$ where the value of the asymmetry
gets frozen. Then the final value of the baryon asymmetry is 
\begin{equation}
Y_{\Delta_B}=3 \times \frac{12}{37}\sum_i  Y_{\Delta_{B/3-L_i}}, \label{yb}
\end{equation}
where the factor $3$ signifies the degrees of freedom of $\Delta_L$. 
\paragraph{}
When the temperature is above $10^{12}$ GeV,  different lepton flavours lose their distinguishability. Therefore the corresponding Boltzmann equations are free
of lepton flavour indices. This variant of leptogenesis is referred to
as the unflavoured leptogenesis and the set of Boltzmann equations in
eq.(\ref{s_b1})-eq.(\ref{s_b3}) 
are modified \cite{Sierra:2014tqa} 
\begin{eqnarray}
&& \dot{Y}_\Sigma=-\Big(\frac{Y_\Sigma}{Y_\Sigma^{eq}}-1 \Big )\gamma_D -2\Big[\Big(\frac{Y_\Sigma}{Y_\Sigma^{eq}}\Big )^2-1 \Big]\gamma_A \label{boltz_uf1}\\ 
&& \dot{Y}_{\Delta_\Delta}=\Big[\frac{Y_{\Delta_\Delta}}{Y_\Sigma^{eq}}-\sum_{k} \Big( B_l C_k^l -B_\phi C_k^\phi \Big)\frac{Y_{\Delta_k}}{Y_l^{eq}}\big]\gamma_D,\label{boltz_uf2}\\ 
&& \dot{Y}_{\Delta_{B-L}}= - \Big[ \Big( \frac{Y_\Sigma}{Y_\Sigma^{eq}} -1 \big) \epsilon^{l}_\Delta -2 \Big( \frac{Y_{\Delta_\Delta}}{Y_\Sigma^{eq}} -
 \sum_k C^l_{k} \frac{Y_{\Delta_k}}{Y_l^{eq}}\Big)B_{l} \Big]\gamma_D \nonumber\\
&&\hspace{1.6cm}-2 \sum_k \Big( C^\phi_k +C^l_k \Big ) \frac{Y_{\Delta_k}}{Y^{eq}_l} \Big (\gamma^{\prime \phi\phi}_{ll} +
\gamma^{\phi l}_{\phi l} \Big )
, \label{boltz_uf3}
\end{eqnarray}
where $\epsilon^{l}_\Delta~(=\sum_i \epsilon^{l_i}_\Delta)$ is the flavour summed or unflavoured CP asymmetry parameter and the asymmetry  vector $\vec{Y}_\Delta$ is now reduced to a 
column vector with only two entries,
$\vec{Y}_\Delta^T=(Y_{\Delta_\Delta},Y_{\Delta_{B-L}})$. Thus, in this
case also, the final baryon asymmetry is computed using the simple
formula of eq.(\ref{yb}).
\subsection{Parameter Space for Leptogenesis} \label{para_space_lepto}
Whether the lepton flavours should be treated separately  depends
entirely on the phenomenon of flavour decoherence which is generally
   assumed \cite{Sierra:2014tqa} to occur   when lepton Yukawa rate
   becomes faster than the Hubble rate at that very temperature.
Some deeper considerations are needed to avoid possibility of
resulting over simplication.  

\begin{figure}[h!]
\begin{center}
\includegraphics[width=6.5cm,height=6.5cm,angle=270]{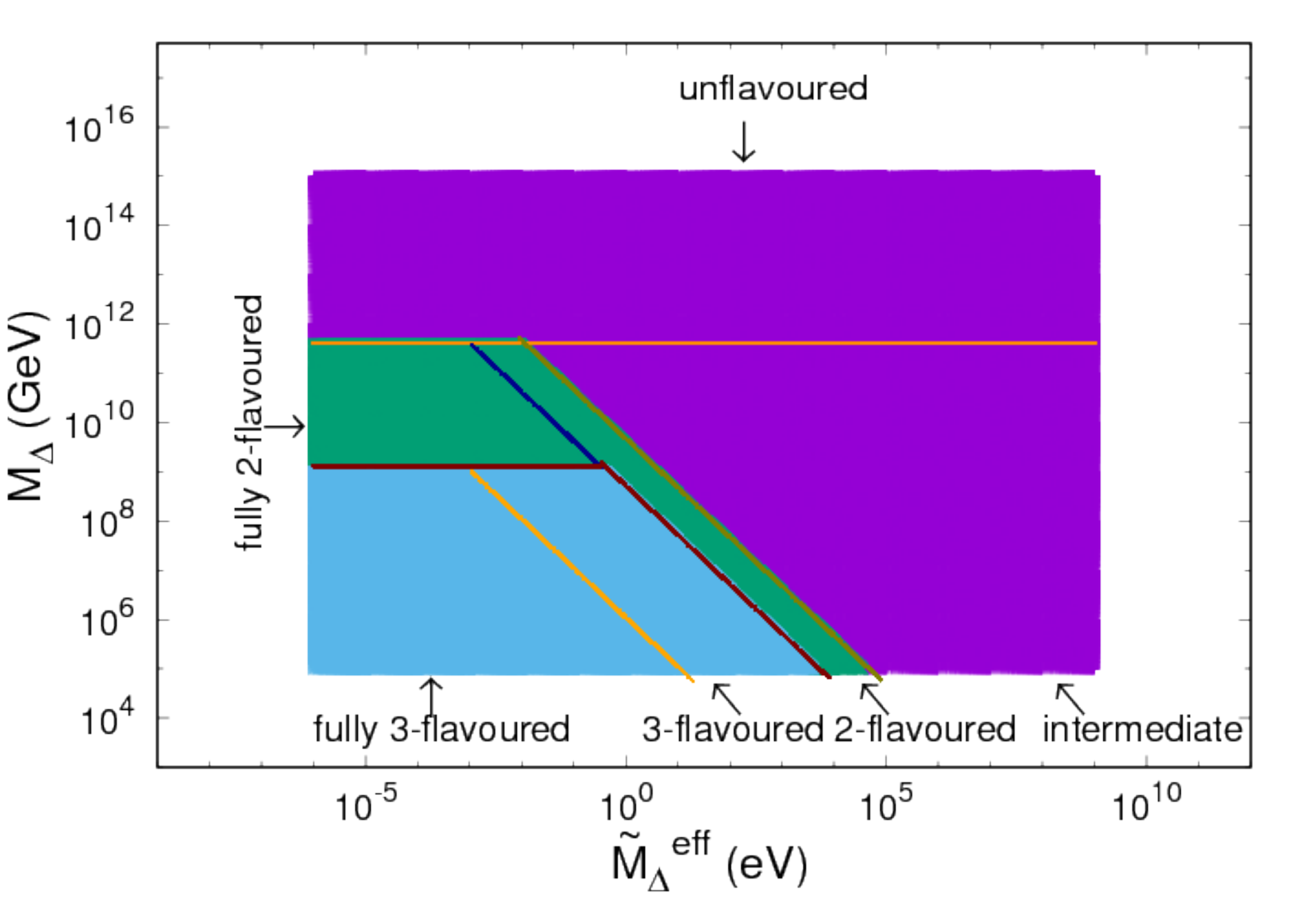}
\caption{Allowed parameter space depicting the region of viability of different kinds of leptogenesis without any constraint on $\tilde{M}_{\Delta}^{eff}$ and $M_\Delta$}
\label{wt2_para}
\end{center}
\end{figure}

In the model with SM + three RHNs + one triplet $\Delta_L$ \cite{Sierra:2014tqa} , the flavour decoherence issue is mainly dictated by the competition of two reactions:
SM lepton Yukawa interaction and inverse decay of lepton to triplet $\Delta_L$. To make this point clear let us suppose that at certain temperature $(T_h)$ during the 
evolution of universe , the lepton Yukawa interaction becomes faster than the Hubble rate, whereas the triplet inverse decay $(ll \rightarrow \bar{\Delta})$ rate
is faster than the lepton Yukawa interaction rate. As a result the charged leptons will inverse decay before they can undergo any charged lepton Yukawa interaction.
Thus separate identity for different lepton flavours still can not be understood. At some lower temperature the inverse decay rate is reduced since it is Boltzmann 
suppressed. At a temperature $T=T_{decoh}$, when the inverse decay rate becomes smaller than the lepton Yukawa interaction rate, the decoherence between the 
lepton flavours is fully achieved. So between the temperature range $(T_h-T_{decoh})$ the flavour decoherence is not fully achieved, i.e within this intermediate 
temperature regime we should not use flavoured leptogenesis formalism. 
\paragraph{}
The decoherence temperature $(T_{decoh})$ is determined by the mass of the scalar triplet $(M_\Delta)$ and effective decay parameter\cite{Sierra:2014tqa} 
\begin{equation}
\tilde{M}_{\Delta}^{eff}= \tilde{M}_\Delta \sqrt{\frac{1-B_\phi}{B_\phi}},\label{mdt_eff}
\end{equation}
where 
\begin{equation}
\tilde{M}_{\Delta}^2 =|\mu_{\Delta}|^2 \frac{v^4}{M_\Delta^4} Tr [Y
  Y^\dagger], \label{mdt_ex} 
\end{equation}
and $B_\phi$ is branching ratio of triplet decay to scalar doublets $\phi\phi$.
\paragraph{}
Decoherence can be fully achieved when our chosen parameter space satisfies the condition that, at a given temperature, lepton triplet inverse decay
rate is slower than the SM lepton Yukawa interaction rate. By imposing this condition we can get an upper limit on $M_\Delta$ as a function 
of $\tilde{M}_{\Delta}^{eff}$
\begin{equation}
\Gamma_{f_i} \geq B_l \Gamma_\Delta^{tot} \frac{Y_{\Sigma}^{eq}}{Y_l^{eq}} ~~~~~~({~f_i=\tau,\mu}).
\end{equation}
This constraint relation can be translated into constraints over $M_\Delta$ and $\tilde{M}_{\Delta}^{eff}$ \cite{Sierra:2014tqa}
\begin{eqnarray}
&& M_\Delta \leq 4 \times \Big ( \frac{10^{-3} {\rm eV}}{\tilde{M}_{\Delta}^{eff}} \Big ) \times10^{11} ~~{\rm GeV} ~~~~({\rm fully ~two ~flavoured}),\label{con_f2f}\\
&& M_\Delta \leq 1 \times \Big ( \frac{10^{-3} {\rm eV}}{\tilde{M}_{\Delta}^{eff}} \Big ) \times10^{9} ~~{\rm GeV} ~~~~({\rm fully ~three ~flavoured})~~. \label{con_f3f}
\end{eqnarray}
In Fig.(\ref{wt2_para}) we have shown the allowed parameter space
for different regimes depending upon
viability of various 
kinds of (flavoured/unflavoured) leptogenesis.
\paragraph{}
 It is essential to state the conditions  which have been used here to generate the parameter space. 
By changing the value of two free parameters  $(\mu_{\Delta},M_\Delta)$ in
every model we are able to vary all such parameters which implicitly or explicitly 
depend upon them. The number of points in the allowed parameter space ($M_\Delta$ vs $\tilde{M}_{\Delta}^{eff}$) is reduced due to imposition of two 
constraint relations as given below.
\paragraph{}
The Type-II seesaw dominance condition which is valid in all SO(10)
models is $\frac{|Y v_L|}{M_D M_N^{-1} M_D^T} \gg 1$. Considering both
real and imaginary parts, 
this condition gives
\begin{eqnarray}
&&Re(m_\nu^I)_{ij} \ll Re(Y v_L)_{ij} \nonumber\\
&&Im(m_\nu^I)_{ij} \ll Im(Y v_L)_{ij}.\label{con_t1}
\end{eqnarray}
\paragraph{}
We estimate reaction densities $\gamma_D$ and $\gamma_{D_{N_1}}$ with only those points which satisfy the condition of eq.(\ref{con_t1}). Then 
we choose only those points which satisfy
\begin{equation}
\gamma_{D_{N_1}} \ll \gamma_D \label{con_n1}
\end{equation}
for each and every value  in the whole range of $z=0-100$.
Fulfillment  of this second constraint relation  allows us to neglect all RHN related quantities
 in eq.(\ref{boltz_big1})-eq.(\ref{boltz_big4}).
Then the estimated leptogenesis is due to the decay of scalar triplet in a type-II
seesaw dominated SO(10) model.\\
\paragraph{}
At first we present the parameter space of $M_\Delta$ vs $\tilde{M}_{\Delta}^{eff}$ depicting different leptogenesis regimes depending upon  distinguishability of lepton flavors
without imposing any constraint on $\tilde{M}_{\Delta}^{eff}$ and $M_\Delta$. We vary $M_\Delta$ (in the range $(10^{5}~-~10^{9})$ GeV)
and $\tilde{M}_{\Delta}^{eff}$ in the range $(10^{-6}~-~10^9)$ eV
independent of each other. Then,  depending upon the constraint
relations in eq.(\ref{con_f2f}) and eq.(\ref{con_f3f}), we place
 the set of points in different leptogenesis regimes and designate them with different colours.
The significance of different regimes shown in the plot of Fig. \ref{wt2_para}
can be explained as follows. The horizontal lines at $(M_\Delta \sim 10^9)$ GeV and $(M_\Delta \sim 10^{12})$ GeV 
basically indicate that above the corresponding values of $M_\Delta$, the $\mu$ and $\tau$ Yukawa interactions can never reach thermal equilibrium which 
simply means that above $M_\Delta \sim 10^{12}$ GeV we can not separate any of the flavours. Similarly, above $(M_\Delta \sim 10^{9})$ GeV, $\mu$ and 
$e$ flavours can never be distinguished. Thus it is not possible to
have fully flavoured
 or three flavoured leptogenesis above $(M_\Delta \sim 10^{9})$ GeV and 2-flavoured leptogenesis
above $(M_\Delta \sim 10^{12})$ GeV. The points in the sky-blue region are obtained  as fully three flavoured using eq.(\ref{con_f3f}) and the points in the green region are obtained as fully two flavoured using  eq.(\ref{con_f2f}). Both the regions are further constrained by the condition  
that $\tau$(or $\mu$) Yukawa interaction is always faster than the $ll
\rightarrow \Delta$ inverse decay rate.\\ 
Tiny patches labeled as three flavoured and two flavoured are obtained by the fulfillment of the condition that 
$\tau$ or $\mu$ Yukawa interaction is faster than the $ll \rightarrow \Delta$ inverse decay rate when $z > z_A$ 
where $z_A=\frac{M_\Delta}{T_A}$,  $T_A$ being the temperature where
the gauge scattering rate is slower than the decay rate.\\
The intermediate regime is little bit tricky. Here depending upon the choice of the parameters $M_\Delta$ and $\tilde{M}_\Delta^{eff}$
at first we have to calculate $T^{\tau}_{decoh}$. If
$T>T^{\tau}_{decoh}$ we would have a unflavoured scenario whereas $T<T^{\tau}_{decoh}$
leads us to two-flavoured scenario. 
\paragraph{}
We now proceed to calculate the quantity $\tilde{M}_{\Delta}^{eff}$
thoroughly following the model under consideration by imposing the Type-II seesaw 
dominance. Then we represent the parameters $\tilde{M}_{\Delta}^{eff}$ and $M_{\Delta}$ graphically again and show the effect of imposition 
of Type-II seesaw dominance constraint. Here we vary $M_\Delta$ in the range
$(10^9~-~10^{15})$ GeV following the gauge coupling unification constraints
 $M_{\Delta} > 10^{9.2}$ GeV ( Model-I ),  $M_{\Delta} > 10^{10.2}$
GeV ( Model-II ), and  $M_{\Delta} > 10^{10.4}$ GeV ( Model-III ).

\begin{figure}[h!]
\begin{center}
\includegraphics[width=6cm,height=6cm]{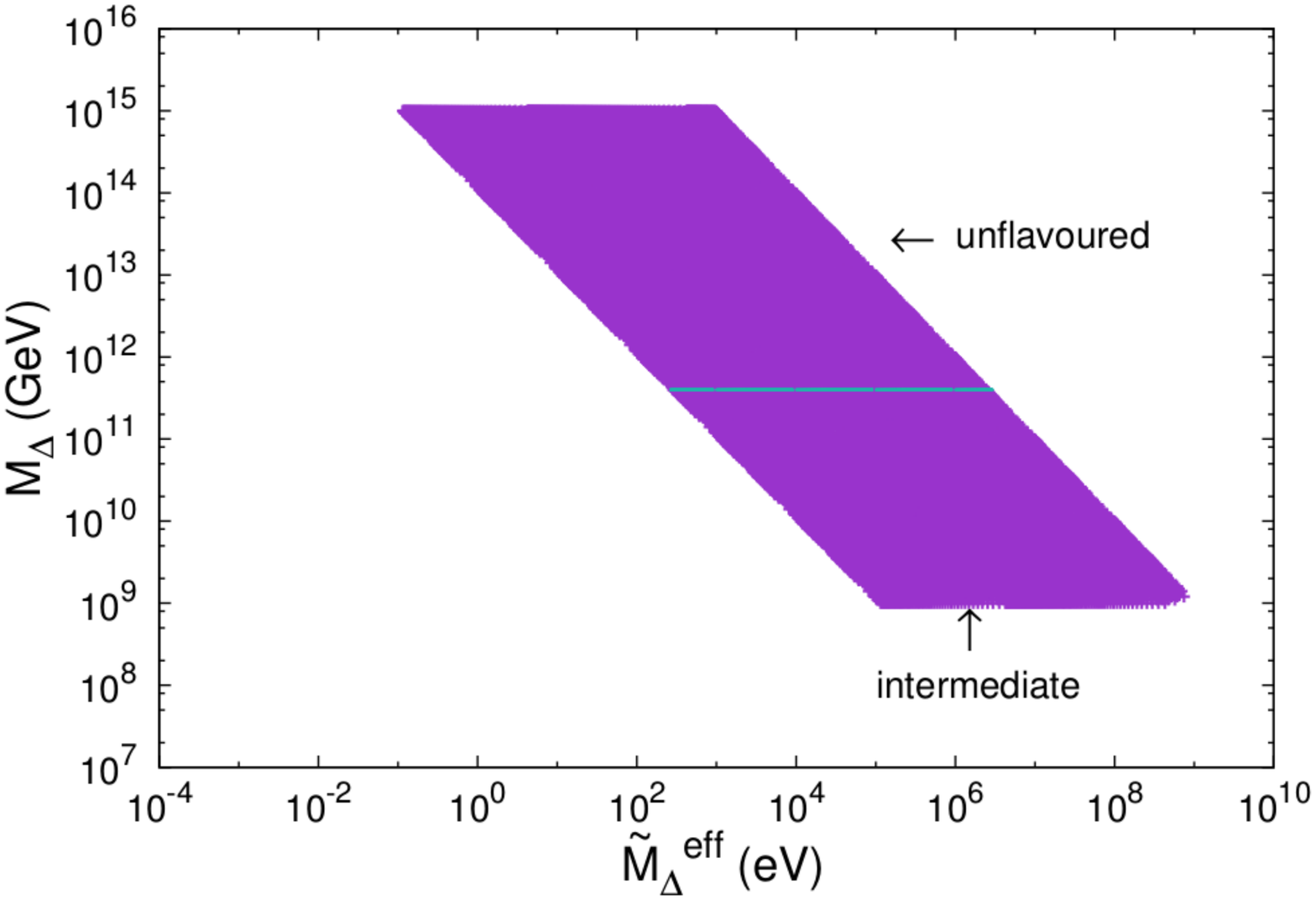}
\hspace{.3cm}
\includegraphics[width=6cm,height=6cm]{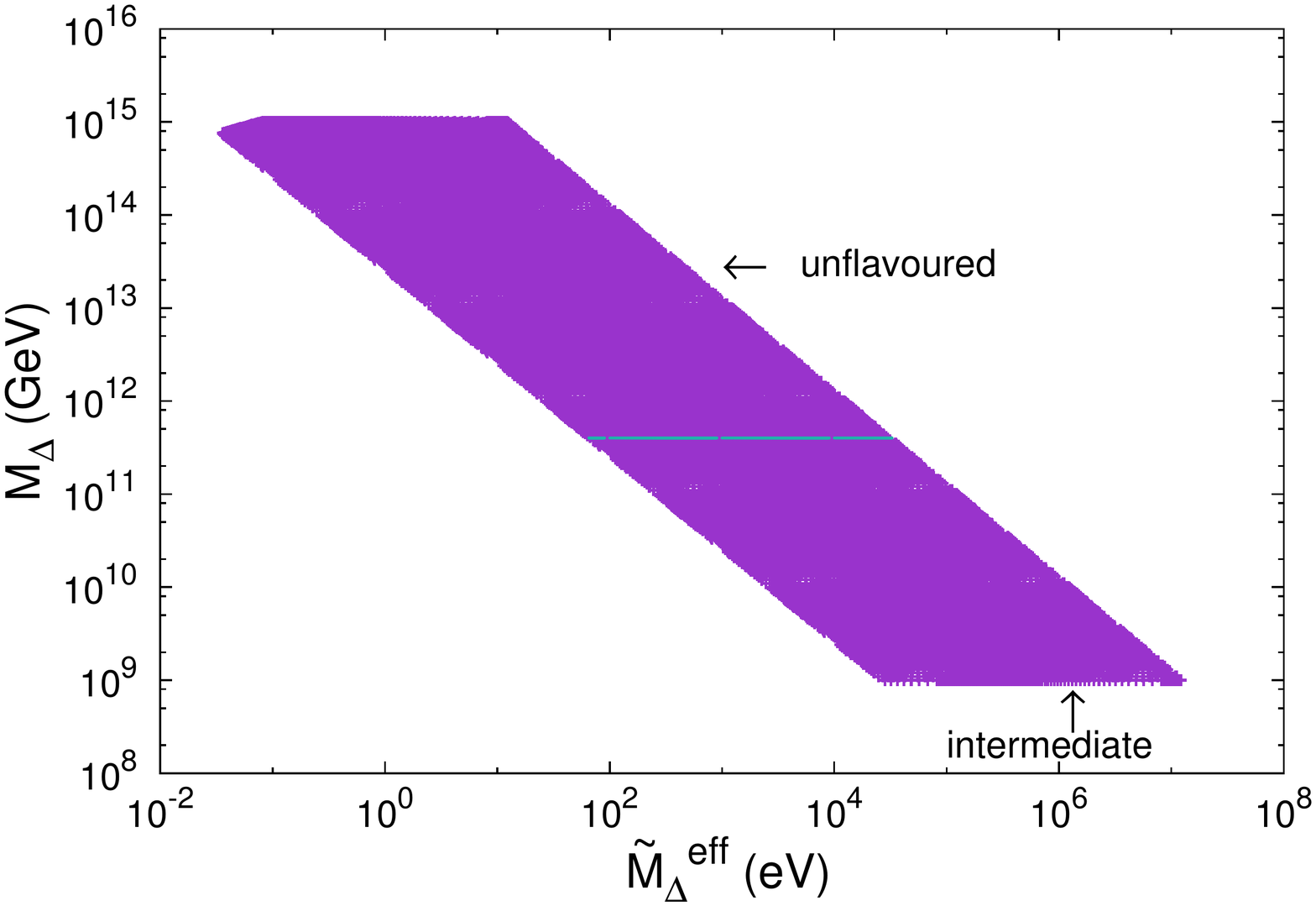}
\caption{Viable regions of
parameter space for different kinds of leptogenesis in the case of
(a) normal hierarchy with $\delta=170^\circ$ (left panel),(b) inverted hierarchy
with $\delta=200^\circ$ (right-panel).}
\label{para_space}
\end{center}
\end{figure}

In order 
to ensure that self-energy correction term does not exceed the tree
level term, we use vlues of  $\mu_{\Delta}\leq M_\Delta$. Thereafter,
imposing Type-II seesaw dominance, we find values of  $\tilde{M}_{\Delta}^{eff}$
corresponding to each set of points $(M_\Delta,\mu_\Delta)$ using
eq.(\ref{mdt_ex}) that also needs the $Y$ matrix which is proportional
to $m_{\nu}$. Choosing the mass eigenvalue of a light neutrino, we
estimate the other two mass eigenvalues. Then using $3\sigma$
values of mixing angles and  any two values of the Dirac
phase $\delta$ we generate a number of sets for $m_{\nu}$ using eq.(\ref{mnu}). Here we present graphically  the 
 parameter space for only those points which at the end produce acceptable values of baryon asymmetry in the $3\sigma$ range. Armed with all these quantities
finally we are able to calculate $\tilde{M}_{\Delta}^{eff}$ for each
set of values of $(M_\Delta,\mu_{\Delta})$.
For a fixed set of oscillation parameters, in this method
$\tilde{M}_{\Delta}^{eff}$ varies with the location of the point $(M_\Delta,\mu_{\Delta})$.
Then using the constraint relations  in
eq.(\ref{con_f2f}) and eq.(\ref{con_f3f}), we subdivide the allowed values of
$(M_\Delta,\mu_{\Delta})$ into different regimes of leptogenesis.\\
We carry out the above mentioned exercise for both normal and inverted
orderings and the corresponding plots are shown in the Fig. \ref{para_space}.
\paragraph{}
In the present context,  a pertinent question to ask is whether it is
acceptable to use a single leptogenesis formalism  such as unflavoured,
or 2-flavoured, or 3-flavoured cases 
and the corresponding single set of Boltzmann equations for the whole $z$ range. Asymmetry is mainly produced at an epoch when $z> z_A$. In this situation if we have 
$z_A>z^\tau_{decoh}$(or $z_A>z^\mu_{decoh}$), then it is justified to use 2-flavoured (or 3-flavoured) leptogenesis formalism for the whole range of $z$.
We have checked  that in the entire violet region  of the parameter space of Fig.\ref{para_space}
we can use the unflavoured leptogenesis formalism. 
\paragraph{}
The plots given in Fig. \ref{para_space} are generated by only those
points which satisfy the two constraint relations, eq.(\ref{con_t1})
and eq.(\ref{con_n1}). Our 
 leptogenesis estimations are then carried out for the points belonging to this parameter space only. Thus in our type-II dominated SO(10)
 where $\Delta_L$ related quantities are important, the corresponding set of Boltzmann equation is 
rightly chosen to be eq.(\ref{boltz_uf1})-eq.(\ref{boltz_uf3})\footnote{Following the same argument 
presented in the paragraph following Fig.3 of Ref\cite{Sierra:2014tqa} 
we also neglect the third term of eq.(\ref{boltz_uf3}) from  numerical computations.}.
\paragraph{}
It is clear from Fig. \ref{para_space} that the allowed parameter space has been reduced  considerably, compared to Fig. \ref{wt2_para}, due to 
imposition of the Type-II dominance constraint. 
Since unification of coupling constants 
forbids us to take $M_\Delta$ below $10^9$ GeV, the 3-flavoured regime is automatically discarded. The reason behind such huge reduction of parameter
space can be explained through some simple mathematical arguments. In our numerical
analysis  the light neutrino mass matrix is due to type-II
 seesaw $m_\nu \simeq Yv_L$. The light neutrino mass matrix is already known to us since we know the mass eigenvalues and mixing angles from 
oscillation data. So $Y$ matrix is known from $Y=(1/v_L)m_\nu$. Again the RH neutrino mass matrix can be expressed
as $M_N \simeq m_\nu \frac{V_{B-L}}{v_L}$. Now we are able to find order of
magnitude values of  type-I remnant  along with dominant Type-II contribution. 
Varying $(M_\Delta,\mu_\Delta)$ we can get different values of $Y$ matrix and $M_N$ matrices. We allow only those set of values of $(M_\Delta,\mu_\Delta)$ which are compatible
with the Type-II dominance constraint. 
We have introduced the Type-II dominance condition in our
analysis by imposing 
\begin{equation}
\frac{(Y v_L)_{ij}}{(M_D^T M_N^{-1}M_D)_{ij}}  >  10 ~.
\end{equation}
Expressing $M_N$ in terms of $m_\nu$ the above equation can be
rewritten to express the Type-II dominance condition 
\begin{equation}
Y_{ij}  >  \frac{10}{V_{B-L}}(M_D^T (m_\nu)^{-1}M_D)_{ij} ~. \label{t2_dom}
\end{equation}
We have used the SO(10) prediction for Dirac neutrinino mass $M_D=M_u=$ up-quark mass.
For NO case
we have taken $V_{B-L}\geq 4\times 10^{17}$GeV. Since numerical values of each quantities in the RHS of above equation are known,
we can have a fair idea about the magnitude of $Y$  matrix needed 
for Type-II seesaw dominance. Exact numerical values are presented in
the Appendix \ref{ap4}.
\paragraph{}
We now try to analyse the effect of constraint  eq.(\ref{con_f2f}) on
$M_\Delta$ and ${\tilde M}_{\Delta}^{eff}$ which has to be 
satisfied in order to be in the fully 2-flavoured regime. Our aim is to express the constraint equation in terms of $Y$ matrix
such that we can infer whether the same $Y$ is simultaneously compatible with type-II dominance eq.(\ref{t2_dom}) and  
the condition to be in the fully 2-flavoured regime. The expression of
$\tilde{M}_{\Delta}^{eff}$ eq.(\ref{mdt_eff}) through eq.(\ref{mdt_ex})
can be simplified by expressing the branching ratio $B_\phi$ in terms of the $Y$ matrix leading to
\begin{equation}
\tilde{M}_{\Delta}^{eff}=\frac{v^2}{M_\Delta} Tr[Y Y^\dagger]\times 10 ^9 ~~{\rm eV}.
\end{equation}
Using this form of ${\tilde M}_{\Delta}^{eff}$ in the constraint relation
of eq.(\ref{con_f2f}) for fully 2-flavoured regime, we get the limiting
values of Yukawa couplings 
\begin{equation}
Tr[ Y Y^\dagger] \leq 6.6\times10^{-6}. \label{2f_fin}
\end{equation}
Thus a point will be in the fully 2-flavoured regime
if the conditions in eq.(\ref{t2_dom}) and eq.(\ref{2f_fin}) are simultaneously satisfied. In  actual 
numerical estimations, as  we have shown
in  Appendix \ref{ap4})  for both the NO and IO cases, it is not at all possible to 
satisfy these two constraint relations given in eq.(\ref{t2_dom}) and eq.( \ref{2f_fin})
simultaneously. Consequently, the
  two- flavoured regime is disallowed
both in NO and IO cases.
In this way we can justify the restricted region  presented in Fig.
\ref{para_space}.\\
\subsection{Remarks on  Baryon Asymmetry Estimation}
We now estimate baryon asymmetry using the points belonging to the parameter space shown in  Fig. \ref{para_space}.
Out of large number of available choices 
we pick some  representative points from each regime for BAU estimation.
Although allowed parameter space has been shown in the 
$M_\Delta$ vs. $\tilde{M}_\Delta^{eff}$ plot, this can be translated into
  $(M_\Delta$ vs. $\mu_{\Delta})$ plot as there is a one-to-one correspondence between the sets:
$(M_\Delta,\mu_\Delta) \rightarrow ( M_\Delta,\tilde{M}_\Delta^{eff}) $.
It is obvious that while demanding that we are taking a point $(M_\Delta,\mu_\Delta)$ from the unflavoured regime implies that the corresponding $( M_\Delta,\tilde{M}_\Delta^{eff})$
will definitely fall in the unflavoured regime of $M_\Delta$ vs $\tilde{M}_\Delta^{eff}$ plot. 
\paragraph{}
 Few important remarks are in order before actual computation. The parameter space shown in Fig. \ref{para_space} are obtained by varying 
$(M_\Delta,\mu_\Delta)$ while keeping $m_\nu$ fixed at its best fit
value as given in eq.(\ref{bfp_nh}) for NO and in eq.(\ref{bfp_ih}) for IO. 
These best fit values include mass squared differences, mixing angles, and the Dirac CP phases. The value of the Dirac CP phase $(\delta)$
has not been measured with desired accuracy till date. Therefore, apart from
the  central value as given in Table \ref{osc}, we have carried
out our analysis also 
for another choice  $\delta=170^\circ$ in the case of NO. 
In  Fig. \ref{para_space} we have shown plots for 
only that value of $\delta$ which at the end produces positive value of baryon asymmetry $(Y_B)$.
\paragraph{}
On the requirement of CP-asymmetry parameters we note that, in the
fully flavoured, 2-flavoured and unflavoured regimes,  we need
  three, two and one CP asymmetry parameters, respectively.
Final value of the generated baryon asymmetry depends crucially upon the sign and magnitude of these CP asymmetry parameters. The magnitude of CP asymmetry parameters 
mainly depends on the mass of the decaying particle which  increases with
 the particle mass. The sign of the CP asymmetry parameters depends on the 
relative phases between the coupling matrices $\lambda$ and $Y$. As the
 phase of $\lambda$ in the chosen up quark diagonal bvasis
 are fixed.
We can tune the phases of $Y$ matrix by taking different values of the
Dirac CP phase $\delta$. As a result  the sign of CP asymmetry
parameter can be changed by changing the value of $\delta$ leading to a positive value of the asymmetry parameter $Y_B$.
\subsection{Baryon Asymmetry for Normal Ordering}
Here at first we calculate  baryon asymmetry taking the points
within the whole violet region  corresponding to unflavoured
+intermediate regime of the left
panel in Fig. \ref{para_space} which is consistent with 
 neutrino data for $\delta=170^{\circ}$.
 When we find unacceptable solutions within best fit values of
 $\delta$, we extend our search region covering   its $3\sigma$ range as quoted in Table \ref{osc}.
\subsubsection{Unflavoured Regime}
\begin{figure}[h!]
\begin{center}
\includegraphics[width=7cm,height=7cm,angle=0]{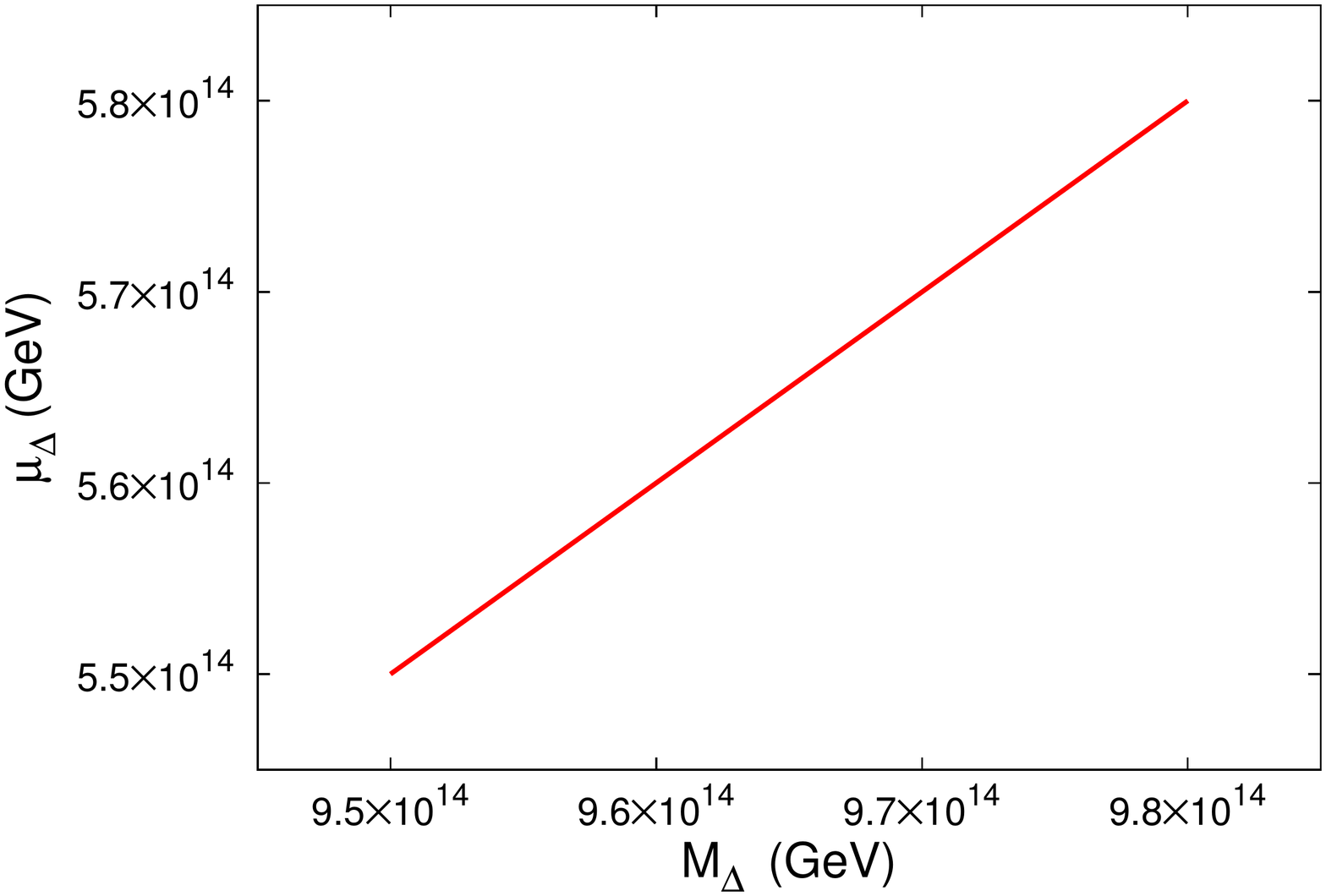}
\hspace{.5cm}
\includegraphics[width=7cm,height=7cm,angle=0]{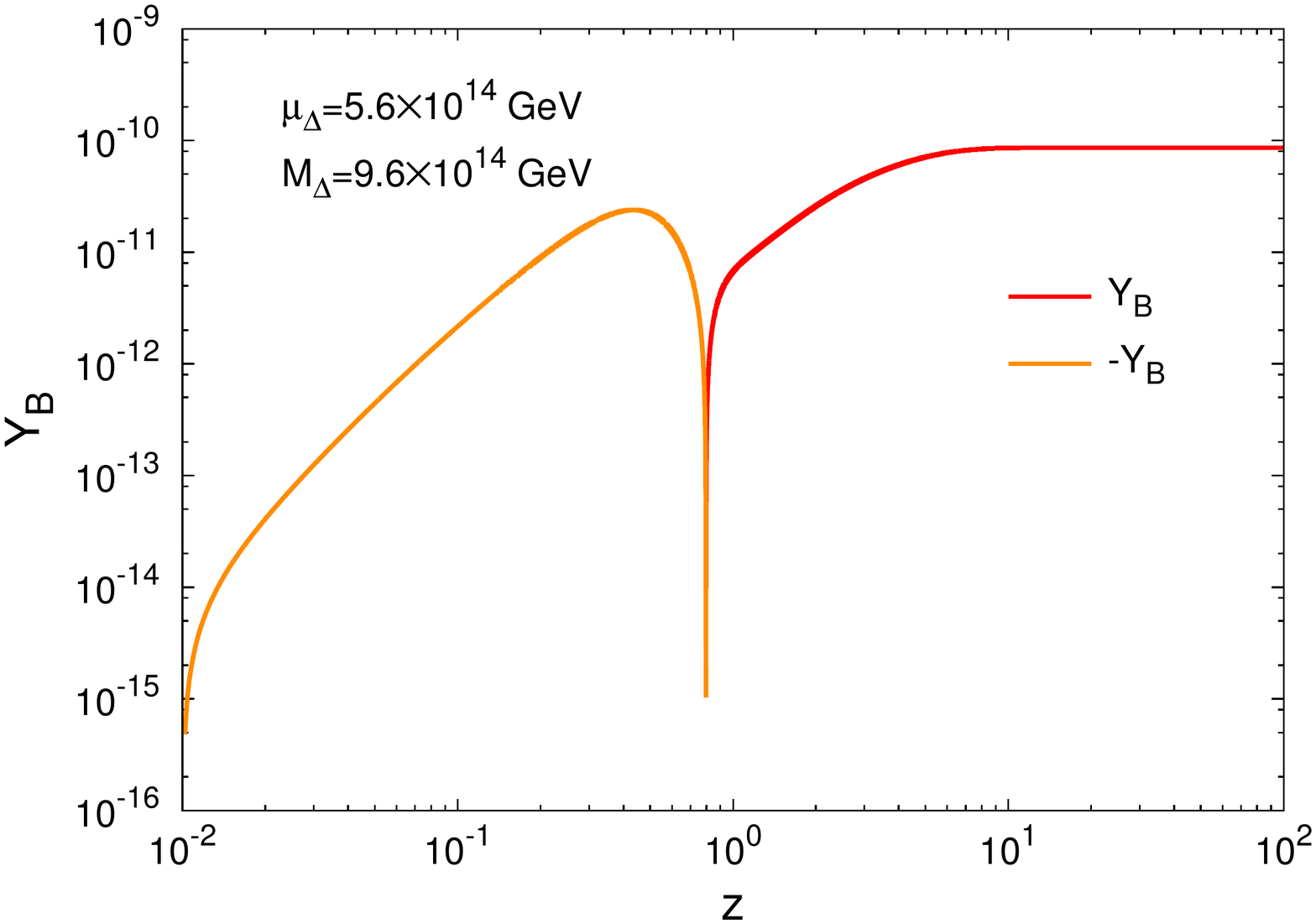}
\caption{Plot of set of pints $(M_\Delta {\rm vs.} \mu_\Delta)$ required to produce $Y_B$ in the experimental range (left panel) and variation of $Y_B$ with $z$ for a definite value of  
$(M_\Delta,\mu_\Delta)$ (right panel). For these NO solutions
 best fit values of the neutrino mass eigenvalues and mixings have
 been used  with phases  
$\delta=170^\circ,\alpha_M=0,\beta_M=0$.}
\label{yb_z1}
\end{center}
\end{figure}

This regime is composed of all the points of the region denoted as unflavoured as well as most of the points from the  region labeled as intermediate.
With these points we calculate the flavour independent CP asymmetry parameters $(\epsilon^{l}_\Delta=\sum_i \epsilon^{l_i}_\Delta)$ and plug them into the 
set of unflavoured Boltzmann equations, eq.(\ref{boltz_uf1})-eq.(\ref{boltz_uf3}), which are then solved to get the final value of $(B-L)$ asymmetry. 
This $(B-L)$ asymmetry is then converted into baryon asymmetry through
sphaleron process through the formula of eq.(\ref{yb}).
Other ingredients required for solving the Boltzmann
equations are $C^l$ and $C^\phi$ matrices discussed  in Appendix Sec.
\ref{ap3}. 
We carry out our analysis with some of the chosen points from the
unflavoured
 regime as depicted in the left panel of Fig. \ref{para_space}.
For  these chosen points the CP asymmetry parameter is found to be
negative and the
$(B-L)$ asymmetry parameter 
$Y_{B}$ also freezes to a positive value for high $z$ in agreement with
Planck satellite data \cite{Planck15} given in eq.(\ref{YB}) and
eq.(\ref{etaB}). We note that  there are a plenty of points in this
regime having such acceptable 
solutions.  
\begin{table}[!h]
\caption{ Numerical values of parameters for NO in the unflavoured regime
  which can produce $Y_B \sim 8.6 \times10^{-11}$ with  
best fit values of the neutrino mass eigenvalues, mixings and
$\delta=170^\circ,\alpha_M=0,\beta_M=0$.}
\label{md_mu1} 
\begin{center}
\begin{tabular}{|c|c|c|c|c|}
\hline
$M_{N_1}/10^{16}$ (GeV)   & $1.084$ & $1.087$ & $1.09$ & $1.09$ \\
\hline
 $M_\Delta/10^{14}$ (GeV) & $9.5$ & $9.6$ & $9.7$ & $9.8$ \\
\hline
 $\mu_\Delta/10^{14}$ (GeV)&  $5.5$ & $5.6$ & $5.7$ & $5.8$ \\
\hline
$\tilde{M}_\Delta^{eff}\times10^{2}$ (eV) &  $0.121$ & $0.120$ & $0.120$ & $0.119$ \\
\hline
$Y_B\times10^{11}$ & $8.53$ & $8.60$ & $8.67$ & $8.71$ \\
\hline
\end{tabular}
\end{center}
\end{table}
In Table \ref{md_mu1} we have shown some of these points. In the left panel
of Fig. \ref{yb_z1} we plot allowed values of $M_\Delta$ vs. $\mu_\Delta$  representing
solutions   $Y_B \sim 8.6 \times10^{-11}$). In the right panel of Fig. \ref{yb_z1} we present the variation of $Y_B$ ( $=Y_{B-L}$) with $z$. Form the plot it is clear
that $Y_B$  indeed freezes to its  experimentally
observed value for large $z$.
\paragraph{}
The whole numerical analysis and successful $Y_B$ solutions have been
obtained for  best fit values
of neutrino data with Dirac phase $\delta=170^\circ$ but with vanishing
  Majorana phases. But in the presence of type-II seesaw dominance in
  SO(10), as there is a one-to-one correspondence between $Y$ and $m_{\nu}$
, the effect of any non-vanishing Majorana phases
or a different Dirac phase can be easily analysed. 
In  Table \ref{t1_ev} we have estimated the type-I seesaw  remnant mass eigen
values $m_{\nu_1}^I,m_{\nu_2}^I,m_{\nu_3}^I$  corresponding to our chosen
parameter space. Their negligible values compared to neutrino data
confirms our type-II seesaw dominance approximation also numerically. 
\begin{table}[!h]
\caption{ Type-I seesaw remnant contribution to neutrino mass
  eigenvalues for some  specific set of parameters in concordance with
  type-II seesaw dominance and BAU with best fit for neutrino data with NO eigen
  values and
$\delta=170^\circ,$ and $\alpha_M=0,\beta_M=0$.}
 \label{t1_ev} 
\begin{center}
\begin{tabular}{|c|c|c|c|c|}
\hline
 $M_\Delta/10^{14}$ (GeV) & $9.5$ & $9.6$ & $9.7$ & $9.8$ \\
\hline
 $\mu_\Delta/10^{14}$ (GeV)&  $5.5$ & $5.6$ & $5.7$ & $5.8$ \\
\hline
$M_{N_1}/10^{16}$ (GeV)   & $1.084$ & $1.087$ & $1.091$ & $1.094$ \\
\hline
$M_{N_2}/10^{16}$ (GeV)   & $9.381$ & $9.409$ & $9.437$ & $9.467$ \\
\hline
$M_{N_3}/10^{17}$ (GeV)   & $5.444$ & $5.460$ & $5.477$ & $5.494$ \\
\hline
$ m_{\nu_1}^I$ (eV) &  $6.15\times10^{-15} $ & $ 6.13\times10^{-15}$ & $6.11\times10^{-15} $ & $6.09\times10^{-15}$ \\
\hline
$m_{\nu_2}^I$ (eV)& $2.13\times10^{-10} $ & $2.13\times10^{-10}$ & $2.12\times10^{-10}$ & $2.11\times10^{-10} $ \\
\hline
 $m_{\nu_3}^I$ (eV)& $ 1.86\times10^{-4}$  & $1.85\times10^{-4} $ & $1.85\times10^{-4} $ & $1.84\times10^{-4} $ \\
\hline  
\end{tabular}
\end{center}
\end{table}

\begin{figure}[h!]
\begin{center}
\includegraphics[width=7cm,height=7cm,angle=0]{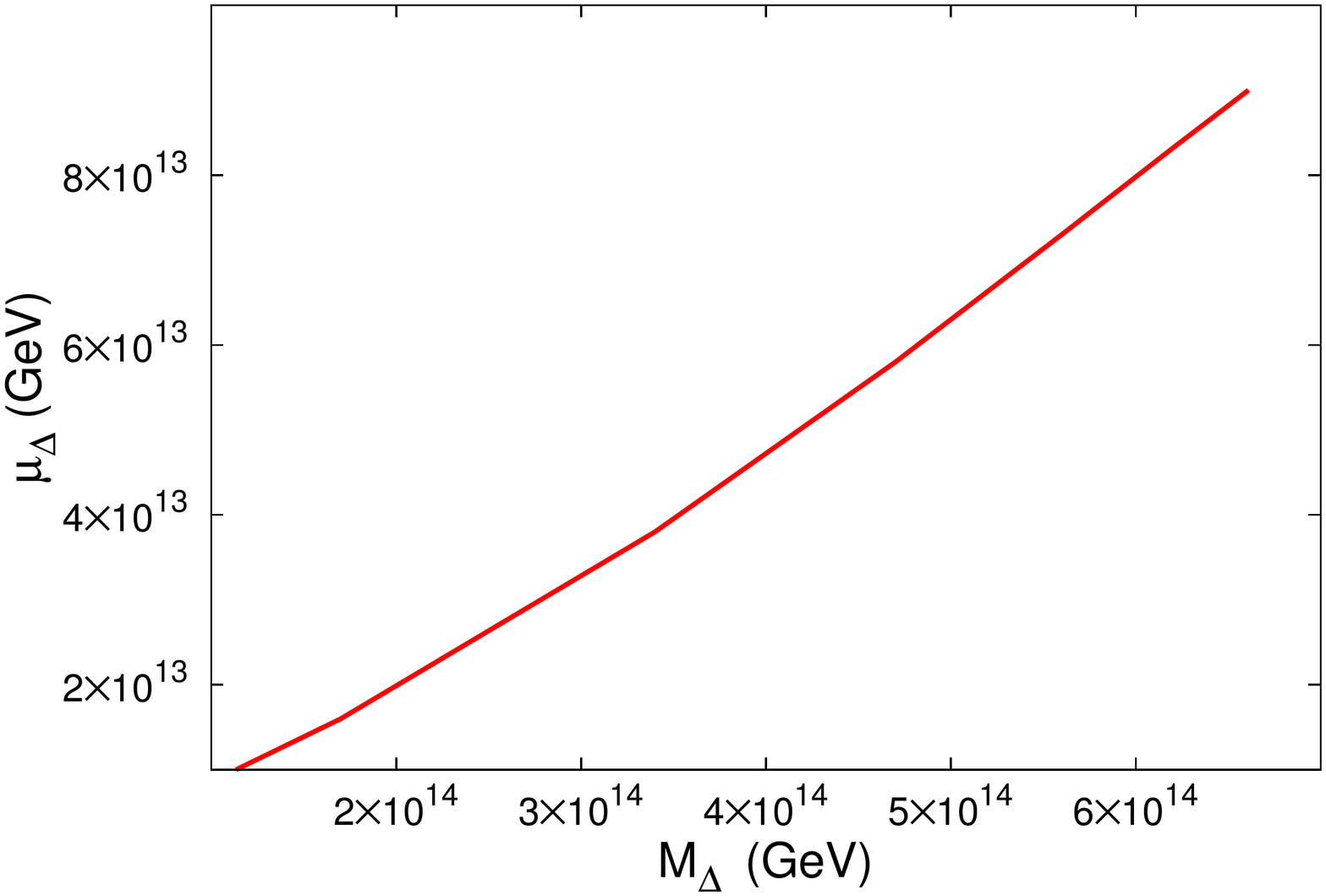}
\hspace{.5cm}
\includegraphics[width=7cm,height=7cm,angle=0]{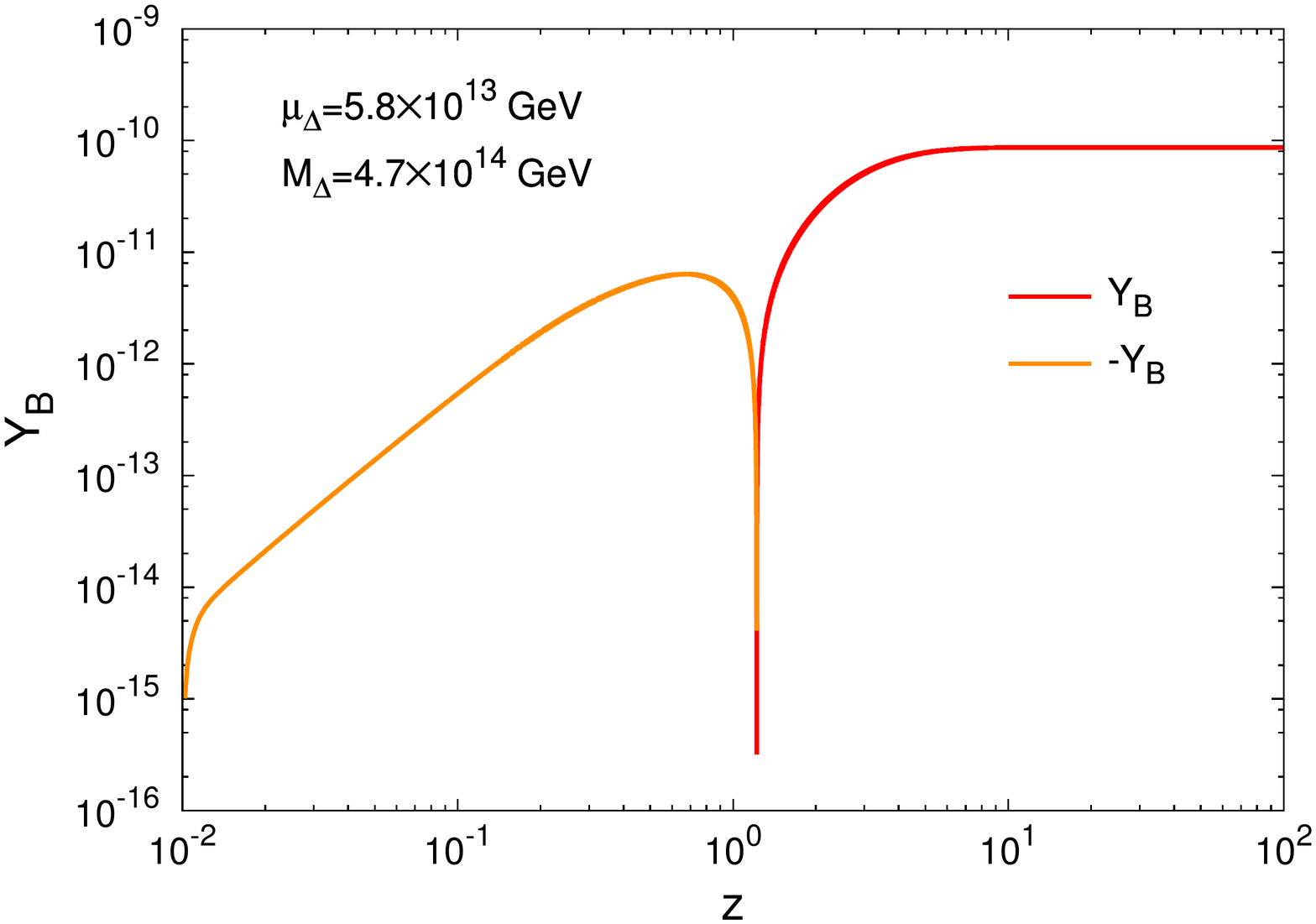}
\caption{Plot of allowed values of $(M_\Delta,\mu_\Delta)$ required to
  produce $Y_B$ in the experimental range (left panel) and variation
  of $Y_B$ with $z$ for a definite set of allowed values of
$(M_\Delta,\mu_\Delta)$ (right panel). In this NO case  best fit to
  neutrino data has been used with phases
$\delta=270^\circ,\alpha_M=45^\circ,\beta_M=45^\circ$. }
\label{yb_z1_majo}
\end{center}
\end{figure}

 As an example we have
repeated the whole procedure
using the best fit values of 
neutrino data but  with a different set of phases
$\alpha_M=45^\circ,\beta_M=45^\circ$ and $\delta=270^\circ$. Numerical solutions are presented in Table
\ref{md_mu1_majo} consistent with  baryon asymmetry value $Y_B
\sim 8.6 \times10^{-11}$. 
In the left panel of Fig.\ref{yb_z1_majo} we present  allowed values
of $M_\Delta$ and $\mu_\Delta$ consistent with
 eq.(\ref{YB}) and in the right panel we show  variation of
 $Y_B$ with $z$ for a specific set of values of $(M_\Delta, \mu_\Delta)$.

\begin{table}[!h]
\caption{ Numerical values of parameters in the unflavoured regime
  consistent with $Y_B \sim 8.6 \times10^{-11}$ and
best fit values of NO neutrino mass eigenvalues, mixings and phases
$\delta=270^\circ,\alpha_M=45^\circ,\beta_M=45^\circ$.}
\label{md_mu1_majo} 
\begin{center}
\begin{tabular}{|c|c|c|c|c|c|c|c|}
\hline
$M_{N_1}/10^{16}$ (GeV)   & $ 0.60$ & $1.19 $ & $2.01 $ & $2.51 $& $ 2.84$ & $3.06 $ & $3.2 $   \\
\hline
 $M_\Delta/10^{14}$ (GeV) & $0.71 $ & $ 1.7$ &  $3.4 $ & $ 4.7$ & $5.6 $ & $6.2 $ & $6.6 $ \\
\hline
 $\mu_\Delta/10^{13}$ (GeV)&  $0.55 $ & $ 1.6$ & $3.8 $ & $0.58 $ & $0.73 $ & $0.83 $& $0.90 $ \\
\hline
$\tilde{M}_\Delta^{eff}\times10^{2}$ (eV) &  $ 0.50$ & $0.82 $ & $1.16 $ & $1.32 $ & $ 1.41$ & $1.48 $ & $1.52 $ \\
\hline
$Y_B\times10^{11}$ & $8.6 $ & $8.65 $ & $8.52 $ & $8.65 $  & $ 8.71$ & $ 8.62$ & $8.61$ \\
\hline
\end{tabular}
\end{center}
\end{table}
\paragraph{}

Few remarks on the choice of the SO(10) (or $U(1)_{B-L}$) breaking VEV  $V_{B-L}=4\times 10^{17}{\rm GeV}$~ are in order. Here we have ensured  type-II seesaw
dominance by comparing each of the type-II matrix element with the
 corresponding element in type-I seesaw. To get successful
leptogenesis with $ Y_B \sim 8\times 10^{-11}$), we have also taken $V_{B-L}
\geq 4 \times 10^{17}$ GeV, but a choice  below this  VEV  may lead to 
$Y_B$ values less tan the Planck\cite{Planck15} data.

Since two-flavoured and three flavoured regimes are not allowed in our
type-II seesaw dominance models as discussed in
Sec.\ref{para_space_lepto}, we do not discuss methodology for
estimation of BAU in these two cases.

\subsection{ Baryon Asymmetry for
  Inverted  Ordering } \label{sec:unfbau}

 In  Fig. \ref{para_space} (right panel)
we have clarified why  the IO case allows only unflavoured regime
which we utilise here to predict BAU.

\begin{figure}[h!]
\begin{center}
\includegraphics[width=7cm,height=7cm,angle=0]{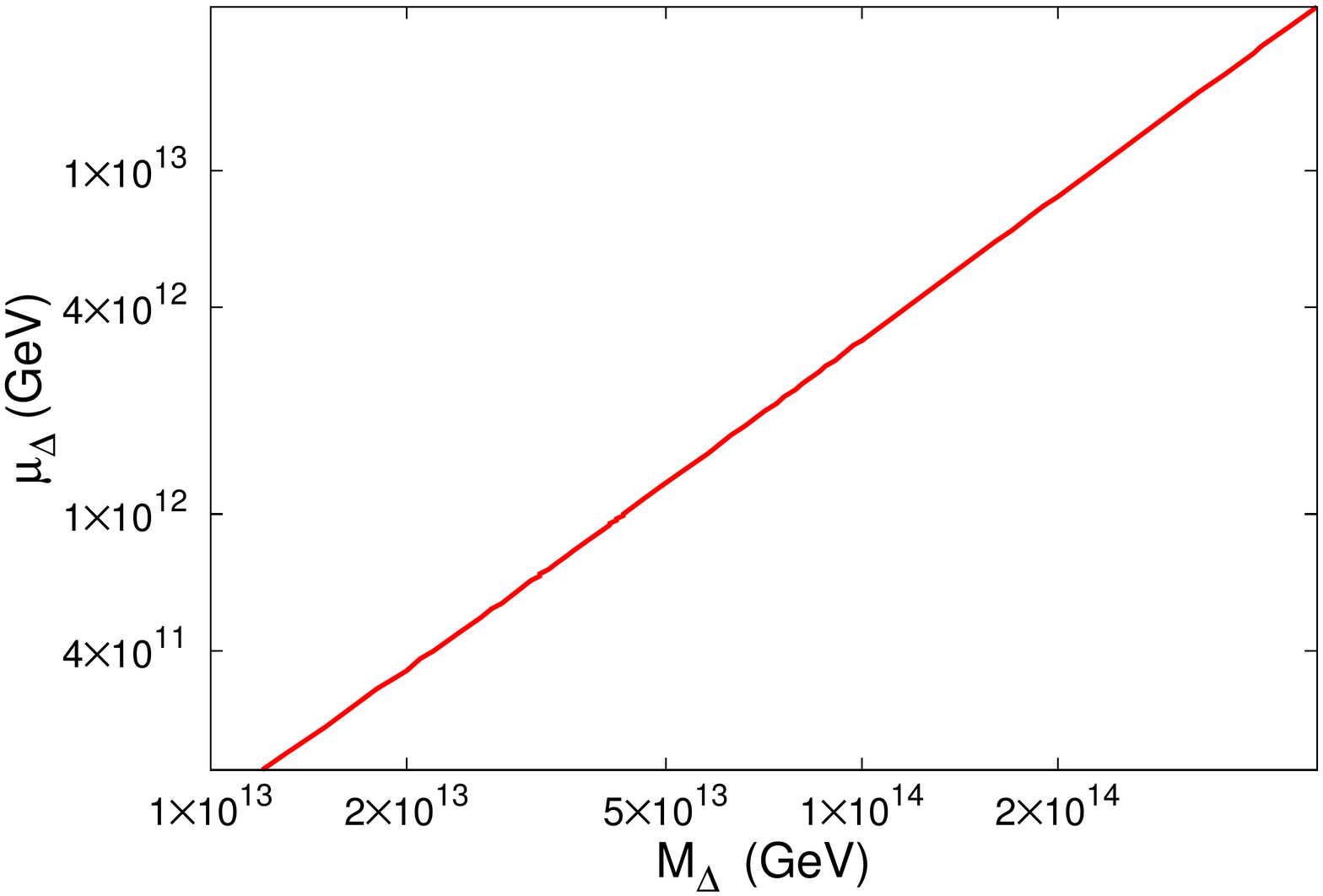}
\hspace{.5cm}
\includegraphics[width=7cm,height=7cm,angle=0]{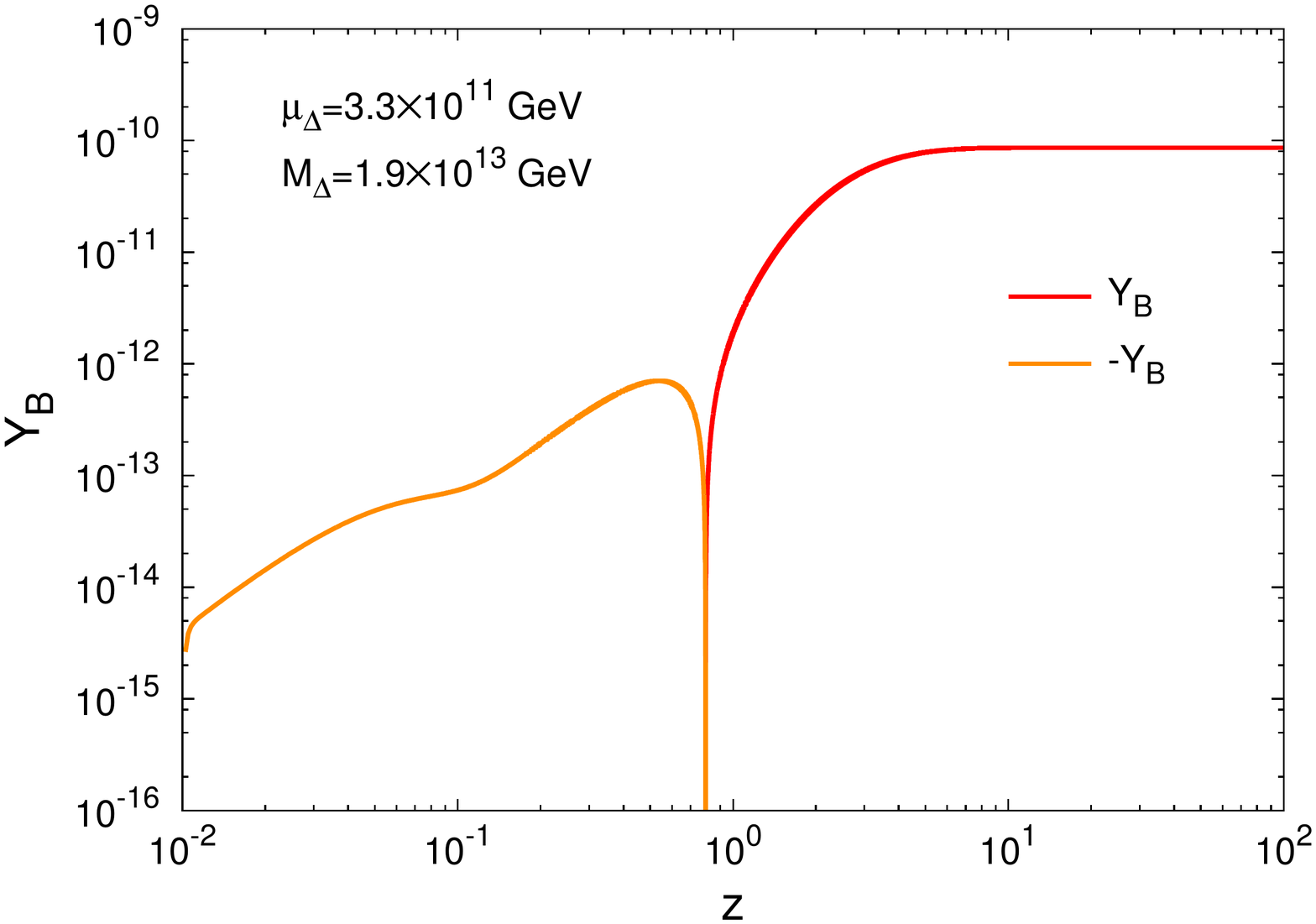}
\caption{Plot of some allowed set of points $(M_\Delta,\mu_\Delta)$ required to produce $Y_B$ in the experimental range (left panel) and variation of $Y_B$ with $z$ for a definite value of  
$(M_\Delta,\mu_\Delta)$ (right panel). In this IO case  best fit
  values of neutrino data have been used 
with phases  
$\delta=200^\circ,\alpha_M=0,\beta_M=0$. }
\label{yb_z2}
\end{center}
\end{figure}

We choose some
representative points from the unflavoured and intermediate regime
depicted as the violet area of right panel in Fig. \ref{para_space}. Then following
 the same steps as NO case, we estimate the asymptotic value of
 $Y_B$ for large enough value of $z$. 
In this case, apart from using best fit values of IO mass eigen values
 and mixings, we have also explored possibility of solutions within
 $3\sigma$ allowed range of Dirac CP phase.
 We have estimated the CP asymmetry parameter for $\delta=284^\circ$ which is
in the fourth quadrant and is the best fit value
\cite{Forero:2014bxa,Esteban:2018azc}, and for
$\delta=200^\circ$ which is in the third quadrant. For $\delta=284^\circ$ the estimated CP asymmetry 
comes out to be positive which yields negative value of baryon
asymmetry $Y_B$ at a large enough value of $z$. On the other hand  using $\delta=200^\circ$ we get negative value of CP
asymmetry parameter and, consequently, we get  positive value of $Y_B$.
in agreement with Planck satellite data.
\begin{table}[!h]
\caption{ Numerical values of parameters in the unflavoured regime
  which can produce $Y_B \sim 8.6 \times10^{-11}$ \label{md_mu2}
  consistent with
best fit values of IO neutrino mass eigenvalues, mixings and phases 
$\delta=200^\circ,\alpha_M=0,\beta_M=0$.}
\begin{center}
\begin{tabular}{|c|c|c|c|c|c|c|c|c|}
\hline
$M_{N_1}/10^{16}$ (GeV)   & $8.92$ & $12.03$ & $14.26$ & $ 17.62$ & $20.55$& $22.29$& $24.53$& $36.20$\\
\hline
 $M_\Delta/10^{13}$ (GeV) & $1.9$ & $3.0$ & $3.9$ & $5.5$ & $7.1$ & $8.10$ & $9.50$& $18.00$\\
\hline
 $\mu_\Delta/10^{11}$ (GeV)&  $3.3$ & $6.1$ & $8.7$ & $14.0$ & $20.0$ & $24.00$ & $30.00$& $73.00$ \\
\hline
$\tilde{M}_\Delta^{eff}$ (eV) &  $5.14$ & $5.92$ & $6.39$ & $6.92$& $7.30$& $7.53$& $7.77$& $8.93$\\
\hline
$Y_B\times10^{11}$ & $8.6$ & $8.65$ & $8.57$ & $8.60$ & $8.66$ & $8.64$ & $8.67$& $8.66$\\
\hline
\end{tabular}
\end{center}
\end{table}
In  Table \ref{md_mu2} we present only those points which produce
$Y_B$  in agreement with  experimental data. In
Fig.\ref{yb_z2} (left panel) we depict the values of $\mu_\Delta$ and
$M_\Delta$ for which  $Y_B \sim 8\times10^{-11}$. We choose one such
combination of $(\mu_\Delta,M_\Delta)$ 
from the $\mu_\Delta$ vs $M_\Delta$ plot and show the variation of
$Y_B$ with $z$. This is presented by  the right panel of Fig.\ref{yb_z2}. 
\paragraph{}
From the systematic analysis of leptogenesis presented in the above sections it is inferred that positive baryon asymmetry in the experimentally observed range can be produced
using best fit values of neutrino oscillation observables with a large
Dirac CP phase in type-II seesaw dominated SO(10) scenario where the lepton asymmetry is generated mainly due 
to the decay of the left handed scalar triplet and this observation
holds both for normal and inverted orderings of neutrino masses. We also
find that predicted values of BAU are in concordance with
$\theta_{23}$ in the second octant.

\section{Summary and Conclusion}\label{sec:sumconc}
In this work we have addressed six  limitations of the standard model: (i) Neutrino mass, (ii) Baryon
asymmetry of the Universe, (iii) Origin of  dark matter  and its stability, (iv) Vacuum stability of SM scalar potential, (iv) Origin
of three gauge forces of SM, and (vi) Experimentally observed proton
stability. For this purpose we have successfully constructed three different
type-II seesaw dominance models using the symmetry breaking pattern SO(10) $\to$ SU(5) $\to$ 
SM which unify gauge couplings and predict the mass of LH triplet scalar
$\Delta_L$ mediating type-II seesaw to be lighter than RHN masses
and the SU(5) unification scale. Manifestly, each of these models allows
 type-II seesaw dominance and triplet
leptogenesis. SO(10) breaking through ${126}_H$ predicts matter parity ($Z_{MP}$) as 
stabilising gauged discrete symmetry for its non-standard fermionic
(scalar) DM
components with even (odd) value of $Z_{MP}$. Neither 
these intrinsic DM properties, nor the coupling unification can  be predicted in SM
extensions \cite{GAMBIT,Sierra:2014tqa}.     
 In our SO(10) models the same Majorana Yukawa coupling matrix $Y$ defines the matricial
 structures of both  the
 type-II seesaw  and the RHN mass matrices. As a result the same PMNS
matrix ($U$) that diagonalises the neutrino mass matrix in the type-II
seesaw dominance approximation also diagonalises the RHN mass matrix. This prevents the
freedom  of choosing a RHN diagonal basis used in \cite{Sierra:2014tqa}. 
Further the underlying quark lepton symmetry of SO(10) predicts the Dirac neutrino Yukawa coupling from the up-quark mass matrix
$\lambda\simeq \sqrt 2 M_{u}/v$ which is unknown in SM extension
\cite{Sierra:2014tqa}. In addition, as the SO(10) breaking
VEV  $V_{B-L}\ge 10^{17}$ GeV and the known  Majorana coupling matrix $Y$
also  determine the RHN mass matrix in our models, the type-I seesaw
remnant occurring in the CP-asymmetry formula is also known.
These predictions and constraints specific to type-II seesaw dominance
in SO(10) are naturally absent in the SM extensions \cite{Sierra:2014tqa}. 
 Thus, for realisation of leptogenesis
in the presence of type-II seesaw dominance in SO(10), we predict a new class of CP-asymmetry
formulas which have been explicitly stated through eq.(\ref{epsn_u}), eq.(\ref{epsd_u}) and eq.(\ref{epsdt2_u}) in the corresponding cases.
 As the
three RHNs are
 fundamental to SO(10) unification, unlike the SM extension \cite{Sierra:2014tqa}, our models do not predict
purely triplet leptogenesis with only LH triplets  excluding
RHNs. Rather they predict RHN loop mediated triplet leptogenesis. Despite these
constraints, all our models predict right value of baryon asymmetry in
agreement with Planck data \cite{Planck15} for both NO and IO neutrino masses through unflavoured
leptogenesis. Another important outcome of these new SO(10)
models  is that their type-II seesaw dominance is capable of
fitting any neutrino data including $\theta_{23}$ in the second octant
and  Dirac and/or Majorana phases. In Model-I and Model-II, we have
further predicted a real scalar DM
candidate of mass $\sim {\cal O} (1)$ TeV  having odd matter parity originating from ${16}_H$ of SO(10) that satisfies observed
 relic density and mass bounds from indirect and direct detection experiments. This real scalar DM 
 also completes the SM vacuum stability. In Model-III an
alternative solution to DM and vacuum stability problems has been noted to be due to
this real singlet scalar plus the fermion triplet $\Sigma_F(3,0,1)$ \cite{Frig-Ham:2010}  having 
masses near $\sim 1$ TeV scale. But compared to non-perturbative treatment of
Sommerfeld enhancement needed for
2.7 TeV fermionic DM, this combined DM of lower masses brings 
the relic density estimation
to the ambit of perturbation theory and SO(10) grand unification where the  SM
vacuum instability is also resolved by the real scalar singlet component.
All the three  SO(10) models predict  unification scales
$M_{SU(5)}=10^{15.2}-10^{15.3}$ GeV and proton lifetimes accessible
to Super-Kamiokande or Hyper-Kamiokande experiments even after
reasonable threshold effects are included.\\
We conclude that non-SUSY SO(10) with  matter parity conserving
spontaneous breaking  through SU(5) route provides a self sufficient theory
for  neutrino mass via type-II seesaw dominance, baryon asymmetry via triplet leptogenesis, intrinsic dark matter and its stability,
vacuum stability of SM scalar potential through TeV scale  dark
matter, precision gauge
coupling unification  with experimentally verifiable proton lifetime,
and explanation of observed parity violation as monopoly of weak interaction via
asymptotic parity restoration. The new class of modified CP-asymmetry formulas  derived in this work are
expected to hold in other type-II seesaw dominated SO(10) models even
in the absence of SU(5) intermediate breaking.   
  
\section{Appendix}
\subsection{Renormalization Group Equations for  Scalar Quartic Couplings}\label{rgeq}
The RG equations for the scalar quartic couplings up to one loop level are given by
\begin{eqnarray}
&&\frac{d \lambda_{\phi }}{d \ln \mu}= \frac{1}{16\pi^2}\left[ ( 12 h_t^2 -3 {g_{1Y}}^2 -9g_{2L}^2) \lambda_\phi -6 h_t^4 +
\frac{3}{8}\{ 2 g_{2L}^4 +({g_{1Y}}^2 +g_{2L}^2)^2 \} +24 \lambda_\phi^2 + 4 \lambda_{\phi \xi}^2 \right]~,\nonumber\\
&&\frac{d \lambda_{\phi \xi}}{d \ln \mu}= \frac{1}{16\pi^2}\left[\frac{1}{2}(12h_t^2 -3 {g_{1Y}}^2 -9 g_{2L}^2)\lambda_{\phi \xi}+
4\lambda_{\phi \xi} (3 \lambda_\phi +2 \lambda_\xi) + 8 \lambda_{\phi \xi}^2 \right]~, \nonumber\\
&&\frac{d \lambda_{ \xi}}{d \ln \mu}= \frac{1}{16\pi^2}\left[ 8 \lambda_{\phi \xi}^2 + 20 \lambda_\xi^2 \right].~\label{rge}
\end{eqnarray}
The RGEs for SM gauge couplings and top quark Yukawa coupling at two loop level are given by
\begin{eqnarray}
{dh_t \over d\ln \mu} &= & {1 \over 16\pi^2}\left({9 \over 2}h_t^2-{17 \over 12}g_{1Y}^2
 -{9 \over 4}g_{2L}^2-8g_{3C}^2 \right)h_t \\ \nonumber
 & +  & {1 \over (16\pi^2)^2}\{-{23 \over 4}g_{2L}^4-{3 \over 4}g_{2L}^2g_{1Y}^2+{1187 \over 216}g_{1Y}^4 + 9g_{2L}^2g_{3C}^2+{19 \over 9}g_{3C}^2g_{1Y}^2-108g_{3C}^4 
\\ \nonumber
 & +& \left({225 \over 16}g_{2L}^2+{131 \over 16}g_{1Y}^2+36g_{3C}^2 \right)h_t^2+6(-2h_{t}^4-2h_{t}^2\lambda_{\phi}+\lambda_{\phi}^2)\},  \\ \nonumber
 {dg_{1Y} \over d \ln \mu} & = & {1 \over 16\pi^2}\left({41 \over 6}g_{1Y}^3\right)+{1 \over (16\pi^2)^2}\left({199 \over 18}g_{1Y}^2+{9 \over 2}g_{2L}^2+{44 \over 3}g_{3C}^2-
{17 \over 6}h_t^2\right)g_{1Y}^3 ,\\ \nonumber
{dg_{2L} \over d \ln \mu} & = & {1 \over 16\pi^2}\left(-{19\over 6}g_{2L}^3\right)+{1 \over (16\pi^2)^2}\left({3 \over 2}g_{1Y}^2+{35 \over 6}g_{2L}^2+12g_{3C}^2-
{3 \over 2}h_t^2\right)g_{2L}^3, \\ \nonumber
{dg_{3C} \over d \ln \mu} & = & {1 \over 16\pi^2}\left(-7g_{3C}^3\right)+{1 \over (16\pi^2)^2}\left({11 \over 6}g_{1Y}^2+{9 \over 2}g_{2L}^2-26g_{3C}^2-
2h_t^2\right)g_{3C}^3,
\end{eqnarray}
where $g_{2L},g_{1Y},g_{3C}$ are the gauge couplings corresponding to $SU(2)_L,U(1)_Y,SU(3)_C$ respectively and $h_t$ is the top quark Yukawa coupling.

\subsection{Formulas for Type-II Seesaw Dominance and Leptogenesis}
\subsubsection{Number Density of  Particle Species} \label{ap1}
Using Maxwell Boltzmann distribution for massless (relativistic) as well as massive particles, the number densities are given by\cite{Sierra:2011ab,Sierra:2014tqa}
\begin{eqnarray}
&& n_\Sigma^{eq} (z)=n_\Delta^{eq} (z) +n_{N}^{eq} (z)^\dagger~, \\
&& n_\Delta^{eq} (z)=\frac{3 M_\Delta^3 K_2(z)}{2 \pi^2 z}, \\
&& n_{N_1}^{eq} (z) =\frac{M_\Delta^3 r^2 K_2(rz)}{\pi^2 z}, \\
&& n_{l,\phi}^{eq} (z)=\frac{2 M_\Delta^3}{\pi^2 z},
\end{eqnarray}
where $r=\frac{M_{N_1}}{M_\Delta}$ and $K_2(z)$ is the modified Bessel function of second kind. The expressions of entropy density and Hubble parameter are listed below.
\begin{eqnarray}
&& s(z)=\frac{4 g^\ast M_\Delta^3}{\pi^2 z^3},\\
&& H(z)=\sqrt{\frac{8 g^\ast}{\pi^2}}\frac{M_\Delta}{M_{\rm Planck} z^2},
\end{eqnarray}
with effective relativistic degrees of freedom $g^\ast=118$ and Planck mass $M_{\rm Planck}=1.22\times10^{19}$ GeV.
\subsubsection{Reaction Densities} \label{ap2}
Decay $(1\rightarrow 2)$ related reaction densities for lightest RH neutrino ($N_1$) and scalar triplet $(\Delta_L)$ are \cite{Sierra:2011ab,Sierra:2014tqa}
\begin{eqnarray}
&& \gamma_{D_{N_1}}=\frac{1}{8 \pi^3} \frac{M_\Delta^3 r^4 K_1(rz) (\lambda \lambda^\dagger)_{11} }{z M_{N_1}},\\
&& \gamma_D= \frac{K_1(z)}{K_2(z)} n_\Sigma^{eq} (z) \Gamma_\Delta^{tot} 
\end{eqnarray}
respectively, where $\lambda$ is the Dirac neutrino Yukawa coupling and $\Gamma_\Delta^{tot}$ is the total triplet decay width.
The generic expression of $(2\leftrightarrow2)$ scattering reaction densities is
\begin{equation}
 \gamma_s= \frac{M_\Delta^4}{64 \pi^4} \int_{x_{min}}^{\infty}  \sqrt{x} \frac{(z\sqrt{x}) \hat{\sigma_s}}{z} dx \label{gen_s}
\end{equation}
where $x=s^\prime/M_{\Delta}^2$ ($s^\prime$ centre of mass energy) and $\hat{\sigma_s}$ is the reduced cross section.
For gauge induced process $x_{min}=4$ and Yukawa induced process it is $x_{min}=0$. The reduced cross sections for the gauge induced processes is \cite{Sierra:2011ab,Sierra:2014tqa}
\begin{eqnarray}
 \hat{\sigma_A} & = &\frac{2}{72 \pi} \Big \{ (15 C_1-3 C_2) \omega + (5 C_2 -11 C_1 ) \omega^3 + 3 (\omega^2 -1)[2 C_1 + C_2 (\omega^2 -1) ] \ln \Big (\frac{1+\omega}{1- \omega} \Big) 
 \Big \} \nonumber \\ & + &
 \Big  (  \frac{50 g_{2L}^4 +41 {g_{1Y}}^4}{48 \pi} \Big ) \omega^{\frac{3}{2}}, \label{gauge_s}
\end{eqnarray}
where $\omega \equiv \omega (x) =\sqrt{1-4/x}$ and $C_1= 12 g_{2L}^4 +3 g_{1Y}^4 +12 g_{2L}^2 g_{1Y}^2,~C_2=6 g_{2L}^4+3 g_{1Y}^4 + 12 g_{2L}^2 g_{1Y}^2$. 
Reduced cross sections for $\Delta L=2$ $s-$ channel and $t-$ channel scatterings  are 
\begin{eqnarray}
&& \hat {\sigma}^{\phi \phi}_{l_i l_j} = 64 \pi B_\phi B_{l_{ij}} \delta^2 \Big[\frac{x}{(x-1)^2 +\delta^2}\Big],\\
&& \hat {\sigma}^{\phi l_j}_{\phi l_j} = 64 \pi B_\phi B_{l_{ij}} \delta^2 \frac{1}{x} \Big[ \ln (1+x) - \frac{x}{1+x} \Big ],
\end{eqnarray}
where $\delta=\Gamma^{tot}_\Delta/M_\Delta$. Similarly the reduced crosssections for lepton flavor violating processes  can be written as
\begin{eqnarray}
&& \hat {\sigma}^{l_n l_m }_{l_i l_j} = 64 \pi B_{l_{nm}} B_{l_{ij}} \delta^2 \Big[\frac{x}{(x-1)^2 +\delta^2}\Big],\\ 
&& \hat {\sigma}^{l_j l_m }_{l_i l_n} = 64 \pi B_{l_{nm}} B_{l_{ij}} \delta^2 \Big[ \frac{x+2}{x+1} -\ln(1+x) \Big] ~.\label{lfv2}
\end{eqnarray}
Reaction densities of different scattering processes
$(\gamma_A,\gamma^{\phi\phi}_{l_i l_j},\gamma^{ l_n l_m}_{l_i
  l_j}~{\rm etc} )$ can be estimated using
 eq.(\ref{gen_s}), and eq.(\ref{gauge_s}) -eq.(\ref{lfv2}). The resonant intermediate state subtracted reaction densities are expressed as 
\begin{eqnarray}
&&\gamma^{\prime \phi\phi}_{l_i l_j} = \gamma^{\phi\phi}_{l_i l_j} -B_{l_{ij}} B_\phi \gamma_D \\
&&\gamma^{\prime l_n l_m}_{l_i l_j} = \gamma^{ l_n l_m}_{l_i l_j} - B_{l_{ij}} B_{l_{nm}} \gamma_D ~.
\end{eqnarray}

\subsubsection{$C^l$ and $C^{\phi}$ Matrices}\label{ap3}
$C^l$ and $C^{\phi}$  are  coupling matrices through which
lepton asymmetry and scalar doublet asymmetry are related to $(B/3
-L_i)$ and triplet decay asymmetry. These  are determined
by solving a constrained set  of equations involving chemical potentials
where the number of active chemical potentials
and  chemical equilibrium conditions strictly depend upon the
temperature regime. After fixing the right set of equations,
constraint relations and nonzero chemical potentials, the corresponding
equations are solved in terms of some common set of variables which are chosen to be $\mu_{B/3 -L_i},\mu_\Delta$. Then this solution enables one to set up a relation between asymmetry of the 
particles in the heat bath and independent parameters $(Y_{\Delta_\Delta},Y_{B/3 -L_i} )$ of the asymmetry vector. Generally it is assumed that lepton flavour decoherence takes place at 
a temperature when the lepton Yukawa interaction rate becomes faster than the Hubble rate, but actually the decoherence is achieved when the SM lepton Yukawa interaction rate becomes faster
than the inverse decay process $(ll \rightarrow
\bar{\Delta})$. Denoting the lepton flavour decoherence temperature  by $T^{f_i}_{decoh}$ where $f_i$ stands for a specific lepton flavour $(e,\mu,\tau)$,
 it turns out that, above the temperature $T^\tau_{decoh}$, all three lepton flavours are indistinguishable and the Boltzmann equation has to be solved for the quantity $(B-L)$. Again the temperature
regime above $T^\tau_{decoh}$ is subdivided into three windows. The structure of $C^l$ and $C^\phi$ matrices are different in those windows since the number of active chemical potentials and 
the governing constraint equations are different in each of these
windows. These issues has been discussed elaborately in Sec.3.1 and Appendix B of \cite{Sierra:2014tqa}. Similarly between
$T^\tau_{decoh} ~ -~ T^\mu_{decoh}$ two lepton flavours $(a(\equiv e+\mu), \tau)$ are effectively active whereas below $T^\mu_{decoh}$, full flavour decoherence is achieved and all three lepton flavours 
are separately active. Thus the flavoured Boltzmann equation has to be
solved in terms of $(B/3-L_e,B/3-L_\mu,B/3-L_\tau)$.  The asymmetry
coupling matrices as stated in \cite{Sierra:2014tqa}
are shown in Table \ref{clcphi}.
\begin{table}[!h]
 \caption{$C^l$ and $C^\phi$ matrices in different temperature regimes}\label{clcphi}
 \begin{center}
  \begin{tabular}{|c|c|c|c|}
  \hline
   $T$ (GeV) & Flavours & $C^l$ & $C^\phi$\\
   \hline
    $\gtrsim 10^{15}$ & single & $\Big (0~\frac{1}{2} \Big )$ & $\Big (3~\frac{1}{2} \Big )$\\
    &  &  & \\
     $[10^{12},10^{15}]$ &  & $\Big (0~\frac{1}{2} \Big )$ & $\Big (2~\frac{1}{3} \Big )$ \\
     &  &  & \\
      $[T_{decoh}^\tau,10^{12}]$ &  & $\Big (0~\frac{3}{10} \Big )$ & $\Big (\frac{3}{4}~\frac{1}{8} \Big )$ \\
     &  &  & \\ 
    \hline
    $[10^9,T_{decoh}^\tau],$ &  two & $\left( \begin{array}{ccc}
              -\frac{6}{359} & \frac{307}{718} & -\frac{18}{359}\cr
              \frac{39}{359} & -\frac{21}{718}  & \frac{117}{359}\cr
             \end{array}\right)$ & $\Big (\frac{258}{359}~\frac{41}{359}~\frac{56}{359} \Big )$\\
     $[T_{decoh}^\mu,T_{decoh}^\tau]$ &  &  & \\
     &  &  & \\
     \hline
    $[10^5,T_{decoh}^\mu]$ &  three & $\left( \begin{array}{cccc}
              -\frac{6}{179} & \frac{151}{358} & -\frac{10}{179} &  -\frac{10}{179}\cr
              \frac{33}{358} & -\frac{25}{716}  & \frac{172}{537} & -\frac{7}{537}\cr
               \frac{33}{358} & -\frac{25}{716} & -\frac{7}{537} & \frac{172}{537}\cr
             \end{array}\right)$ & $\Big (\frac{123}{179}~\frac{37}{358}~\frac{26}{179} ~\frac{26}{179}\Big )$\\
       &  &  &  \\
       $\lesssim 10^{5}$ &  & $\left( \begin{array}{cccc}
              -\frac{9}{158} & \frac{221}{711} & -\frac{16}{711} &  -\frac{16}{711}\cr
              \frac{9}{158} & -\frac{16}{711}  & \frac{221}{711} & -\frac{16}{711}\cr
               \frac{9}{158} & -\frac{16}{711} & -\frac{16}{711} & \frac{221}{711}\cr
             \end{array}\right)$ &  $\Big (\frac{39}{79}~\frac{8}{79}~\frac{8}{79} ~\frac{8}{79}\Big )$\\ 
             & & & \\
             \hline      
  \end{tabular}
 \end{center}
\end{table}

\subsubsection{$Y$ Matrix in Type-II Seesaw Dominated Scenario} \label{ap4}
Using the up-quark diagonal basis, the Dirac neutrino mass matrix is
chosen to be equal to the up quark mass matrix \cite{pnsa:2016}
$M_D \simeq {\rm diag}( .00054, .26302, 81.99)$ GeV.\\\\
{\bf Normal Ordering (NO)}\\
Diagonal elements of light neutrino mass matrix for NO case is given by\\ 
$(m_\nu)_{diag}={\rm diag}(.001, .00865, .0502)\times10^{-9}$ GeV.\\ 
Then
using $V_{B-L}=4\times10^{17}$ GeV in eq.(\ref{t2_dom}) we get the lowest value of $Y$ needed for type-II seesaw dominance
\begin{eqnarray}
Y & = & \Big (\frac{10^{-16}}{4} \Big ){\rm GeV}^{-1} M_D^T (m_\nu)^{-1} M_D \nonumber\\
  & = & \Big (\frac{10^{-7}}{4} \Big ) \left(
\begin{array}{ccc}
 0.000207\, -4.45\times10^{-8}i & -0.02608\, +0.00207 i & 16.5879\, + 0.5502 i \\
 -0.02608\, +0.00207 i & 8.5984\, -0.5027 i & -3568.99\, + 95.89 i \\
 16.5879\, + 0.5502 i & -3568.99\, + 95.89 i & 2.017\times10^6\, + 86276.3 i \\
\end{array}
\right), \nonumber\\
\end{eqnarray}
which gives
\begin{eqnarray}
Tr[ Y Y^\dagger] & = & 2.54\times10^{-3}. \label{eq:yyNH}
\end{eqnarray}
 As this condition is in direct contradiction with  the 
fully 2-flavoured regime case for which  $Tr[Y Y^\dagger] \leq 6.6\times10^{-6}$, it is not possible to get points in the fully 2-flavoured regime in our Type-II seesaw dominated scenario.\\\\
{\bf Inverted Ordering (IO)}\\\\
Light neutrino mass eigen values for inverted ordering are  
$(m_\nu)_{diag}={\rm diag}(0.04938,0.0501,0.001)\times10^{-9}$~ GeV. 
Then
using $V_{B-L}= 10^{17}$ GeV in eq.(\ref{t2_dom}),  the lowest value
of $Y$ needed for Type-II seesaw dominance is
\begin{eqnarray}
Y & = &  10^{-16}{\rm GeV}^{-1} M_D^T (m_\nu)^{-1} M_D \nonumber\\
  & = & 10^{-7} \left(
\begin{array}{ccc}
 -7.78\times10^{-8} -3.09\times10^{-6} i & 0.0038-0.0159 i & 1.0118 -4.206 i \\
  0.0038-0.0159 i & 40.007+0.0099 i & 10174.3 +2.205 i \\
  1.0118 -4.206 i &  10174.3 +2.205 i & 2.81\times10^6 +469.504 i \\
\end{array}
\right), \nonumber\\
\end{eqnarray}
which gives 
\begin{eqnarray}
Tr[ Y Y^\dagger]  =  7.93\times10^{-2}~. \label{eq:yyIH} 
\end{eqnarray}
Thus for inverted ordering, as this value is far bigger than
 the two-flavored case, we conclude that desired values of BAU
 in  2-flavoured regime is not allowed in this case also. 

\subsection{Renormalization Group Solutions for Mass Scales and
  Threshold Effects}\label{sec:AppTh}
For the SM gauge group ($\equiv
G_{213}$)  we denote
the three respective gauge couplings by $g_{2L},g_{1Y}$ and $g_{3C}$
and the top quark Yukawa coupling by $y_{top}$. The renormalisation
group equations (RGEs) for gauge couplings are
\ba
\mu\frac{{\partial g}_i}{{\partial
    \mu}}&=&\frac{a_i}{16\pi^2}+\frac{1}{(16\pi^2)^2}\sum_j{g_i^3(b_{ij}g_j^2-k_i\delta_{ij}y_{top}^2)}, \label{eq:tl}
\ea 
where $a_i(b_{ij})$ are one (two)-loop beta function coefficients and $k_i=(17/6,3/2,2)$.
Details of RGEs for individual couplings  are 
\begin{align}
{dy_{top} \over d \ln \mu}= & {1 \over 16\pi^2}\left({9 \over 2}y_{top}^2-{17 \over 12}g_{1Y}^2
 -{9 \over 4}g_{2L}^2-8g_{3C}^2 \right)y_{top} \\ \nonumber
  +  & {1 \over (16\pi^2)^2} [-{23 \over 4}g_{2L}^4-{3 \over 4}g_{2L}^2g_{1Y}^2+{1187 \over 216}g_{1Y}^4 + 9g_{2L}^2g_{3C}^2+{19 \over 9}g_{3C}^2g_{1Y}^2-108g_{3C}^4 
\\ \nonumber
+& \left({225 \over 16}g_{2L}^2+{131 \over 16}g_{1Y}^2+36g_{3C}^2 \right)y_{top}^2+6(-2y_{top}^4-2y_{top}^2\lambda_{\phi}+\lambda_{\phi}^2) ], \\ 
 {dg_{1Y} \over d \ln \mu}= & {1 \over 16\pi^2}\left({41 \over 6}g_{1Y}^3\right)+{1 \over (16\pi^2)^2}\left({199 \over 18}g_{1Y}^2+{9 \over 2}g_{2L}^2+{44 \over 3}g_{3C}^2-
{17 \over 6}y_{top}^2\right)g_{1Y}^3,  \\
{dg_{2L} \over d \ln \mu}= & {1 \over 16\pi^2}\left(-{19\over 6}g_{2L}^3\right)+{1 \over (16\pi^2)^2}\left({3 \over 2}g_{1Y}^2+{35 \over 6}g_{2L}^2+12g_{3C}^2-
{3 \over 2}y_{top}^2\right)g_{2L}^3 , \\
{dg_{3C} \over d \ln \mu }= & {1 \over 16\pi^2}\left(-7g_{3C}^3\right)+{1 \over (16\pi^2)^2}\left({11 \over 6}g_{1Y}^2+{9 \over 2}g_{2L}^2-26g_{3C}^2-
2y_{top}^2\right)g_{3C}^3.\label{eq:RGSM}
\end{align}
\vspace{.25cm}
Defining the mass scale dependent fine-structure constants
$\alpha_i(\mu)=\frac{g_i^2(\mu)}{4\pi}(i=2L,Y,3C)$, we use the integral form of
eq.(\ref{eq:tl}) in different ranges of mass scales
\be
\frac{1}{\alpha_i(\mu)}=\frac{1}{\alpha_i(M_Z)}-\frac{a_i}{2\pi}\ln(\frac{\mu}{M_Z})+... , 
\ee
where  ellipses stand for possible two-loop and threshold effects. 
 Near the GUT scale $\mu \sim M_U$, the matching formulas for different gauge
 couplings($\alpha^{-1}_i,i=2L,Y,3C$) are defined as
 \cite{Weinberg:1980,Hall:1981,Ovrut:1981,mkp:1987,RNM-mkp:1993,lmpr:1995,Lang:1994}  
 \be
\alpha^{-1}_i (M_U)=\alpha^{-1}_G-\frac {\lambda_i(M_U)}{12\pi},\label{eq:matching1} 
\ee
 where  $\lambda_i,(i=2L,Y, 3C)$ are the three matching functions due to superheavy scalars (S), Majorana fermions (F) and gauge bosons (V),

\ba
\lambda_{i}^S(M_U) & =&\sum_{j}Tr\left(t_{iSj}^2\hat{p}_{Sj}\ln{M_j^S \over M_U}\right),\nonumber \\
 \lambda_{i}^F(M_U) & =&\sum_{k} 4 Tr\left(t_{iFk}^2\ln{M_k^F \over
  M_U}\right), \nonumber \\ 
\lambda_{i}^V(M_U) & =&\sum_{l} Tr\left(t_{iVl}^2\right)
-21\sum_lTr\left(t_{iVl}^2\ln{M_l^V \over M_U}\right),\label{eq:matching2}
\ea
Here $t_{iS}$, $t_{iF}$ and  $t_{iV}$ represent the matrix representations of 
broken generators for scalars, Majorana fermions, and gauge bosons, respectively. The term $\hat{p}_{Sj}$ denotes the projection operator  that removes the Goldstone components from the scalar that contributes to spontaneous symmetry breaking. 

We note that only the  SU(5) remnants
${75}_H,{50}_H,{24}_H,{15}_H,{5}_H, {\rm and} ~{24}_F$ of their
respective parent SO(10) representations
 ${210}_H,{126}_H,{45}_H,{126}_H,{10}_H,{\rm and} ~{45}_F$ are either
near $M_{SU(5)}$ or their SM sub-multiplets are below $M_{SU(5)}$. Other components of respective SO(10)
representations being at much higher mass scale $\mu=M_{SO(10)}\simeq
100 M_{SU(5)}$ decouple from all the predictions of this work. 
Decomposition of these SU(5) representations  under SM gauge group $G_{213}$  are given in Table.{\ref{tab:decomp}}. 
\begin{table}[h!]
\caption{Superheavy components of  
SU(5) representations under the SM gauge group $G_{213}$ used to
estimate GUT threshold effects. }
\vskip 0.5cm
\begin{tabular}{| p{13 cm} | }
\hline
 $5_H \supset H_1(1,-1/3,3) $ 
 \\ \hline 
$24_H \supset S_1(3,0,1)+S_2(1,0,8)$ 
 \\ \hline  
$ 50_H  \supset  H_1^{\prime}(1,-1/3,3)+H_2^{\prime}(1,-2/3,6)+H_3^{\prime}(2,-1/6,\bar{3})+H_4^{\prime}(2,-1/2,8)$ 
\\ \hline
 $75_H  \supset S_1^{\prime}(1,5/3,3)+S_2^{\prime}(1,-5/3,\bar{3})+S_3^{\prime}(1,0,8)+S_4^{\prime}(2,5/6,6)+S_5^{\prime}(2,-5/6,\bar{6})$
\\ \hline 
 $15_H \supset H_1^{\prime\prime}(3,-1,1)+H_2^{\prime\prime}(2,1/6,3)+H_3^{\prime\prime}(1,2/3,6)$    
\\ \hline 
 $24_F \supset  F_1(2,-5/6,3)+ F_2(2,5/6,\bar{3})$ 
\\ \hline  
 $24_V \supset  V_1(2,-5/6,3)+ V_2(2,5/6,\bar{3})$ 
\\ \hline  
\end{tabular}
\label{tab:decomp}
\end{table}


We have used Table.{\ref{tab:decomp}} and utilized appropriate decompositions
of SU(5) representations ${75}_H,{24}_H,{50}_H,{5}_H, {24}_F,{24}_V$ in the
respective models  to estimate threshold
effects. Different scalar(H), fermionic(F), and gauge
boson(V) components and their respective beta function coefficients
contributing to these matching functions are estimated as shown in
Table {\ref{tab:bth}}. The constant factor in the gauge-boson threshold
effect has been already taken into account in Sec.\ref{sec:kgu}. 

\begin{table}[!h]
\caption{Beta function coefficients for matching functions defined through
eq.(\ref{eq:matching2}) due to
  superheavy components of scalars(H). fermions (F), and gauge bosons (V)
  contributing to GUT threshold effects at the SU(5) unification scale.}
\begin{center}
\begin{tabular}{|c|c|c|}
\hline
{SU(5) representations } & {G213 submultiplets} &{($\lambda_{2L}$,$\lambda_{Y}$,$\lambda_{3C}$)}\\
\hline
 $5_H$ & $H_1(1,-1/3,3)$ & $(0,2/5,1)$ \\
\hline
$24_H $ & $S_1(3,0,1) $ & $(2,0,0) $ \\
& $S_2(1,0,8) $ & $(0,0,3) $\\
\hline
$50_H $ & $H_1^{\prime}(1,-1/3,3) $ & $(0,2/5,1) $ \\
& $H_2^{\prime}(1,-2/3,6) $ & $(0,16/5,5) $\\
& $H_3^{\prime}(1,-1/6,\overline{3}) $ & $(3,1/5,2) $\\
& $H_4^{\prime}(2,-1/2,8) $ & $(8,24/5,12) $\\
\hline
$75_H $ & $S_1^{\prime}(1,5/3,3) $ & $(0,5,1/2) $ \\
& $S_2^{\prime}(1,-5/3,\overline{3}) $ & $(0,5,1/2) $\\
& $S_3^{\prime}(1,0,8) $ & $(0,0,3) $\\
& $S_4^{\prime}(2,5/6,6) $ & $(3,5,5) $\\
& $S_5^{\prime}(2,-5/6,\overline{6}) $ & $(3,5,5) $\\
\hline
$15_H $ & $H_1^{\prime\prime}(3,-1,1) $ & $(4,18/5,0) $ \\
& $H_2^{\prime\prime}(2,1/6,3) $ & $(3,1/5,2) $\\
& $H_3^{\prime\prime}(1,2/3,6) $ & $(0,16/5,5) $\\
\hline
$24_F $ & $F_1(2,-5/6,3) $ & $(6,10,4) $ \\
& $F_2(2,5/6,\overline{3}) $ & $(6,10,4) $\\
\hline
$24_V $ & $V_1(2,-5/6,3) $ & $(-63/2,-105/2,-21) $ \\
& $V_2(2,5/6,\overline{3}) $ & $(-63/2,-105/2,-21) $\\
\hline
\end{tabular}
\end{center}
\label{tab:bth}
\end{table}
   
As we find below the effect of the full
representation ${15}_H$ of SU(5) surfaces in the GUT coupling value but cancels out from
mass scale predictions in all the three models.  

\subsubsection{Minimal Model-I}
We have used the most recent  electroweak
precision data \cite{PDG:2016}
\ba   
\alpha_S(M_Z)&=& 0.1182\pm 0.0005, \, \nonumber\\
\sin^2\theta_W (M_Z)&=&0.23129\pm 0.00005,\,\nonumber\\
\alpha^{-1}(M_Z)&=& 127.94 \pm 0.02.\, \label{eq:inputpara}
\ea
Using RGEs and  the combinations ${1 \over \alpha(M_Z)}-{8 \over 3}{1
  \over \alpha_{2L}(M_Z)}$ and  ~~ ${1 \over \alpha(M_Z)}-{8 \over
  3}{1 \over \alpha_{3C}(M_Z)}$, we have derived analytic formulas for
the SU(5) unification scale ($M_U$)  and the intermediate
$\kappa(3,0,8)$ mass scale ($M_{\kappa}$) treating  the heavy triplet
scalar mass scale($M_{\Delta}$) as constant. We also analytically
estimate the SU(5) GUT fine-structure constant
$\alpha_G=g_G^2/{(4\pi)}$.
The beta function coefficients in three different mass ranges
 $\mu=M_{Z}\to M_{\kappa}$, $\mu= M_{\kappa}-M_{\Delta}$ and $\mu=M_{\Delta}-M_U$ are 
\par\noindent{\large\bf {\underline {$\mu = M_{Z} \to M_{\kappa}:$}}}\\ 
\begin{equation}
 a_Y=\frac{41}{10},\,\, a_{2L}=-\frac{19}{6},\,\, a_{3C}=-7, \label{eq:smai}
\end{equation}
\par\noindent{\large\bf {\underline {$\mu=  M_{\kappa}\to M_{\Delta}:$}}}\\
\begin{equation}
 a_Y^{\prime}=\frac{41}{10},\,\,a_{2L}^{\prime}=-\frac{1}{2},\,\,
 a_{3C}^{\prime}= -\frac{11}{2},
\end{equation}
\par\noindent{\large\bf {\underline {$\mu= M_{\Delta}\to M_{U}:$}}}\\ 
\begin{equation}
 a_Y^{\prime\prime}=\frac{79}{15},\,\, a_{2L}^{\prime\prime}=\frac{2}{3},\,\,
  a_{3C}^{\prime\prime}=-{13 \over 3}.
\end{equation}
Using the standard  procedure \cite{RNM-mkp:1993}  we get
\begin{align}
\ln {M_U \over M_Z} &= \frac{16\pi}{187\alpha}\left({7\over8}-\frac{10\alpha}
{3\alpha_{3C}}+s_W^2 \right)+\Delta_{I}^U, \nonumber\\
\ln {M_{\kappa}  \over M_Z}& = \frac{4\pi}{187\alpha}\left(15+\frac{23\alpha}
{3\alpha_{3C}}-63s_W^2 \right)+\Delta_{I}^{\kappa},  \nonumber\\
{1 \over \alpha_G}& = {3 \over 8\alpha}+{1 \over 187\alpha}\left({347 \over 8}
 +{466\alpha \over 3\alpha_{3C}}-271s_W^2 \right)+\frac{7}{12\pi}\ln
 {M_{\Delta}\over M_Z}
+\Delta_{I}^{\alpha_G}. \label{eq:guform1} 
\end{align}
We note that the effect of $M_{\Delta}$ naturally cancels from one
loop analytic expressions for $\ln {M_U \over M_Z}$ and $\ln
{M_{\kappa} \over M_Z}$. 
Analytic formulas for GUT threshold effects on the unification scale, intermediate scale and GUT fine structure constant are  

\begin{align}
\Delta\ln{M_U \over M_Z}& ={5\over 3366}(7\lambda_Y+9\lambda_{2L}-16\lambda_{3C}) \nonumber\\
\Delta\ln{M_{\kappa}\over M_Z}& ={1\over 561}(-48\lambda_{2L}+25\lambda_Y+23\lambda_{3C}) \nonumber \\
\Delta({1 \over \alpha_G})&={1\over 20196\pi}(1130\lambda_Y-675\lambda_{2L}+ 1264\lambda_{3C}), 
\end{align}
Estimated threshold effects on different mass scales due to superheavy masses 
 near the SU(5) unification scale are 
\begin{align}
\Delta\ln{M_{\kappa} \over M_Z}& = 0.0588235\eta_{5}+0.951872\eta_{75} \nonumber \\
\Delta\ln{M_U \over M_Z}& = -0.0196078\eta_{5}-0.0445633\eta_{75} \nonumber \\
\Delta({1 \over \alpha_G})& = 0.0270459\eta_{5}+0.571275\eta_{75}
+0.189653\eta_{15} \nonumber \\
\end{align}
 Denoting $\eta_{S}=\ln ({M_{S} \over M_U})$ and maximising uncertainty in $M_U$ gives 
\ba
\Delta \ln ({M_U \over M_Z})&=&\pm 0.06417\eta_{S} ,~~\nonumber\\
\Delta \ln ({M_{\kappa} \over M_Z})&=& \pm 1.0107\eta_{S} , ~~\nonumber\\
\Delta \left({1 \over \alpha_G}\right)&=&\pm 0.408667\eta_{S}.
\ea 
We also note that the degenerate superheavy gauge bosons contribute
quite significantly to threshold 
  correction on the unification scale \cite{Hall:1981,mkp:1987,RNM-mkp:1993}
\begin{equation}
 \left(\frac{M_U}{M_U^0}\right)_{V}=10^{\pm 0.9358\eta^{\prime}_V}.
\end{equation} 
 Adding all corrections together we obtain
\begin{equation}
M_U=10^{15.24\pm 0.11\pm 0.0642\eta^{\prime}_S\pm
  0.9358\eta^{\prime}_V} {\rm GeV}, \label{MUtotal1}
\end{equation}   
where $\eta_i^{\prime}(i=S,V)=\log_{10}(M_i/M_U)$.
The first uncertainty appearing in the exponent as $\pm 0.11$ is 
 due to input
parameters of eq.(\ref{eq:inputpara}).

\subsubsection{Minimal Model-II} 
The beta function coefficients in two different mass ranges
 $\mu=M_{Z}\to M_{\eta}$, $\mu= M_{\eta}-M_{\Delta}$ and $\mu=M_{\Delta}-M_U$ are 
\par\noindent{\large\bf {\underline {$\mu = M_{Z} \to M_{\eta}:$}}}\\ 
\begin{equation}
 a_Y=\frac{41}{10},\,\, a_{2L}=-\frac{19}{6},\,\, a_{3C}=-7, 
\end{equation}
\par\noindent{\large\bf {\underline {$\mu=  M_{\eta}\to M_{\Delta}:$}}}\\
\begin{equation}
 a_Y^{\prime}=\frac{45}{10},\,\,a_{2L}^{\prime}=-\frac{5}{6},\,\, a_{3C}^{\prime}= -\frac{9}{2},
\end{equation}
\par\noindent{\large\bf {\underline {$\mu= M_{\Delta}\to M_{U}:$}}}\\ 
\begin{equation}
 a_Y^{\prime\prime}=\frac{17}{3},\,\, a_{2L}^{\prime\prime}=2,\,\,
  a_{3C}^{\prime\prime}=-{10 \over 3}.
\end{equation}
Using the standard procedure  we get
analytic formulas for the two mass scales $M_{\rm U}$ and $M_{\eta}$ 
\begin{align}
\ln {M_U \over M_Z} &=\frac{18\pi}{247\alpha}\left(1+{4 \over 3} s_W^2-4{\alpha \over \alpha_{3C}} \right)+\Delta_{II}^U, \nonumber\\
\ln {M_{\eta}  \over M_Z}& =\frac{4\pi}{247\alpha}\left(16+\frac{55}{3}\frac{\alpha}{\alpha_{3C}}-61s_W^2 \right)+\Delta_{II}^{\eta}, \nonumber\\
{1 \over \alpha_G}& ={1 \over 494\alpha}\left(241-502s_W^2+{1060 \over
  3}{\alpha \over \alpha_{3C}}\right)
+\frac{7}{12\pi}\ln
 {M_{\Delta}\over M_Z}
+\Delta_{II}^{\alpha}. \label{eq:guformnm} 
\end{align}
The effect of mass scale $M_{\Delta}$ is noted to cancel out from the expressions of 
$\ln(M_U/M_Z)$ and $\ln(M_{\eta}/ M_Z)$.

Excellent unification of gauge couplings is found for
\ba
M^0_U  &=& 10^{15.248 +0.0445}\, {\rm GeV},  \nonumber\\
M^0_{\sigma} &=& 10^{3.0}\, {\rm GeV}, \nonumber\\
\alpha^{-1}_{G_0} &=& 33.78. \, \label{eq:massscaleseta}  
\ea
 Analytic formulas for GUT threshold effects on the unification scale, intermediate scale and GUT fine structure constant are  
\begin{align}
\Delta\ln{M_{\eta}\over M_Z}& ={5\over 2223}(27\lambda_{2L}-16\lambda_Y-11\lambda_{3C}), \nonumber \\
\Delta\ln{M_U \over M_Z}& ={1\over 494}(5\lambda_Y+7\lambda_{2L}-12\lambda_{3C}), \nonumber\\
\Delta({1 \over \alpha_G})&={1\over 17784\pi}(1205\lambda_Y-783\lambda_{2L}+ 1060\lambda_{3C}).
\end{align}
Estimated threshold corrections for different mass scales due to superheavy masses 
 are 
\begin{align}
\Delta\ln{M_{\eta} \over M_Z}& = -0.0391363\eta_{5}+0.0472335\eta_{24}
-0.136302\eta_{50}, \nonumber \\
\Delta\ln{M_U \over M_Z}& = -0.02024\eta_{5}-0.0445344\eta_{24}
-0.312551\eta_{50}, \nonumber \\
\Delta({1 \over \alpha_G})& = 0.00862716\eta_{5}+0.0288884478\eta_{24}
+0.185680767\eta_{15}+0.4107744\eta_{50}. \nonumber \\
\end{align}
 Denoting $\eta_{S}=\ln ({M_{S} \over M_U})$ and maximising  uncertainty in $M_U$ leads to  
\ba
\Delta \ln ({M_U \over M_Z})&=&\pm 0.3773\eta_{S} ,~~\nonumber\\
\Delta \ln ({M_{\eta} \over M_Z})&=& \pm 0.1279\eta_{S} , ~~\nonumber\\
\Delta \left({1 \over \alpha_G}\right)&=&\pm 0.2626\eta_{S}.
\ea  
We also note that the degenerate superheavy gauge bosons contribute
quite significantly to threshold corrections on the unification scale \cite{Hall:1981,mkp:1987,RNM-mkp:1993}
\begin{equation}
 \left(\frac{M_U}{M_U^0}\right)_{V}=10^{\pm 0.9352\eta^{\prime}_V}.
\end{equation} 
 Adding all corrections together we obtain
\begin{equation}
M_U=10^{15.28\pm 0.1334\pm 0.3773\eta^{\prime}_S\pm 0.9352\eta^{\prime}_V} {\rm GeV} \label{MUtotal2}
\end{equation}   
The first uncertainty appearing in the exponent as $\pm 0.1334$ is due
to input parameters of eq.(\ref{eq:inputpara}).
\subsubsection{Triplet Fermion Dark Matter Model-III}  
In this model, instead of type-I seesaw ansatz for neutrino masses in
\cite{Frig-Ham:2010}, we have implemented type-II seesaw dominance.
The one-loop beta function coefficients  in three different mass ranges
 $\mu=M_{Z}\to M_{\Sigma}$, $\mu= M_{\Sigma}-M_{O}$, $\mu= M_{O}-M_{\Delta}$ and $\mu=M_{\Delta}-M_U$ are 
\par\noindent{\large\bf {\underline {$\mu = M_{Z} \to M_{\Sigma}:$}}}\\ 
\begin{equation}
 a_Y=\frac{41}{10},\,\, a_{2L}=-\frac{19}{6},\,\, a_{3C}=-7, 
\end{equation}
\par\noindent{\large\bf {\underline {$\mu=  M_{\Sigma}\to M_{O}:$}}}\\
\begin{equation}
 a_Y^{\prime}=\frac{41}{10},\,\,a_{2L}^{\prime}=-\frac{11}{6},\,\, a_{3C}^{\prime}= -7,
\end{equation}
\par\noindent{\large\bf {\underline {$\mu=  M_{O}\to M_{\Delta}:$}}}\\
\begin{equation}
 a_Y^{\prime}=\frac{41}{10},\,\,a_{2L}^{\prime}=-\frac{11}{6},\,\, a_{3C}^{\prime}= -5,
\end{equation}
\par\noindent{\large\bf {\underline {$\mu= M_{\Delta}\to M_{U}:$}}}\\ 
\begin{equation}
 a_Y^{\prime\prime}=\frac{79}{15},\,\, a_{2L}^{\prime\prime}=-\frac{2}{3},\,\,
  a_{3C}^{\prime\prime}=-{23 \over 6}.
\end{equation}
Following the standard procedure we get
\begin{align}
\ln {M_U \over M_Z} &= \frac{12\pi}{91\alpha}\left(1-\frac{5\alpha}
{3\alpha_{3C}}-s_W^2 \right)-\frac{20}{91}\ln{M_O \over M_Z}+\Delta_U, \nonumber\\
\ln {M_{\Sigma}  \over M_Z}& = \frac{3\pi}{182\alpha}\left(19+\frac{178\alpha}
{3\alpha_{3C}}-110s_W^2 \right)+\frac{89}{91}\ln{M_O \over M_Z} +\Delta_{\Sigma},  \nonumber\\
{1 \over \alpha_{G}}& = {23 \over 91\alpha}\left(1 +{158\alpha \over 69\alpha_{3C}}-s_W^2\right)+\frac{158}{273\pi}\ln{M_O \over M_Z}
+\frac{7}{12\pi}\ln{M_{\Delta} \over M_Z} +\Delta_{\alpha_G}, \label{eq:guform2} 
\end{align}
Analytic formulas for GUT threshold effects are
\begin{align}
\Delta_{\Sigma}&=\Delta\ln{M_{\Sigma}\over M_Z}={1\over 2184}(-273\lambda_{2L}+95\lambda_Y+178\lambda_{3C}), \nonumber \\
\Delta_U&=\Delta\ln{M_U \over M_Z} ={5\over 273}(\lambda_Y
-\lambda_{3C}), \nonumber\\
\Delta_{\alpha_G}&=\Delta({1 \over \alpha_G})={1\over 3276\pi}(115\lambda_Y+ 158\lambda_{3C}). 
\end{align}
Estimated threshold corrections on different mass scales due to superheavy masses 
 are 
\begin{align}
\Delta\ln{M_{\Sigma} \over M_Z}& = 0.0989\eta_{5}-0.005495\eta_{24}
-0.81318\eta_{24_F}, \nonumber \\
\Delta\ln{M_U \over M_Z}& = -0.010989\eta_{5}-0.054945\eta_{24}
-0.131868\eta_{24_F}, \nonumber \\
\Delta({1 \over \alpha_G})& = 0.01982\eta_{5}+0.04605\eta_{24}
+0.18568\eta_{15}+0.131754\eta_{24_F}. \nonumber \\
\end{align}
 
Maximising the uncertainty in $M_U$ leads to  ~~~~~~~~
\ba
\Delta \ln ({M_U \over M_Z})&=&\pm 0.065934\eta_{S}+\pm 0.131868\eta_{F} ,~~\nonumber\\
\Delta \ln ({M_{\Sigma} \over M_Z})&=& \pm 0.0934\eta_{S}+\pm 0.81318\eta_{F} , ~~\nonumber\\
\Delta \left({1 \over \alpha_G}\right)&=&\pm 0.11981\eta_{S}+\pm 0.13175\eta_{F},
\ea 
where $\eta_{S}=\ln ({M_{S} \over M_U})$,~~$\eta_{F}=\ln ({M_{F} \over M_U})$ . 

Degenerate superheavy gauge bosons are also noted to contribute
quite significantly to threshold effects on $M_U$ \cite{Hall:1981,mkp:1987,RNM-mkp:1993}
\begin{equation}
 \left(\frac{M_U}{M_U^0}\right)_{V}=10^{\pm 1.1538\eta^{\prime}_V},
\end{equation} 
where $\eta^{\prime}_V=\log_{10}({M_V \over M_U})$.\\
 Adding all corrections together we obtain
\begin{equation}
M_U=10^{15.31\pm 0.03\pm 0.0659\eta^{\prime}_S\pm 1.1538\eta^{\prime}_V\pm 0.13186\eta^{\prime}_F} {\rm GeV}. \label{MUtotal3}
\end{equation}   
The first uncertainty  $\pm 0.03$ is due to
input parameters in eq.(\ref{eq:inputpara}).
 Expression for $\Delta_U$ has been also reported in Sec.3 under Model
 III. We note that the effects of mass scale $M_{\Delta}$ for the
 complete multiplet ${15}_H$ cancels out from one loop effects.
Possibility of  D- Parity
broken TeV scale
left-right-gauge theory \cite{cmp-PRL:1984} and $SU(2)_L-$ fermionic
triplet DM with a stabilising $Z_2$ symmetry external to non-SUSY SO(10)  have been  discussed in \cite{RNM:2015}.  

\subsection{Renormalization Corrections for ${\rm Dim.} 6$ Operator}\label{sec:dim6}
We now provide  analytic formulas for short
distance ($A_{SR}$) and long distance ($A_{LR}$) renormalisation
factors occurring in the proton lifetime formula  discussed in
Sec.\ref{sec:taup}  and their numerical values as shown in 
Table \ref{tab:SRLR}.
 The long distance renormalisation factor $A_L=A_{LR}$  represents evolution
 of the dim.6 operator for mass scales $\mu=Q=2.3$ GeV to
 $\mu=M_{Z}$. 
\be
A_L=A_{LR}=\left(\frac{\alpha_3(M_b)}{\alpha_3(M_{Z})}\right)^{6
  \over 23}\left(\frac{\alpha_3(Q)}{\alpha_3(M_{b})}\right)^{6 \over
  25}. \label{AL} 
\ee
Numerically we find $A_L=A_{LR}=1.15$.

Analytic expressions for  ($A_{SR}$) and  numerical
 values of the  product $A=A_{LR}A_{SR}$ are presented for each model  
 in  Table \ref{tab:SRLR}.
 
\begin{table}[!h]
\caption{Estimation of short distance renormalisation factor ($A_{SR}$) and the product
  $A=A_{SR}A_{LR}$  where $A_{LR}=1.15$.}
\begin{center}
\begin{tabular}{|c|c|c|c|}
\hline
{Models } & { Formulas for $A_{SR}$} &{$A_{SR}$} & {$A$} \\
\hline
 Model-I & $\left(\frac{\alpha_3(M_Z)}{\alpha_3(M_{\kappa})}\right)^{-2 \over a_3}\left(\frac{\alpha_3(M_{\kappa})}{\alpha_3(M_{\Delta})}\right)^{-2 \over a_3^{\prime}}\left(\frac{\alpha_3(M_{\Delta})}{\alpha_3(M_{U})}\right)^{-2 \over a_3^{\prime\prime}} $ & $2.238$ &$2.576$ \\
&$\times \left(\frac{\alpha_2(M_Z)}{\alpha_2(M_{\kappa})}\right)^{-9 \over 4a_2}\left(\frac{\alpha_2(M_{\kappa})}{\alpha_2(M_{\Delta})}\right)^{-9 \over 4a_2^{\prime}}\left(\frac{\alpha_2(M_{\Delta})}{\alpha_2(M_{U})}\right)^{-9 \over 4a_2^{\prime\prime}} $ & & \\
&$\times \left(\frac{\alpha_1(M_Z)}{\alpha_1(M_{\kappa})}\right)^{-11 \over 20a_1}\left(\frac{\alpha_1(M_{\kappa})}{\alpha_1(M_{\Delta})}\right)^{-11\over 20a_1^{\prime}}\left(\frac{\alpha_1(M_{\Delta})}{\alpha_1(M_{U})}\right)^{-11 \over 20a_1^{\prime\prime}} $ & & \\
\hline
 Model-II & $\left(\frac{\alpha_3(M_Z)}{\alpha_3(M_{\eta})}\right)^{-2 \over a_3}\left(\frac{\alpha_3(M_{\eta})}{\alpha_3(M_{\Delta})}\right)^{-2 \over a_3^{\prime}}\left(\frac{\alpha_3(M_{\Delta})}{\alpha_3(M_{U})}\right)^{-2 \over a_3^{\prime\prime}} $ & $2.17$ & $2.495$ \\
&$\times \left(\frac{\alpha_2(M_Z)}{\alpha_2(M_{\eta})}\right)^{-9 \over 4a_2}\left(\frac{\alpha_2(M_{\eta})}{\alpha_2(M_{\Delta})}\right)^{-9 \over 4a_2^{\prime}}\left(\frac{\alpha_2(M_{\Delta})}{\alpha_2(M_{U})}\right)^{-9 \over 4a_2^{\prime\prime}} $ & & \\
&$\times \left(\frac{\alpha_1(M_Z)}{\alpha_1(M_{\eta})}\right)^{-11 \over 20a_1}\left(\frac{\alpha_1(M_{\eta})}{\alpha_1(M_{\Delta})}\right)^{-11\over 20a_1^{\prime}}\left(\frac{\alpha_1(M_{\Delta})}{\alpha_1(M_{U})}\right)^{-11 \over 20a_1^{\prime\prime}} $ & &   \\
\hline
Model-III & $\left(\frac{\alpha_3(M_Z)}{\alpha_3(M_{\sigma})}\right)^{-2 \over a_3}\left(\frac{\alpha_3(M_{\sigma})}{\alpha_3(M_{O})}\right)^{-2 \over a_3^{\prime}}\left(\frac{\alpha_3(M_{O})}{\alpha_3(M_{\Delta})}\right)^{-2 \over a_3^{\prime\prime}}\left(\frac{\alpha_3(M_{\Delta})}{\alpha_3(M_{U})}\right)^{-2 \over a_3^{\prime\prime\prime}} $ & $1.78$ & $2.05$ \\
&$ \times \left(\frac{\alpha_2(M_Z)}{\alpha_2(M_{\sigma})}\right)^{-4 \over 9a_2}\left(\frac{\alpha_2(M_{\sigma})}{\alpha_2(M_{O})}\right)^{-4 \over 9a_2^{\prime}}\left(\frac{\alpha_2(M_{O})}{\alpha_2(M_{\Delta})}\right)^{-4 \over 9a_2^{\prime\prime}}\left(\frac{\alpha_2(M_{\Delta})}{\alpha_2(M_{U})}\right)^{-4 \over 9a_2^{\prime\prime\prime}} $  & &\\
&$\times \left(\frac{\alpha_1(M_Z)}{\alpha_1(M_{\sigma})}\right)^{-11 \over 20a_1}\left(\frac{\alpha_1(M_{\sigma})}{\alpha_1(M_{O})}\right)^{-11 \over 20a_1^{\prime}}\left(\frac{\alpha_1(M_{O})}{\alpha_1(M_{\Delta})}\right)^{-11 \over 20a_1^{\prime\prime}}\left(\frac{\alpha_1(M_{\Delta})}{\alpha_1(M_{U})}\right)^{-11 \over 20a_1^{\prime\prime\prime}} $
 & &   \\
\hline
\end{tabular}
\end{center}
\label{tab:SRLR}
\end{table}

It would be interesting to address the question of charged fermion
masses in these models consistent with \cite{GJ:1979} and its variants
\cite{Altarelli:2011,Joshipura:2011}.\\
\vskip .3cm
\par\noindent{\bf ACKNOWLEDGMENT}\\
M. K. P. acknowledges financial support under the project 
SB/S2/HEP-011/2013 awarded by the Department of Science and Technology,
Government of India. M. C. acknowledges the award of a
Post-Doctoral Fellowship and B. S. acknowledges a Ph.D. Scholarship  by Siksha 'O' Ausandhan, Deemed to be
University.

\end{document}